\documentclass[12pt,a4paper,twoside]{book}
\usepackage[utf8]{inputenc}
\usepackage{setspace}
\usepackage{amsmath,amsfonts,amssymb}
\usepackage{cite}
\usepackage{algorithm,algpseudocode}
\usepackage{setspace}
\usepackage{tikz}
\usepackage[bottom]{footmisc}
\usepackage[a4paper,
twoside,
bindingoffset=1cm,
inner=2.5cm,
outer=2cm,
top=3.5cm,
bottom=3.5cm,
headsep=1cm
]{geometry}

\usepackage{charter}
\usepackage{tcolorbox}
\usepackage{mathtools}
\usepackage[sort,numbers]{natbib}
\usepackage{notoccite}
\usepackage{booktabs,caption,multicol,hhline,tikz,multirow,array}
\usetikzlibrary{arrows.meta}
\usepackage{colortbl}
\usepackage{arydshln}
\usepackage{url}
\usepackage{subcaption}
\usepackage{caption}
\usepackage{graphicx, multirow, array,amsmath}
\usepackage{tabularx}
\usepackage{hyperref}
\usepackage{enumitem}
\usepackage{slashed}
\usepackage[acronym, nopostdot]{glossaries}
\usepackage{tikz}
\newcolumntype{M}[1]{>{\centering\arraybackslash}m{#1}}
\newcolumntype{N}{@{}m{0pt}@{}}
\usepackage{fancyhdr}
\fancypagestyle{plain}{%
	\fancyhead{}%
	\fancyfoot[C]{\bfseries\thepage}}
\fancypagestyle{mainmatter}{%
	\fancyhf{}
	\fancyhead[LE,RO]{\bfseries\thepage}
	\fancyhead[LO]{\bfseries\nouppercase\rightmark}
	\fancyhead[RE]{\bfseries\nouppercase\leftmark}}
\fancypagestyle{frontmatter}{%
	\fancyhead{}%
	\fancyfoot[C]{\bfseries\thepage}}

\makeglossaries

\begin{document}
%\UseRawInputEncoding

\setstretch{1.1}

%\nobibliography*
%\newcounter{cont}

% Title

\thispagestyle{empty}

\begin{center}
%	\vfill
	\vspace*{0.25cm}
	
	\begin{spacing}{2}
		{\Huge \textbf{Holographic neutron stars at finite temperature}}
	\end{spacing}
	
	\vspace*{2cm}
	
	{\Large PhD Thesis}
	
%	\vspace{2.5cm}
	\vfill
	
	{\Large Tobías Canavesi}
	
	\vspace{0.25cm}
	
	{\Large Doctoral advisor: Nicolás Grandi}
	
%	\vspace{2cm}
	\vfill
	
	{\large  La Plata Physics Institute - IFLP}
	
	\vspace{0.25cm}
	
	{\large Faculty of Sciences}
	
	\vspace{0.25cm}
	
	{\large University of La Plata}
	
%	\vspace{4cm}
	\vfill
	
	\includegraphics[width=0.3\linewidth]{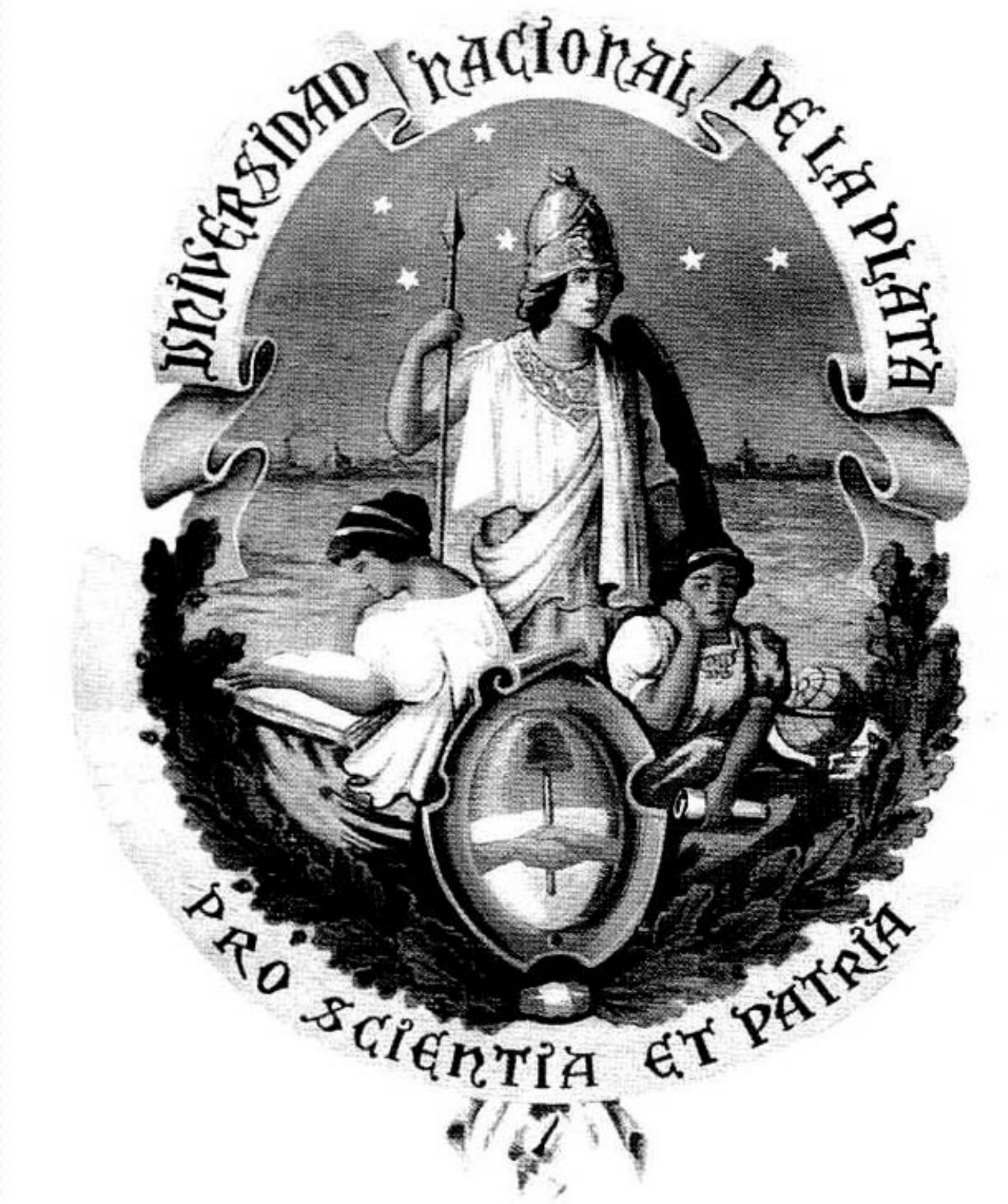}
	
	\vfill
	
	{\large La Plata
		
		 2023}
	
\end{center}

\newpage
\thispagestyle{empty}
\mbox{}
% \newpage
% \thispagestyle{empty}
% \begin{flushright}
% \parbox[t]{10cm}{\textit{"Scientists study the world as it is,}\\ \textit{\hphantom{"}engineers  create the world that never has been."}} \\
% \vspace{1em}
% \parbox[t]{5cm}{\small(Kármán Tódor)}
% \end{flushright}
 %%%%%%%%%%%%%%%%%%%%%%%%%%%%%%%%%%%%%%%%%%%%%%%%%%%%%%%%%%%%%%%%%%%
% sugawara's macro
%%%%%%%%%%%%%%%%%%%%%%%%%%%%%%%%%%%%%%%%%%%%%%%%%%%%%%%%%%%%%%%%%%%
\newcommand{\Om}{\Omega}
\newcommand{\df}{\stackrel{\rm def}{=}}
\newcommand{\co}{{\scriptstyle \circ}}
\newcommand{\de}{\delta}
\newcommand{\lb}{\lbrack}
\newcommand{\rb}{\rbrack}
\newcommand{\rn}[1]{\romannumeral #1}
\newcommand{\msc}[1]{\mbox{\scriptsize #1}}
\newcommand{\dsp}{\displaystyle}
\newcommand{\scs}[1]{{\scriptstyle #1}}

\newcommand{\ket}[1]{| #1 \rangle}
\newcommand{\bra}[1]{| #1 \langle}
\newcommand{\vac}{| \mbox{vac} \rangle }

\newcommand{\e}{\mbox{{\bf e}}}
\newcommand{\va}{\mbox{{\bf a}}}
\newcommand{\bc}{\mbox{{\bf C}}}
\newcommand{\br}{\mbox{{\bf R}}}
\newcommand{\bz}{\mbox{{\bf Z}}}
\newcommand{\bq}{\mbox{{\bf Q}}}
\newcommand{\bn}{\mbox{{\bf N}}}
\newcommand {\eqn}[1]{(\ref{#1})}

\newcommand{\cp}{\mbox{{\bf P}}^1}
\newcommand{\n}{\mbox{{\bf n}}}
\newcommand{\sbz}{\msc{{\bf Z}}}
\newcommand{\sn}{\msc{{\bf n}}}

\newcommand{\be}{\begin{equation}}\newcommand{\ee}{\end{equation}}
\newcommand{\bea}{\begin{eqnarray}} \newcommand{\eea}{\end{eqnarray}}
\newcommand{\ba}[1]{\begin{array}{#1}} \newcommand{\ea}{\end{array}}

\newcommand{\sign}{\text{sgn}}

\newcommand{\N}{\mathbb{N}} 
\newcommand{\Q}{\mathbb{Q}} 
\newcommand{\R}{\mathbb{R}} 
\newcommand{\Z}{\mathbb{Z}}

\newcommand{\cleqn}{\setcounter{equation}{0}}
\makeatletter
\@addtoreset{equation}{section}
\def\theequation{\thesection.\arabic{equation}}
\makeatother

\def\np{Nucl. Phys. {\bf B}}
\def\pl{Phys. Lett. {\bf B}}
\def\mpl{Mod. Phys. {\bf A}}
\def\ijmp{Int. J. Mod. Phys. {\bf A}}
\def\cmp{Comm. Math. Phys.}
\def\prd{Phys. Rev. {\bf D}}

\def\ds{dS_{d_-,d_+}}
\def\ads{AdS_{d_-,d_+}}
\def\min{{\cal M}_{d_- +1,d_+}}
\def\ds{dS_{d_-,d_+}}
%% OJO!!!
\def\g{G}

\def\va{\vec a}
\def\vb{\vec b}
\def\vu{\vec u}
\def\vv{\vec v}
\def\vt{\vec t}
\def\vn{\vec n}
\def\ve{\vec e}
\def\vx{{\vec x}}
\def\vxM{{\vec x}_{+}}
\def\vxm{{\vec x}_{-}}
\def\vwM{{\vec w}_{+}}
\def\vwm{{\vec w}_{-}}
\def\vnM{{\check n}_{+}}
\def\vnm{{\check n}_{-}}
\def\dM{{d_{+}}}
\def\dm{{d_{-}}}
\def\ro{r_{0}{}}
\def\vS{\vec {S}}
\def\vsuno{{{\vec s}_1}}
\def\vsdos{{{\vec s}_2}}
% OJO!!!
\def\ym{y}
\def\hX{{\hat X}}
\def\hJ{{\hat J}}
\def\hP{{\hat P}}
\def\hK{{\hat K}}
\def\hD{{\hat D}}
\def\va{\vec a}
\def\vb{\vec b}
\def\vu{\vec u}
\def\vv{\vec v}
\def\vt{\vec t}
\def\vn{\vec n}
\def\ve{\vec e}
\def\vx{{\vec x}}
\def\vxM{{\vec x}_{+}}
\def\vxm{{\vec x}_{-}}
\def\vwM{{\vec w}_{+}}
\def\vwm{{\vec w}_{-}}
\def\vnM{{\check n}_{+}}
\def\vnm{{\check n}_{-}}
\def\dM{{d_{+}}}
\def\dm{{d_{-}}}
\def\ro{r_{0}{}}
\def\vS{\vec {S}}
\def\vsuno{{{\vec s}_1}}
\def\vsdos{{{\vec s}_2}}
% OJO!!!
\def\ym{y}
\def\hX{{\hat X}}
\def\hJ{{\hat J}}
\def\hP{{\hat P}}
\def\hK{{\hat K}}
\def\hD{{\hat D}}

%%%%%%%%%%%%%% MACROS MONO %%%%%%%%%%%%%%

\def\vr{{\check e}_n}
\def\vteta{{\check e}_\theta}
\def\vfi{{\check e}_\varphi}
\def\vro{{\check e}_\rho}
%%%%%%%%%%%%%% MACROS MONO %%%%%%%%%%%%%%

\def\vr{{\check e}_r}
\def\vteta{{\check e}_\theta}
\def\vfi{{\check e}_\varphi}

%%%%%%%%%%%%%%%%%%%%%%%%%%%%%%%%%%%%%%%%%
\newcommand{\matriz}[4]{\left(
	\begin{array}{cc}#1 \\#3 \end{array}\right)}

% \newpage
% \thispagestyle{empty}
% \mbox{}

\frontmatter
\renewcommand{\chaptermark}[1]{\markboth{#1}{}}
\renewcommand{\sectionmark}[1]{\markright{\thesection\ #1}}
\pagestyle{frontmatter}

% Acknowledgement
\chapter*{Acknowledgments}
\markboth{}{}

First of all, I would like to thank my supervisor, Nicolás Grandi, for directing my PhD studies and for helping me achieve this great goal. I take with me many talks and blackboards.

I would like to thank my colleagues at  IFLP and friends who helped me to achieve the results presented here. In particular, I would like to highlight the interesting discussions that I had, both on a personal and scientific level, with the following people: Pablo Pisani, Carlos Arg\"uelles, Diego Correa, Mart\'\i n Schvellinger, Guillermo Silva, and Adrian Lugo. 

Last but not least, I want to thank Miranda for joining me on this adventure. And, of course, I want to thank my family and friends for helping me along this long but enjoyable journey.

%NICOLAS repetís mucho "I would like", fijate otros agradecimientos como hacen para no repetir. También repetís "achieve", por ahí podés mejorar eso también.

\tableofcontents

% Empty pages

%\newpage
%\thispagestyle{empty}
%\mbox{}

% \newpage
% \thispagestyle{empty}
% \mbox{}

% Main

\mainmatter
\renewcommand{\chaptermark}[1]{\markboth{#1}{}}
\renewcommand{\sectionmark}[1]{\markright{\thesection\ #1}}
\pagestyle{mainmatter}

% List of Figures
%\listoffigures
% List of Tables
%\listoftables

% Acronym List
\printglossary[type=\acronymtype, title=Abbreviations]

% Intro:
\chapter*{Preface}
\doublespacing

This thesis is the result of my work at La Plata National University. It reflects the studies carried out during my PhD in collaboration with my thesis advisor, Nicolás Grandi, and other colleagues at the AdS/CFT group at La Plata Physics Institute.

The main focus of this thesis is to study self-gravitating fermion systems in the context of gauge/gravity duality. This thesis is divided into three parts. The first part concentrates on an introduction to some of the topics and definitions that will be used later. In Chapter\ref{chap:gauge-grav} we provide a brief introduction to the duality. In Chapter \ref{chap:cond-mat}, we describe some elements of Condensed Matter related to quantum critical points and explain their utility for the description of High $T_c$ superconductors. In Chapter \ref{chap:astro}, we explain some elements of Astrophysics useful to our work, introducing the concept of neutron stars as well as the turning point and Katz criteria, which will be used later to check the stability of our models (section \ref{sec:neutronstars}). We also describe the work done by Argüelles et al. \cite{cita1} regarding self-gravitating systems of fermions as a possible candidate for dark matter (section \ref{sec:carlos}).

In the second part of the thesis, we will explain the contributions made in this area of study, starting with Chapter \ref{ch:carlos_nico}, where we explain the work \cite{cita2, cita16} that focuses on a system of neutral fermions in hydrostatic equilibrium at zero and finite temperature, respectively, in an asymptotically AdS space-time. In the following Chapter \ref{ch:nmodes}, we analyze scalar modes on this background and also explore Fermionic modes in Chapter \ref{ch:fermions}. Later in Chapter \ref{ch:holo_neutron_stars}, we extend the analysis on such a model, exploring the instability regions of the corresponding phase diagram.

In the last part, Chapter \ref{ch:summary}, we show a summary of all contributions made in this thesis and also possible paths for further research.

In this essay, there are two colored boxes with hints. Red boxes contain known literature, while blue boxes contain new contributions from this thesis. We use natural units where $\hbar=c=k_B=1$.

% Thesis:

\part{Introduction}
%Intro
\chapter{Some elements from gauge/gravity duality}
\label{chap:gauge-grav}
\doublespacing

This chapter provides a brief review of the main aspects of the gauge/gravity duality, also known as Holography, AdS/CFT, or Maldacena Conjecture \cite{cita3}. For a more extensive introduction to this topic, see for example  \cite{cita4, cita5}. We take a bottom-up approach to the duality, not providing an explicit string embedding of our models. Most of the content in this chapter is based on the book \cite{cita6}.

\section{Basics of gauge/gravity duality}
{So, what is gauge/gravity duality?} It is a way to change a difficult problem into a simpler one, described by completely different degrees of freedom in a absolutely distinct background. It is conjectured that the two descriptions are dual and, therefore, the results obtained from the ``simple'' side can be mapped back in terms of the language of the other, to obtain the desired observables. We use the duality to solve strongly coupled field theories, for which no other reliable method is available, in terms of weakly coupled degrees of freedom. Although there are many results that indicate that the duality assumption is correct \cite{cita7,cita8}, it not rigorously proven.

One of the most intriguing features of the gauge/gravity duality is the fact that it is holographic. This means that the weakly coupled theory do not live in the same number of space-time dimensions than the field theory, but rather it is defined in a space-time with one more dimension. Although this seems somewhat strange, such holographic characteristic had already been observed by 't Hooft and Susskind \cite{cita9,cita10}. Bekenstein and Hawking made a famous discovery regarding the holographic principle. They proved that the entropy of a black hole can be written as  
\begin{equation}
S_{BH}=\frac{A}{4 {l}^2_p}\,,
\label{eq:entropy.blackhole}
\end{equation}
where $A$ is the area of the event horizon and $l^2_p$ is the Planck length. The equation above shows that a gravitational object in $3+1$ dimensions with size $r$ whose volume scales as $V\sim r^{3}$, has an entropy that grows with the area $A\sim r^2$. So, it implies that we can represent quantum gravity inside a volume in terms of degrees of freedom on its surface. This suggests that the dual weakly coupled higher dimensional theory must include gravitation. 

To see where the additional dimension comes from, let us introduce an important concept called renormalisation group (RG). In the context of field theory, this refers to the fact that integrating out short-distance degrees of freedom induces a flow, which describes how the theory changes as one goes to longer wavelength. This can be rewritten in terms of differential equations expressing the running of the coupling constants of the model as the scale is changed. In the context of the gauge/gravity duality, such scaling ``direction'' turns into an extra space-time dimension in the gravitational dual. The scaling flow in the field theory gets encoded into purely geometric properties of this higher dimensional gravitational space-time. 
\begin{figure}[ht]
	\centering
	\includegraphics[width=.9\textwidth]{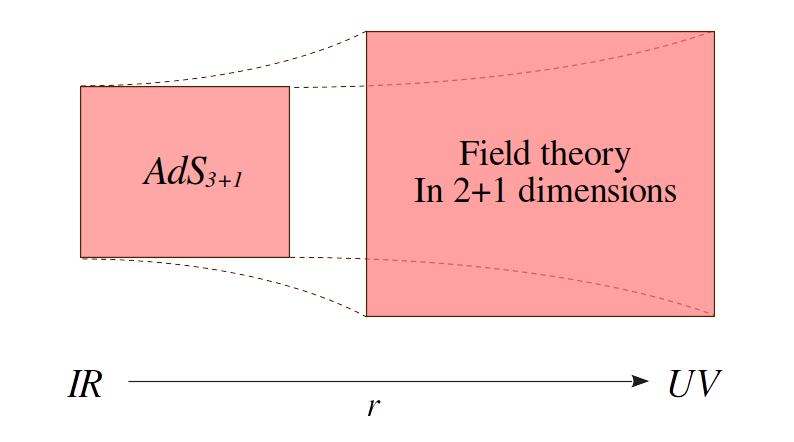}
	\caption{
 \small Schematic representation of the gauge/gravity duality.  \normalsize
		\label{fig:duality}
	}
\end{figure}

To make these ideas more concrete, let's explore the conjecture starting from the field theory side, and taking a relativistic $(2+1)$-dimensional conformal field theory (CFT) as our starting point. The natural description of such CFT is that it lives in a flat $(2+1)$-dimensional non-dynamical Minkowski space-time. We assume that this theory is manifestly invariant under global Lorentz transformations and under time and spatial translations, resulting in the conservation of energy and momentum. So, the corresponding bulk must have $3+1$ dimensions, the extra dimension is often called ``radial direction'' Fig.~\ref{fig:duality}. 

A generic field theory is subject to renormalisation. We have a differential equation, local in the RG scale $u$,  that provides a description of how the coupling constants $g_{i}$ of the theory change under scale transformations,
\begin{equation}
u\frac{\partial g_{i}}{\partial u}=\beta_i(g_{j}(u))\,.
\label{eq:renormalization.group}
\end{equation}
At a \emph{critical point}, the beta functions vanish by definition  $\beta_i=0$ and the physics becomes scale-invariant. The combined space and time scale transformation $x^{\nu}\rightarrow \lambda x^{\nu}$ is now also a symmetry. For a relativistic Lorentz-invariant theory, scale invariance together with unitarity implies invariance under the full set of conformal transformations, {\em i.e.} all transformations that preserve angles but not necessarily lengths. These include the so-called special conformal transformations, that combine with scale and Lorentz transformations to form the conformal group $SO(3,2)$. 

Since the field theory we wish to describe is conformal at such critical point, we must insist that its holographic gravitational dual encondes such invariance. We must therefore find a metric in such a way that the total $(3+1)$-dimensional space-time has the scaling symmetry $x^{\nu}\rightarrow \lambda x^{\nu}$ as an isometry. It is known \cite{cita6} that a space-time that fulfills this symmetry condition is given by the $(3+1)$-dimensional anti-de Sitter space. This is an explicit connection between AdS and CFT. The isometries of the AdS metric form  the same group, namely $SO(3,2)$. The AdS metric is 
\begin{equation}
\label{eq:AdS.metric}
ds^2=g_{AB}dx^Adx^B=\frac{r^2}{L^2}\eta_{\mu\nu} dx^{\mu}dx^{\nu}+\frac{L^2}{r^2}dr^2\,.
\end{equation}
Where $\eta_{\mu\nu}$ is the Minkowski metric, and the free lenght parameter $L$ %with dimension of lenght
is called the ``AdS radius''. We call the UV region $r\rightarrow \infty$ the boundary %\footnote{A light-like object can reach in a finite time.} 
of the space-time (where the CFT ``lives'') 
and the IR region at $r\rightarrow 0$ the AdS interior Fig.~\ref{fig:duality}.

It is important to remember that modifying the boundary theory also means altering our background space-time. This implies that we need to give some kind of dynamics to it.  The simplest dynamical gravitational theory with the minimal number of derivatives, that fulfill invariance under arbitrary differentiable coordinate transformations and also has the AdS space-time as a solution, is given by the Einstein-Hilbert action with a negative cosmological constant,
\begin{equation}
\label{eq:action.Einstein.Hilbert}
S_{\sf EH}=\frac{1}{16\pi G}\int d^{4}x\sqrt{-g}\left(R+\frac6{L^2}+\dots \right)\,,
\end{equation}
here, $g={\rm det} (g_{AB})$, $R$ is the Ricci scalar, $G$ is the gravitational coupling constant, and the cosmological constant is  $-3/L^2$.

Since Newton's constant is the coupling strength of General Relativity controlling fluctuations of space-time, and the Planck length is given by $l_{p}\sim\sqrt{G}$, we need $L\gg l_{p}$ to avoid large quantum gravity effects. This allows us to maintain the limitation to the minimal number of derivatives. In case this condition is relaxed, the AdS gravity theory would correspond to the full String Theory including all higher derivative corrections. 

The simplification to the classical and weakly coupled Einstein gravity \eqref{eq:action.Einstein.Hilbert} occurs only when the number of degrees of freedom at any point in the CFT is large. For instance, for a $(2+1)$-dimensional CFT dual to a $(3+1)$-dimensional AdS space, we find that the AdS curvature $L$ in units of the Planck length $l_{p}$ is related to the effective number of degrees of freedom in the CFT $C$,
%\textbf{https://www.sciencedirect.com/science/article/pii/S0550321318302074}
\begin{equation}
    \label{eq:central.c}
    \frac{L^2}{l^2_{p}}\sim C\,.
\end{equation}
Thus we need $C \gg 1$ in order to have a reliable dual classical gravitational description. The constant $C$ is called the \emph{central charge} of the CFT theory. % and proportional to the total number of degrees of freedom.
% %
% In the String Theory context it is the string length $l_s$ that sets the scale at which deviations from Einstein gravity become important, rather than the Planck length. A more precise version of Eq.\ref{dofrelation1} is therefore
% %
%  \begin{equation}\label{dofrelation2}
%  \frac{L^4}{l^4_{s}}\sim g_s N \sim g^2_{CFT}N \gg 1,
%  \end{equation}
%  where $g_s$ and $g_{CFT}$ are the string and conformal field theory couplings constants, respectively. 
%  \newpage

If these conditions are met, gravity is classical and we can solve the equations of motion derived from the Einstein-Hilbert action
\begin{equation}
G_{AB}-\frac{3}{L^2} g_{AB}= 8\pi G\, T_{AB}\,.
\label{eq:Einstein.equations}
\end{equation}
where $T_{AB}$ is the energy-momentum tensor.
 %
 %
% Thus, to rely on classical gravity in the bulk one has to make the number of degrees of freedom $N$ in the field %theory very large, so large that the combination $g^2_{CFT} N$ with $g^2_{CFT}\ll 1$ is also large, this completes %the crucial idea of duality.
\newcounter{boxcounter}
\setcounter{boxcounter}{1}
\begin{tcolorbox}[colback=red!5!white,colframe=red!75!black,title=Hint \arabic{boxcounter}] The gauge/gravity duality maps a strongly coupled, non-gravitational system with a large number of degrees of freedom per point to a weakly coupled, perturbative gravitational problem in one additional dimension.
\end{tcolorbox}
\stepcounter{boxcounter}

\section{Holographic correlators}
\label{intro:correlator}
\subsection{Conformal scalar correlator}
Let us begin by interpreting from the gravitational point of view a concept of the field theory very important  for this thesis: the two-point correlation function. In a Lorentz-invariant  conformal field theory at zero temperature, its form is fixed by conformal and Lorentz invariance, with the scaling dimension of the operator as the only free parameter. Consider two points separated by a distance $x$ in the Euclidean space-time. Given a UV scalar operator $\mathcal{O}(x)$ of the conformal field theory with conformal dimension $\Delta$, meaning that under a scale transformation it transforms as $\mathcal{O}(\lambda x)=\lambda^{-\Delta}\mathcal{O}(x)$, the two-point propagator is
\begin{equation}
    \label{eq:conformal.correlator}
    \langle\mathcal{O}(x) \mathcal{O}(0)  \rangle=\frac{1}{|x|^{2\Delta}}\,.
\end{equation}
Here we have fixed the arbitrary normalization to one. The correlation function can only depend on the Euclidean invariant  distance $|x|=\sqrt{x_\mu x^\mu}=\sqrt{t_E^2+\vec x^2}$ with $t_E$ the Euclidean time. 

We are usually interested in this propagator as a function of energy and momentum, so we can Fourier transform this expression to write it as a function of the four momentum $k_\mu=(\omega,\textbf{k})$. After a Wick rotation from imaginary to a real time with the appropriate prescription, we obtain the retarded correlation function 
\begin{equation}
    \label{eq:conformal.correlator.Fourier}
    \langle\mathcal{O}(-k) \mathcal{O}(k)  \rangle\sim k^{2\Delta-3} \,.
\end{equation}
The imaginary part of Eq.~\eqref{eq:conformal.correlator.Fourier}, known as the spectral function $A(\vec{k},\omega)$, obeys at zero momentum a pure power law $A(\vec k,\omega)\sim \omega^{2\Delta-3}$. 

\subsection{Holographic reconstruction of the scalar correlator}
So, how can we reconstruct such two-point correlators from objects that we can identify in the AdS space-time? Let us start considering a scalar field $\phi$ in AdS, and find out what can be done with it. The simplest action we can write down for its dynamics is a minimally coupled Klein-Gordon action,
\begin{equation}
\label{eq:action.Klein.Gordon}
S_{\sf KG}=\int d^{4}x\sqrt{-g}\left(-\frac{1}{2}g^{AB}\partial_{A}\phi\partial_{B}\phi-\frac{1}{2}{\sf m}^2\phi^2+...\right)\,,
\end{equation}
Considering a first approximation we can take this action as a classical field theory in the curved AdS space-time, with the metric in Eq.~\eqref{eq:AdS.metric}.
Thus the equation of motion for $\phi$ is
\begin{equation}
\frac{1}{\sqrt{-g}}\partial_{A}\left(\sqrt{-g} g^{AB}\partial_{B}\phi\right)-{\sf m}^2\phi=0\,.
\label{eq:EOM.Klein.Gordon}
\end{equation} 
We now Fourier transform and Wick rotate to Euclidean signature 
\begin{eqnarray}
\label{eq:scalar.Fourier}
\phi(x^{\mu},r)&=&
i \int \frac{
d^3k
}{(2 \pi)^{3}}f_{k}(r)e^{-ik_\mu x^\mu}.
\end{eqnarray}
So we can obtain an equation of motion for the dependence on the radial direction of the Fourier components $f_{k}$
\begin{equation}
\label{eq:EOM.scalar.radial}
\left( \frac{\partial^2}{\partial r^2}+\frac{4}{r}\frac{\partial}{\partial r}-\left(\frac{k^2 L^4}{r^4}-\frac{{\sf m}^2 L^2}{r^2}\right)\right)f_{k}(r)=0\ \,.
\end{equation}
As we stated before, we can think that the field theory lives at the boundary of the AdS space-time $r\to \infty$. Imposing that $f_k(r%/k
)$ %=f_k^{(0)}(r/k)^{\alpha}+...$ 
 behaves as a power law near the boundary, we can solve this ordinary differential equation as a series expansion, finding two independent solutions 
\begin{equation}
\label{eq:scalar.asymptotics}
f_k(r%/k
)=r^{-3+\Delta} \left(A(k)+{ O}\left(\frac{k}{r}\right)\right)
+r^{-\Delta} \left(B(k)+{ O}\left(\frac{k}{r}\right)\right)\,,
\end{equation}
where $\Delta=\frac{3}{2}+\sqrt{\frac{9}{4}+{\sf m}^2L^2}$ and the coefficients $A(k)=c_{1} k^{3-\Delta}$ and $B(k)=c_{2} k^{\Delta}$ have a  power-law dependence on $k$ which is fixed due to dimensional reasons. Since our field is real, the exponents must be real, which implies what is known as the Breitenlohner-Freedman (BF) bound \cite{cita11},
\begin{equation}
\label{eq:BF}
{\sf m}^2 L^2\ge -\frac{9}{4}\,.
\end{equation}
It is important to stress that this bound results in a  mass squared that can be negative. This is a special property of the AdS space-time:  as long as the BF bound is satisfied, the scalar is stable even in presence of a negative mass-squared.

Returning to the original question on how to determine the two-point correlation function in the AdS dual of a conformal field theory, we note that the information available near the boundary is contained in the universal asymptotes of the solutions to the equation of motion Eq.~\eqref{eq:scalar.asymptotics}. The boundary field theory should not contain information regarding the direction $r$, but we note that the leading coefficients $A(k)$ and $B(k)$ of the two independent solutions each have a simple algebraic monomial dependence on the frequency and momentum of the field theory. Therefore using the bulk solution $\phi(r)$, we can try to reconstruct the known answer in the CFT, in the form of the correlator Eq.~\eqref{eq:conformal.correlator.Fourier}. Up to a constant, we obtain 
\begin{equation}
\label{eq:holographic.scalar.correlator}
\frac{B(k)}{A(k)}\sim k^{2\Delta-3}\,.
\end{equation}
This simple result, which is coincident with the two-point correlation function of a Euclidean conformal scalar operator $\mathcal{O}$ with conformal dimension $\Delta=\frac{3}{2}+\sqrt{\frac{9}{4}+{\sf m}^2L^2}$ in momentum space Eq.~\eqref{eq:conformal.correlator.Fourier}, is in fact the the right dictionary rule. 

It is important to notice that the present calculations %done on this section 
work equally well in a metric which takes the AdS form \eqref{eq:AdS.metric} only in the asymptotic region $r\to\infty$. 
 
\begin{tcolorbox}[colback=red!5!white,colframe=red!75!black,title=Hint \arabic{boxcounter}] The two-point correlation function for scalar operators in a CFT can be expressed as a ratio of coefficients $A(k)$ and $B(k)$ of the leading and sub-leading asymptotes of the corresponding bullk massive scalar $\phi(r,k)$. The near-boundary region provides 
$$\langle\mathcal{O}(-k)\mathcal{O}(k) \rangle=\frac{B(k)}{A(k)}\,.$$ 
The mass of the bulk scalar sets the scaling dimension $\Delta$, which is given by $\Delta=\frac{3}{2}+\sqrt{\frac{9}{4}+{\sf m}^2L^2}$.
\end{tcolorbox}
 \stepcounter{boxcounter}
\subsection{Worldline approximation for the scalar correlator}
\label{sec:worldline}
In chapter \ref{ch:holo_neutron_stars}, we will use a different method to estimate the two-point correlator of a scalar operator $\mathcal{O}$ with conformal dimension $\Delta$. We will achieve this by using the classical trajectory of a particle of mass ${\sf m}$ that connects two points on the boundary. We will demonstrate how to use this approximation in this section. 
 
We use the WKB approximation to solve the Klein-Gordon equation for $\phi$ in the limit of large mass ${\sf m}$, defining
\be
\label{eq:WKB.expansion}
\phi = \exp\left({\sf m} S_{1}+ S_0 +{\cal O}\left({\frac 1{\sf m} }\right)\right)\,.
\ee
Replacing this expansion into  \eqref{eq:EOM.Klein.Gordon} and separating the orders in ${\sf m}$, we get to the lowest order
\be 
g^{AB}\left(\partial_A S_1 \right)
\left(\partial_B S_1\right) -1=0\,, 
\label{eq:WKB.first.order} 
\ee
In the AdS metric, we can solve this equation as 
\be
\label{eq:WKB.solution}
{\sf m}S_1^\pm = k_\mu x^\mu \pm {\sf m}\int_{r^*}^r dr'\, \frac L{r'} \sqrt{1-\frac{L^2k^2}{r'^2}} \,,
\ee
where $r^*$ is the turning point at which the square root vanishes.
Close to the boundary $r\to\infty$, there are two approximated solutions
\be 
{\sf m}S_1^\pm 
=k_\mu x^\mu \pm {\sf m}{L}\log{r}+{\rm cte}\,,
\label{eq:WKB.AdS}
\ee
which when replaced into $\phi$ give to the expansion \eqref{eq:scalar.asymptotics} in the large mass limit.  
The coefficients $A(k)$, $B(k)$ can then be extracted from $S_1^\pm$ by making
\be 
A(k) = \lim_{r\to \infty} r^{ -{\sf m}L} e^{{\sf m}S_1^{+}-k_\mu x^\mu} \,, \hspace{1cm} B(k) = \lim_{r\to \infty} r^{{\sf m}L} e^{{\sf m}S_1^{-}-k_\mu x^\mu} \,.
\label{eq:WKB.asymptotics}
\ee
This allows us to write
\be
\langle\mathcal{O}(-k)\mathcal{O}(k) \rangle=\frac{B(k)}{A(k)}=\lim_{r\to \infty}r^{2 {\sf m}L} e^{\,{\sf m}(S_1^--S_1^+)}\,.
\label{eq:WKB.correlator.Fourier}
\ee
By Fourier transforming we get 
the Matsubara two point correlator of a boundary scalar operator as a function of the boundary span of the points on the boundary $x^\mu=x^\mu_--x^\mu_+$ in the form
\begin{equation}
\langle
{\cal O}(x){\cal O}(0)
\rangle
=
\lim_{r\to\infty}r^{2{\sf m} L}
e^{\,({\sf m}S_1^--k_\mu x^\mu_--{\sf m}S_1^++k_\mu x^\mu_+)}\,,
\label{eq:WKB.correlator}
\end{equation}

\begin{figure}[t]
\vspace{-.7cm}	\centering
	\includegraphics[width=.7\textwidth]{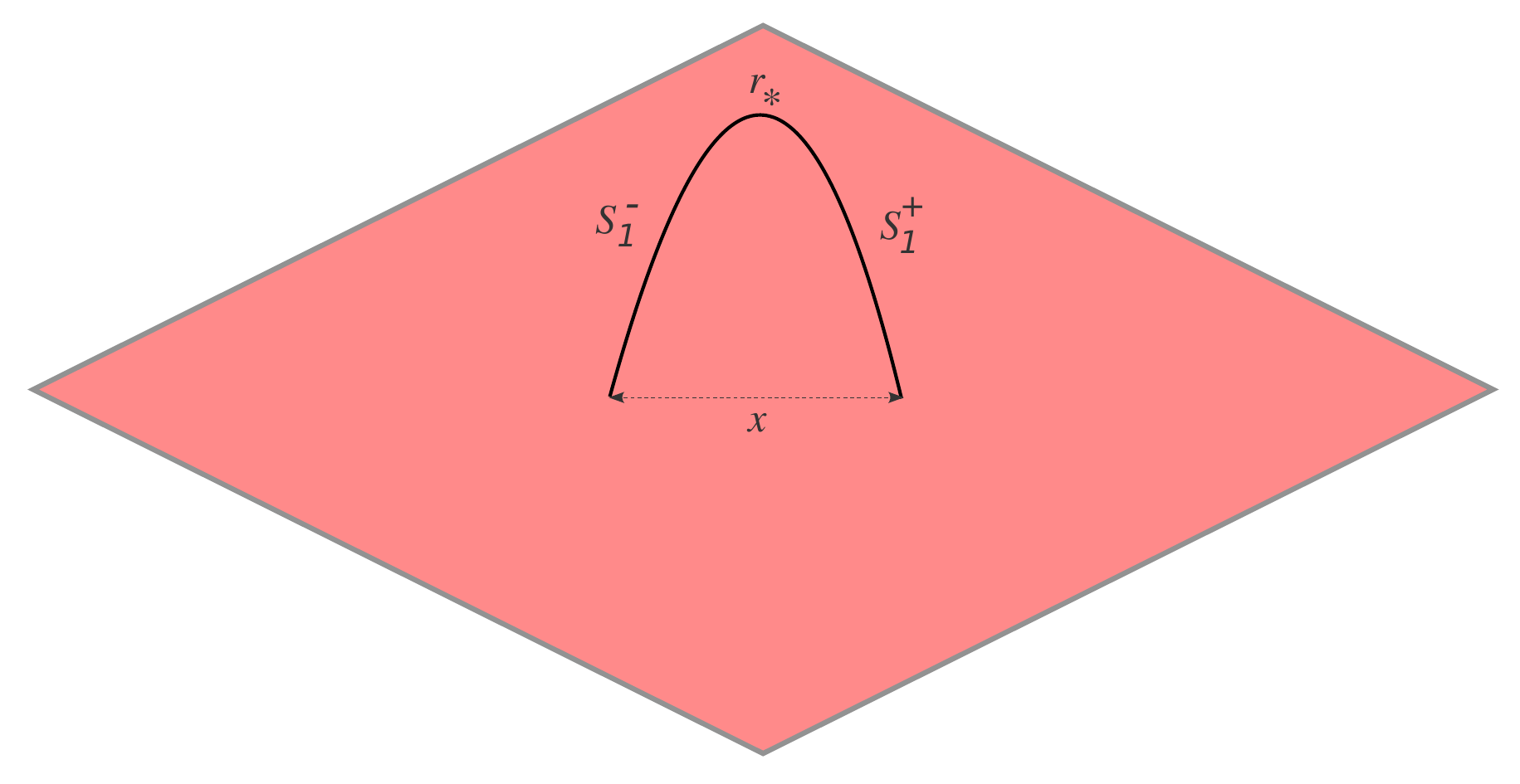}
 \vspace{-.3cm}
 	\caption{\small Schematic representation of a trajectory that starts at the boundary, dives into the bulk following the ``$-$'' branch up to a radius $r_*$, and then comes back to the boundary though the ``$+$'' branch. \normalsize 
		\label{fig:adsbulk}
}
\end{figure}

This construction can be generalized to metrics which take the AdS form \eqref{eq:AdS.metric} only  asymptotically, by noticing that equation \eqref{eq:WKB.first.order} is the Hamilton-Jacobi equation for a particle moving in the bulk with an Euclidean worldline action
\be
S_{\sf EWL}=  {\sf m}\int
d\tau \,\sqrt{g_{AB}\,{x'}^A {x'}^B}  \,. \label{eq:worldline.action} 
\ee
where $\tau$ is the Euclidean proper time, and $x'^A=\partial_\tau x^A$. This implies that ${\sf m}S_1$ is the corresponding Jacobi principal function, which is given by the action $S_{\sf EWL}$ evaluated on-shell, on the particle trajectories which connect the tip $r_*$ to the boundary. We have two solutions $S_1^{\pm}$ because there are two such trajectories, the difference of the actions evaluated on each of them $S_{\sf EWL}^{\sf on \,shell}={\sf m}S_1^--k_\mu x^\mu_--{\sf m}S_1^++k_\mu x^\mu_+$ is the classical action of a trajectory that starts at the boundary at $x_-^\mu$, dives into the bulk following the ``$-$'' branch up to a radius $r_*$, and then comes back to the boundary though the ``$+$'' branch up to the point $x_+^\mu$, see Fig.~\ref{fig:adsbulk}. We can then write
\be
\langle
{\cal O}(x){\cal O}(0)
\rangle=\lim_{r\to \infty}r^{2 {\sf m}L} e^{-S_{\sf EWL}^{\sf on\,shell}}\,.
\label{eq:correlator.worldline}
\ee
This expression gives  the correlator on the large mass limit, in terms of the classical action of a bulk Euclidean particle. The worldline approximation was used in \cite{cita41} to study thermalization. For details on its implementation see \cite{cita42}.

\begin{tcolorbox}[colback=red!5!white,colframe=red!75!black,title=Hint \arabic{boxcounter}] The two-point correlator for a scalar operator with large conformal dimension $\Delta\approx {\sf m}L$ is given by the exponential of the on shell Euclidean worldline action for a particle of mass ${\sf m}$.
\end{tcolorbox}
\stepcounter{boxcounter}

\subsection{Holographic spinor correlator}
A similar mass-scaling relation to the one existing for an scalar operator \eqref{eq:holographic.scalar.correlator} also exists for operators with spin. In this subsection we will describe how to calculate the two-point correlation function for operators with spin $1/2$ in the context of the gauge/gravity duality. Later in chapter \ref{ch:fermions} we will apply this to a particular case.

Let us start from the full covariant Dirac equation,
\begin{equation}
\label{eq:Dirac}
\left(e_{a}^{A}\Gamma^{a}\left( \partial_{A} + \frac{1}{4}\omega_{\mu b c}\Gamma^{bc} 
%-iq A_{A}
\right)-m \right) \Psi = 0 \,,
\end{equation}
where $\Gamma^{ab}=\frac{1}{4 i}[\Gamma^a,\Gamma^b]$, being $\Gamma^a$ the $(3+1)$-dimensional Dirac gamma matrices. Moreover $e_{a}^{A}$ is the vielbein $g_{AB} = \eta_{a b} e_{a}^{A} e_{b}^{B}$ and $\omega_{A b c}$ is the spin connection that satisfies $D_{A}e_{B}^{a}=0=\partial_{A}e_{B}^{a}-\Gamma^{C}_{A B}e_{C}^{a} + \omega_{A b}{}^{a} e^{b}_B$. 
For the AdS metric \eqref{eq:AdS.metric}, we
can make a redefinition of the Dirac field $\Psi = (-g)^{ -\frac{1}{4} }\sqrt{r/L} \,\Phi$ obtaining %so the above equation can be rewritten as
\begin{equation}
     \left(e_{a}^{A}\Gamma^{a}%\left( 
     \partial_{A}
     %- i q  A_{A}\right) 
     -m \right)\Phi=0\,.
\end{equation}
Using a projection onto transverse helicities where the spin is orthogonal to the direction of the boundary momentum and to the radial direction, and choosing conveniently the Dirac matrices along the tangent-space directions, is easy to show that
\begin{equation}
    \frac{L^2}{r^2} \left(i \sigma_2 \partial_r - \frac{m}{L}\,r\,\sigma_1 \right)\Phi_{ k}^{\alpha} = \left((-1)^{\alpha}k_x \sigma_3 -\omega \right)\Phi^{\alpha}_{k} \,,
\end{equation}
where $\Phi_{k}^{\alpha}$ with $\alpha=1,2$ are two-compontent spinors, and the boundary momentum points along the $x$ direction. %As for the scalar operator, 
Near the AdS boundary %the two-component spinor has 
we get
an asymptotic expansion
\small
 \begin{equation}
     \Phi^{\alpha}_k(r)=
     \left(A^\alpha(k)+O\left(\frac kr\right)\right)\left(\begin{array}{c}1\\0\end{array}\right) r^{m L} +      \left(B^\alpha(k)+O\left(\frac kr\right)\right) \left(\begin{array}{c}0\\1\end{array}\right) r^{- mL}+ ...\,,
 \end{equation}
 \normalsize
which allows us to calculate a two-point function. %as
\begin{tcolorbox}[colback=red!5!white,colframe=red!75!black,title=Hint \arabic{boxcounter}] The two-point correlation function for operators with spin %$G^{\alpha}(\omega;k)$ 
is calculated as a ratio of coefficients $A^{\alpha}(k)$ and $B^{\alpha}(k)$ of the leading and sub-leading asymptotes of the corresponding bulk fermion $\Phi_{k}^{\alpha}$.
$$
\langle{\cal O}^\beta(-k){\cal O}^\alpha(k)\rangle
%G^{\alpha\beta}(\omega;k)
= \frac{B^{\beta}(k)}{A^{\alpha}(k)}, \hspace{1cm}\alpha,\beta =1,2.
$$
\end{tcolorbox}
\stepcounter{boxcounter}
\vspace{-1cm}
\subsection{Holographic correlator in global coordinates}
\label{sec:global}
A coordinate system that covers the entire space is called \emph{global}. In the case of AdS, a set of global coordinates allows the parametrization
\begin{equation}\label{eq:globalAdS}
%ds^2=L^2\left(-\cosh^2\! r\,dt^2+dr^2+\sinh^2\!{r}\,d\Omega^2_2\right)\, ,
ds^2=L^2\left(-(1+r^2)dt^2+\frac{dr^2}{1+r^2}+r^2\,d\Omega^2_2\right)\, ,
\end{equation}
here  $d\Omega^2_2=d\vartheta^2+\sin^2\!\vartheta\, d\varphi^2$ is the metric on the unit 2-dimensional sphere. Therefore, the conformal boundary of $AdS_4$ is the product of a timeline and a 2-sphere, namely the cylinder $\mathbb{R}\times S^2$. Thus we now have a conformal field theory defined on a finite-volume spatial manifold, which explicitly breaks the conformal invariance %of the boundary theory.
\begin{tcolorbox}[colback=red!5!white,colframe=red!75!black,title=Hint \arabic{boxcounter}] The formulas for calculating the correlators remain the same, except that the quantum number $k$ and $\omega$ are replaced by the quantum numbers $\ell$, $m$, and $\omega$.% The correlators are denoted as 
$$\langle{\cal O}_{- m\ell}(-\omega){\cal O}_{ m\ell}(\omega)\rangle%=g_{\ell m}(\omega)
= \frac{B_{\ell m}(\omega) }{A_{\ell m}(\omega)}$$
%\end{tcolorbox}
%\stepcounter{boxcounter}
%\begin{tcolorbox}[colback=red!5!white,colframe=red!75!black,title=Hint \arabic{boxcounter}] 
%We can reproduce the calculations leading to 
Regarding the worldline form %of the correlator \eqref{eq:correlator32} in the present case, obtaining
it is a function of the angular span of the points on the boundary $\vartheta,\varphi$ and the elapsed Euclidean time $t_E$ %in the form
$$
\langle
{\cal O}(\vartheta, \varphi,  t_E){\cal O}(0)
\rangle
=
\lim_{r\to\infty}r^{2{\sf m} L}e^{-S_{\sf EWL}^{\sf on\,shell}(\vartheta, \varphi, t_E)}\,.
$$
\end{tcolorbox}
\stepcounter{boxcounter}
\section{Holographic thermodynamics}
\subsection{Black hole temperature}
One of the most iconic, and also the first, non-trivial solution of Einstein's equations \eqref{eq:Einstein.equations} in the case of vanishing cosmological constant $L\to\infty$, is the well-known  Schwarzschild black hole. In $3+1$ dimensions its metric reads
\begin{equation}
\label{eq:schwarzchild.metric}
ds^2=\ell^2
\left(
-\left(1-\frac{2 M}{r}\right)e^\chi\, dt^2+\left(1-\frac{2 M}{r}\right)^{-1}dr^2+r^2 d\Omega^2_2
\right) \,,
\end{equation}
where $M$ is the black hole mass, and we added two further ingredients for later convenience: a constant $\ell$ with units of length to have dimensionless time $t$ and radial $r$ variables, and a constant $\chi$ which sets the time scale.
%where $d\Omega^2_2=d\vartheta^2+\sin^2\!\vartheta\, d\varphi^2$ is the metric on the two sphere, % $f(r)=1-\frac{2 G M}{r}$, where $G$ is the Newton constant, and $M$ is the mass of the black hole. 
We can recover flat Minkowski space-time imposing $r\rightarrow \infty$ or $M=0$, so this space-time is asymptoticaly flat. The horizon sits where $g_{tt}=0$ or in other words at $r_h=2M$.%GM$

Inside the black hole, the time and the radial direction exchange roles. The horizon is a coordinate singularity, to reach $r=\infty$ any object located beyond $r_{h}$ %$=2GM$
needs a escape velocity larger that the speed of light. To an observer at infinity, objects that fall towards the horizon never cross through it, since for them time comes to a stop at horizon. This infinite time dilation also applies to the case of a collapsing star, so an observer at infinity will see the star frozen over the event horizon. On the other hand,  an observer at free fall would not notice anything strange when crossing the horizon\footnote{Even if the tidal forces turn him or her into spaghetti!}.

It has been known since the 1970's, and thanks to Hawking's work, that black holes generate black body radiation. A temperature known as the Hawking temperature can be associated with it, given by
\begin{equation}\label{eq:hawkingtemp}
	T=\frac{1}{8 \pi% G 
 M}\,.
\end{equation}

The presence of a temperature is a generic feature of space-times with a horizon. To compute the Hawking temperature from a more general static black hole metric, we follow the Gibbons and Hawking approach. Rewriting the black hole metric \eqref{eq:schwarzchild.metric} in the form
\begin{equation}\label{eq:generic.form}
ds^2=\ell^2
\left(-e^{\nu}dt^2+e^{\lambda}dr^2+r^2 d\Sigma ^2\right)\,.
\end{equation}
%
%
%\begin{equation}\label{eq:BlackHoleMetric}
%ds^2=-g_{tt}(r)dt^2+\frac{dr^2}{g^{rr}(r)}+g_{\Sigma \Sigma}(r)d\Sigma^2,
%\end{equation}
%
we can generalize the horizon metric to $d\Sigma^2$ and the functions $e^\nu$ and $e^{-\lambda}$ with the only requirement that they have a single zero at the black hole radius $r_h$.
%where $g_{tt}(r)$ and $g^{rr}(r)$ have a single zero at the horizon $r_{h}$, and $d\Sigma^2$ is the horizon metric. 
We perform a Wick rotation to Euclidean signature $t_E= i t$ , to obtain
\begin{equation}%\label{eq:generic.form.Eucildean}
ds^2=\ell^2\left(e^{\nu}dt_E^2+e^{\lambda}dr^2+r^2 d\Sigma ^2\right)\,.
\end{equation}
%
%
%\begin{equation}%\label{eq:BlackHoleMetricWick}
%ds_{E}^2=g_{tt}(r)dt_E^2+\frac{dr^2}{g^{rr}(r)}+g_{\Sigma\Sigma}(r)d\Sigma^2.
%\end{equation}
%
Let us focus in the metric near the horizon $r=r_{h}$, making a series expansion for $e^\nu$ and  $e^{-\lambda}$  and replacing it into  the metric
\begin{equation}\label{eq:generic.form.Eucildean.Expanded}
ds_{E}^2=(e^{\nu})'|_{r=r_h}(r-r_{h})dt_E^2+\frac{dr^2}{(e^{-\lambda})'|_{r=r_h}(r-r_{h})}+r_h^2\,d\Sigma^2+...\,.
\end{equation}
In terms of a new variable $\zeta=2\sqrt{(e^{-\lambda})'|_{r=r_h}(r-r_{h})}$ the metric becomes
\begin{equation}\label{eq:generic.form.Eucildean.disc}
ds_{E}^2=\frac{1}{4}\zeta^2 (e^{\nu})'|_{r=r_h}(e^{-\lambda})'|_{r=r_h}\,dt_E^2+d\zeta^{2}+r_h^2\,d\Sigma^2+...\,.
\end{equation}
\begin{figure}[ht]
	\centering
	\includegraphics[width=.65\textwidth]{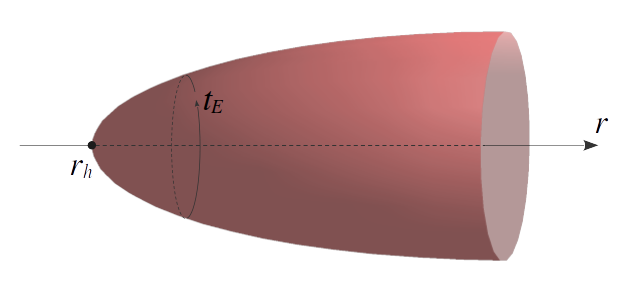}
	\caption{\small 
 The Euclidean geometry of $(t_E,r)$ plane on a black hole space-time, where $e^\nu$ vanishes at the horizon $r=r_{h}$. Ir order for geometry to be smooth at $r=r_{h}$, the Euclidean time $t_E$ has a period $\beta=4\pi/\sqrt{(e^{\nu})'|_{r=r_h}(e^{-\lambda})'|_{r=r_h}}$. \normalsize 
		\label{fig:duality2}}
\end{figure}
This is just the metric of a plane in polar coordinates, with $t_E$ acting as the compact angular direction. In the limit $\zeta\rightarrow 0$, we see that the factor of $dt_E^2$ vanishes, this means that the Euclidean time direction shrinks to a point. However since the horizon is not a special point, we should not allow this point to be singular. Smoothness at the horizon can be achieved by insisting that $\zeta=0$ is the center of a Euclidean polar coordinate system, and this implies that $t_E$ is periodic with period $\beta=4\pi/\sqrt{(e^{\nu})'|_{r=r_h}(e^{-\lambda})'|_{r=r_h}}$. Fig.~\ref{fig:duality2} illustrates Euclidean space-time with its periodic imaginary time direction looks like as a function of the radial direction. This periodicity is directly identified with the inverse temperature of the black hole,
\begin{equation}
\label{eq:blackhole_temp}
    T=\frac{\sqrt{(e^{\nu})'|_{r=r_h}(e^{-\lambda})'|_{r=r_h}}}{4\pi}\,.
\end{equation}
For the particular case of a spherical black hole $d\Sigma^2=d\Omega^2_2$ in flat space \eqref{eq:schwarzchild.metric}, this formula get us back to the result \eqref{eq:hawkingtemp}. 
Let us now use it to calculate the Hawking temperature for a black hole in AdS.

 \begin{tcolorbox}[colback=red!5!white,colframe=red!75!black,title=Hint \arabic{boxcounter}] For a black hole to be in equilibrium with a thermal bath, the bath temperature has to be the one given by the Hawking formula, guaranteeing a smooth Euclidean continuation of the geometry. 
\end{tcolorbox}
\stepcounter{boxcounter}

\subsection{Holographic temperature}
\begin{figure}[t]
	\centering 
	\includegraphics[width=0.8\textwidth]{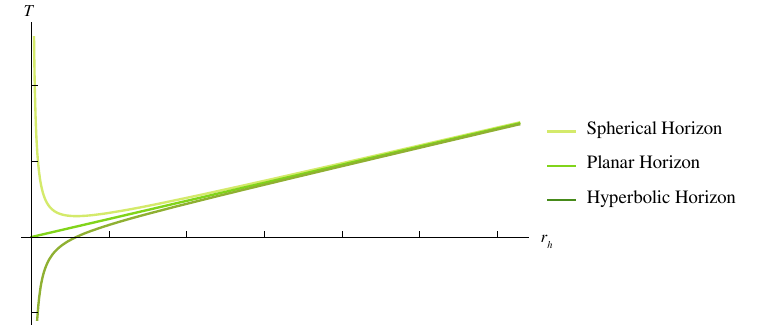}
	\caption{\small Black hole temperature as a function of the horizon radius for the three cases that have been widely studied in the literature, %the case of the spherical horizon $k=1$, the planar horizon $k=0$ and the hyperbolic horizon $k=-1$ and taking $L=1$,
 for the AdS-Schwarzschild black hole. 
		\normalsize
  \label{fig:duality3}}
\end{figure}

Starting from Einstein equations in $(3+1)$-dimensional Minkowski-space-time with negative cosmological constant \eqref{eq:Einstein.equations},
%
%\begin{equation}\label{eq:eisteinAdS}
%	R_{\mu\nu}-\frac{1}{2}g_{\mu\nu}R-\frac{d(d+1)}{2 L^2}g_{\mu\nu}=0,
%\end{equation}
%
the solution for an AdS-Scwarz\-schild black hole has the metric
%
%\begin{equation}\label{eq:AdSBlackHole}
%ds^2=\frac{r^2}{L^2}\left(-f(r)dt^2+d\Sigma_{k}^2\right)+\frac{L^2}{r^2 f(r)}dr^2,
%\end{equation}
\begin{equation}\label{eq:AdSBlackHole}
ds^2=L^2\left(-\left(k-\frac{2M}{r}+{r^2}\right)dt^2+\frac{1}{\left(k-\frac{2M}{r}+{r^2}\right)}\,dr^2+{r^2}d\Sigma_{k}^2\right)\,,
\end{equation}
with $d\Sigma_{1}^2=  d\Omega^2_2$  %(where $d\Omega^2_2$ is the unit metric on the sphere $S^{2}$)
representing a spherical horizon, $d\Sigma_{0}^2=\sum_{i=1}^{d}dx_{i}^2$ giving rise to  a planar horizon, and $d\Sigma_{-1}^2= dH^2_2$ (with $dH^2_2$ is the unit metric on the two dimensional hyperbolic space $\mathbb{H}^2$) entailing for a hyperbolic horizon. The 
metric \eqref{eq:AdSBlackHole} has the generic form \eqref{eq:generic.form} with the functions $e^\nu$ and $e^{-\lambda}$ suitably identified. 
%function $f(r)$ is known as the blackening factor, and takes the form
%
%\begin{equation}\label{eq:ffactor}f(r)=1-\frac{M}{r^{3}}+\frac{kL^2}{r^2}\,.
%\end{equation}
%
The condition $e^\nu=e^{-\lambda}=0$ gives us the radial position of the horizon in terms of the mass, as
\begin{equation}\label{eq:masshorizon}
2M=r_{h}^{3}+kr_{h}\,.
\end{equation}
%\begin{equation}\label{eq:masshorizon}
%M=r_{h}^{3}+kL^2 r_{h}.
%\end{equation}
%
Using equation \eqref{eq:blackhole_temp} we can now obtain the temperature for the AdS Schwarzschild black hole, as
\begin{equation}
\label{eq:temperatureAdSglobal}%\label{eq:temperatureAdS}
T=
%\frac{\sqrt{g^{'}_{tt}(r_{0}) g^{rr'}(r_{0})}}{4\pi}=\frac{r^{2}_{0}}{4\pi L^2}\frac{df(r)}{dr}\Bigg|_{r=r_{0}}
\frac{1}{4\pi }\left(3 r_{h}+\frac{k}{r_{h}}\right)\,.
\end{equation}
In this thesis we will   focus on the case $k=1$. % so we can rewrite the temperature for this case in this form
%\begin{equation}\label{eq:temperatureAdSglobal}
%T=\frac{3 r_h}{4 \pi  L^2}+\frac{1}{4 \pi  r_h}.
%\end{equation}
%
In Fig.~\ref{fig:duality3} we can observe the black hole temperature as a function of the horizon radius for the three cases. It grows linearly for large black holes.

 \begin{tcolorbox}[colback=red!5!white,colframe=red!75!black,title=Hint \arabic{boxcounter}] The Hawking temperature of a black hole in the AdS bulk is equal to the temperature in the field theory. %It grows linearly with respect to the horizon radius at high enough temperatures.
\end{tcolorbox}
\stepcounter{boxcounter}

\subsection{Holographic free energy}
\label{sec:grand.canonical.potential}
It is also important to know how to calculate the holographic thermodynamic potential in the context of the duality. We will use an equality that we will not prove: The grand canonical potential of a field theory in equilibrium reads %at finite temperature is% given by
% Zaanen (191) F_AdS = -S_E / β de ahi sale el menos que anula el otro.
% F=-k_{B}T\ln Z_{CFT}=k_{B}T S^{\rm On\mbox{-}shell}_{E}[g_{E}]
\begin{equation}\label{eq:freeEnergy}
\Omega=-T\ln Z_{CFT}=T S^{\sf on\,shell}_{\sf E}  \,,
\end{equation}
where %$g_{E}$ is the Euclidean metric of the gravitational dual, and
$S^{ \sf on\,shell}_{\sf E}$ is the on-shell  Euclidean action, see below. There are two things to take into account in order to be able to estimate the free energy properly,
\begin{itemize}\label{num:recipe}
	\item To have a well-posed variational problem, Gibbons, Hawking and York \cite{cita12} showed that we have to add to the action  a suitably defined boundary term.
	\item After we add the Gibbons–Hawking–York term, the on-shell action is still divergent, and has to be renormalized by adding suitable boundary counterterms. 
\end{itemize} 
So the total action is equal to
\begin{equation}\label{eq:actiontotal}
S_{\sf E}=S_{\sf EEH}+S_{\sf GHY}+S_{\sf CT}+S_{\sf EM}\,,
\end{equation}
where $S_{\sf EEH}$ is the Euclidean version of he Einstein-Hilbert action \eqref{eq:action.Einstein.Hilbert}, the term $S_{\sf EM}$ represents the Euclidean form of the bulk matter action, and the boundary terms take the form
\begin{eqnarray}
\label{eq:actionbulk}
%S_{\sf EEH}&=&-\frac{1}{16\pi G}\int_{0}^{\beta}d\tau\int_{r_{0}}^{\infty}dr\int d^{2}x\sqrt{g_{\sf E}} \left( R+\frac{6}{L^2}\right),
%\\
\label{eq:actionGHY}S_{\sf GHY}&=&-\frac{1}{8\pi G}\int_{0}^{1/T}dt_E\int_{r\rightarrow\infty}d^{2}x\,\sqrt{h_{\sf E}}\,K,
\\
\label{eq:actionct2}
S_{\sf CT}&=&\frac{1}{16\pi G}\int_{0}^{1/T}dt_E \int_{r\rightarrow\infty}d^{2}x  \,\sqrt{h_{\sf E}}\,\frac{4}{L}+\dots\,
\end{eqnarray}
Here $g_{E}$ is the Euclidean metric of the gravitational dual, and ${h_{\sf E}}_{AB}={g_{\sf E}}_{AB}- n_A n_B$ is the induced Euclidean metric on the boundary with $n^A$ an outward unit vector normal to it, %the boundary ($n^A$ is a null eigenvector of $h_{AB}$,  implying that the determinant $h$ is taken only in the directions orthogonal to $n^{A}$), 
and $K$ is the trace of the extrinsic curvature $K= h_{\sf E}^{AB}\nabla_{A}n_{B}$. % We explain calculation for the particular case under study in this thesis in Appendix \ref{AppendixA}.

\begin{tcolorbox}[colback=red!5!white,colframe=red!75!black,title=Hint \arabic{boxcounter}]  
The thermodynamic potential of the dual strongly coupled field theory is given by the bulk Euclidean action evaluated on-shell, including the Gibbons-Hawking and holographic renormalization terms. 
\end{tcolorbox}
\stepcounter{boxcounter}

%Critical systems
\chapter{Some elements from Condensed Matter}
\label{chap:cond-mat}
Why study condensed matter physics using the gauge/gravity duality? There are several reasons. Firstly, studying the gravitational dual is a simple and direct way to understand the physics of many strongly-coupled systems that are of great technological interest. Secondly, studying these systems is also fruitful for the area of high-energy physics, as it motivates useful theoretical ideas to explain complex processes.
Finally, these laboratory systems are real and can greatly aid in understanding duality, they can serve as a guide to construct gravitational duals. 

This section provides a brief description of essential concepts related to condensed matter to establish a connection with holography. The information presented in the following sections is based on \cite{cita6} and \cite{cita13}.

\newpage
\section{Phase transitions}
As we change some thermodynamic parameter, such as temperature, a system can undergo a phase transition to a more stable macroscopic state. During such transition, a thermodynamic potential, like the free energy or enthalpy, becomes non-analytic. In an $n$th-order phase transition, such potential loses analyticity at the $n$th derivative.

\begin{figure}[ht]
%\vspace{-1cm}
\centering
	\includegraphics[width=0.7\textwidth]{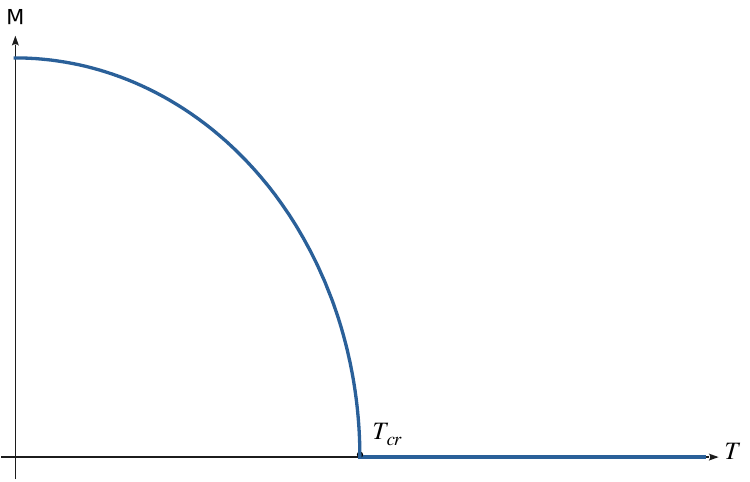}
	\caption{ \small Temperature-dependence of the order parameter.   
\normalsize		\label{fig:ferromagnet}}
\end{figure}

As an example, we can think on a ferromagnet whose magnetization ${\sf M}$ gets smaller as the temperature decreases, disappearing at a certain  critical value  $T = T_{cr}$, see Fig.~\ref{fig:ferromagnet}. 
We can consider the free energy for a magnetic system as a function $\Omega(T,{\sf M})$ of the temperature and the magnetization. As ${\sf M}$ is continuous at the transition, the phase transition is at least second-order. A macroscopic variable such as ${\sf M}$ that characterizes the difference between two phases is called an order parameter. 
\begin{tcolorbox}[colback=red!5!white,colframe=red!75!black,title=Hint \arabic{boxcounter}] At a phase transition, the thermodynamic potential of the system $\Omega$ is non-analytic. At an $n$th-order phase transition, the $n$th derivative of the thermodynamic potential becomes discontinuous or divergent.
\end{tcolorbox}
\stepcounter{boxcounter}
%\newpage

\section{Quantum criticality}
In the context of quantum field theory, the connection with condensed matter is based on the quantum dynamics of order parameters. Is starts from Landau's great intuition that the collective behavior of a system with an large number of interacting microscopic degrees of freedom can be captured by the order parameter, a rather simple object. Most of the content to motivate this connection is taken from \cite{cita58}.

As we approach a critical point in the phase diagram, the ground state defining a certain phase becomes less stable. This implies that there are large fluctuations with low energy, making them as probable as small ones. At the critical point, the fluctuations become scale invariant and spread throughout the system, triggering the phase transition. Conventional phase transitions occur at nonzero temperature, being the growth of random thermal fluctuations as we approach the critical point what leads to a change in the physical state of a system. Quantum phase transitions instead take place at zero temperature, and the fluctuations that enable them have a quantum origin.

The use of holography is promising because quantum critical theories are difficult to study using traditional methods. Athough in $1+1$ dimensions there are solvable theories with a critical point, as for example the Ising model in a transverse magnetic field, the situation is quite different for the physically more interesting $(2+1)$-dimensional case. Outside the context of holography, there are no models of strongly coupled quantum criticality in $2+1$ dimensions, in which analytic results for processes such as transport can be obtained. 
Typically, the critical theory is strongly coupled and so the action is not directly useful for the computation of many quantities of interest. So, the holographic principle gives us the opportunity to study strongly coupled quantum critical systems of physical interest.

\subsection{The Wilson-Fisher fixed point}
In the present subsection we will discuss a typical example of a system that displays quantum criticality, which gives us more intuition on how holographic tools could help to understand more about the underlying physics.

Let $\vec \Phi$ be an $N$ dimensional vector, on a $(2+1)$-dimensional theory with action 
\begin{equation}\label{eq:wf}
	S[\vec \Phi]=\int d^3 x \left(\partial\vec \Phi\cdot \partial\vec \Phi+q\,\vec \Phi\cdot \vec \Phi + u (\vec \Phi\cdot \vec \Phi)^2\right)\,.
\end{equation}
This model becomes quantum critical as $q\to q_c$ and is known as the Wilson-Fisher fixed point. At finite $N$ the relevant coupling $u$ flows to large values and the critical theory is strongly coupled. %The derivative in \ref{eq:wf} is the Lorentzian 3-derivative with signature $(-,+,+)$ and we set the velocity $v=1$.
Let us discuss two lattice models for which the theory \eqref{eq:wf} describes the vicinity of a quantum critical point.

\begin{figure}[t]
	\centering
	\includegraphics[width=0.63\textwidth]{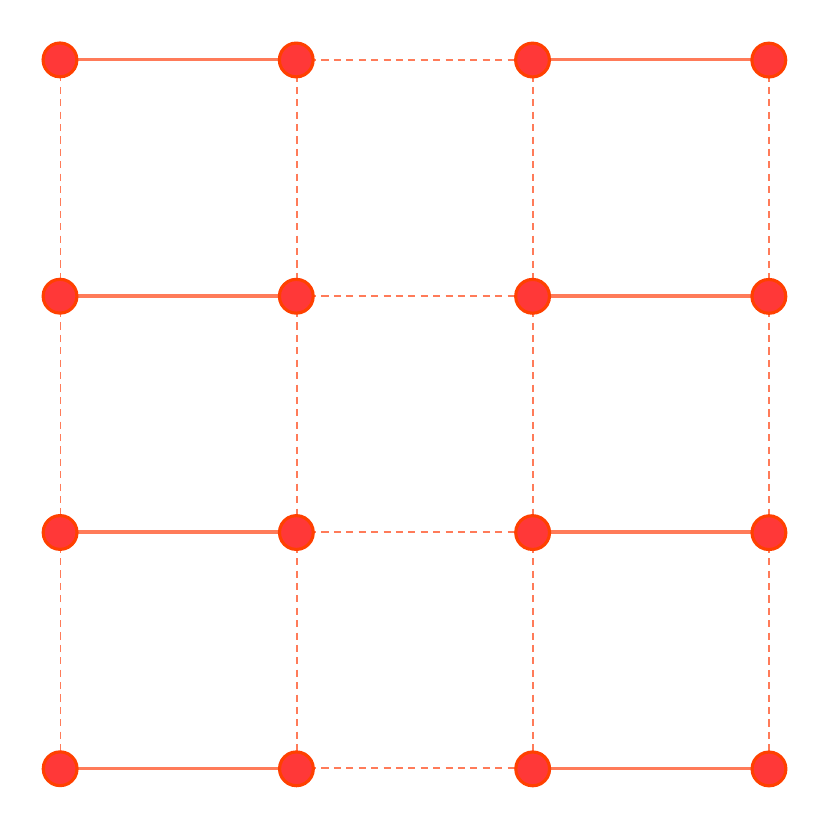}
	\caption{\small At large $g$, the dashed couplings $J/g$ are weaker than the solid ones $J$. This favors paring into spin singlet dimers (pairs of spins with antiferromagnetic exchange)
\normalsize		\label{fig:fm}}
\end{figure}

The first model is an insulating quantum magnet. Consider spin  degrees of freedom $\vec S_i$ living on a square lattice with the Hamiltonian
\begin{equation}\label{eq:qm}
	H_{}=\sum_{\langle i j\rangle }J_{ij}\,\vec S_i \cdot\vec  S_j \,,
\end{equation}
where $\langle i j\rangle$ denotes nearest neighbor interactions and we will consider the antiferromagnetic case $J_{ij}>0$. We choose the couplings $J_{ij}$ to take one of two values, $J$ or $J/g$ as shown in Fig.~\ref{fig:fm}, where the parameter $g$ takes values in the range $[1,\infty)$.
The ground state of this model is different in the limits $g \to 1$ and $g \to \infty$. 

At $g=1$ all couplings between spins are equal, this is the isotropic antiferromagnetic Heisenberg model, and the ground state has Néel order characterized by
%\begin{equation}\label{eq:spinneel}
$\langle\vec  S_i \rangle=(-1)^i\vec  \Phi(x_i)$\,,
%\end{equation}
%
where $(-1)^i$ alternates in value between adjacent lattice sites. The picture that we can imagine here for the classical ground state is a set of neighbouring spins anti-aligned. Here we are considering $\vec \Phi$ as a three component vector that changes slowly as we move through the lattice. The low energy excitations around this ordered state are spin waves described by the action \eqref{eq:wf} with $N=3$ and $\vec \Phi\cdot\vec \Phi$ fixed to a finite value $-u/q$. Spin rotation symmetry is broken in this phase.

In the limit of large $g$, in contrast, the ground state is given by decoupled dimers. That is, each pair of neighboring spins with coupling $J$ (rather than $J/g$) form a spin singlet. At finite but large g, the low energy excitations are triplons. These are modes in which one of the spin singlet pairs is excited to a triplet state.  The triplons have three polarizations and are again described by the action \eqref{eq:wf} with $N=3$ but expanded around a vacuum with $\vec \Phi=0$.

These two limits suggest that the low energy dynamics of the coupled-dimer antiferromagnet \eqref{eq:qm} is captured by the action \eqref{eq:wf} across its phase diagram, and that there will be a quantum critical point at an intermediate value of $g$ described by $N=3$ Wilson-Fisher fixed point theory. 

\begin{figure}[t]
	\centering
	\includegraphics[width=0.45\textwidth]{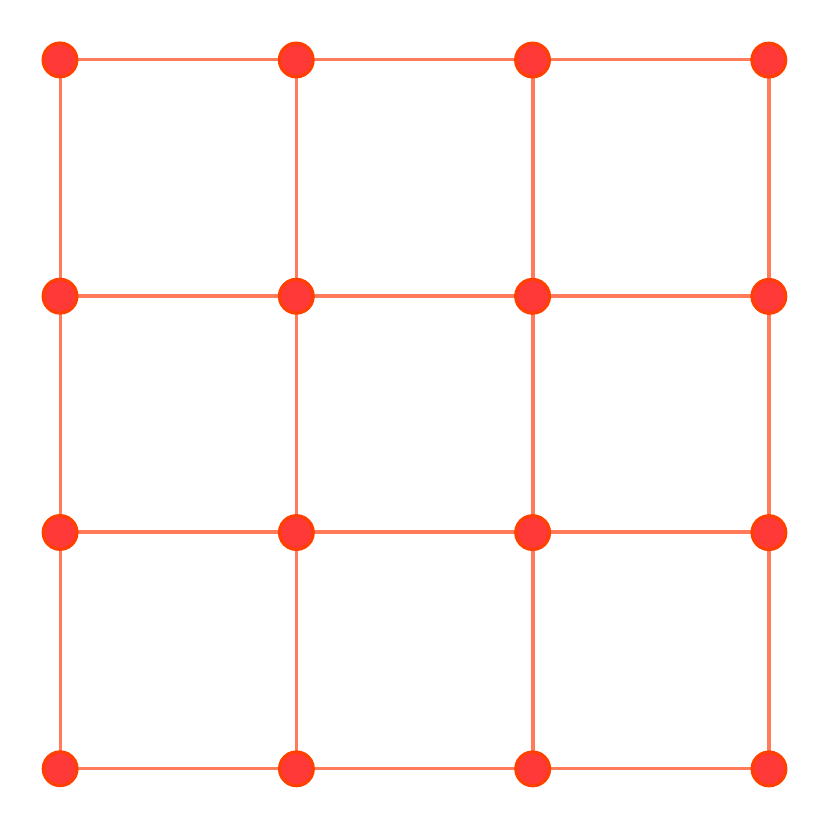}\hfill 
	\includegraphics[width=0.48\textwidth]{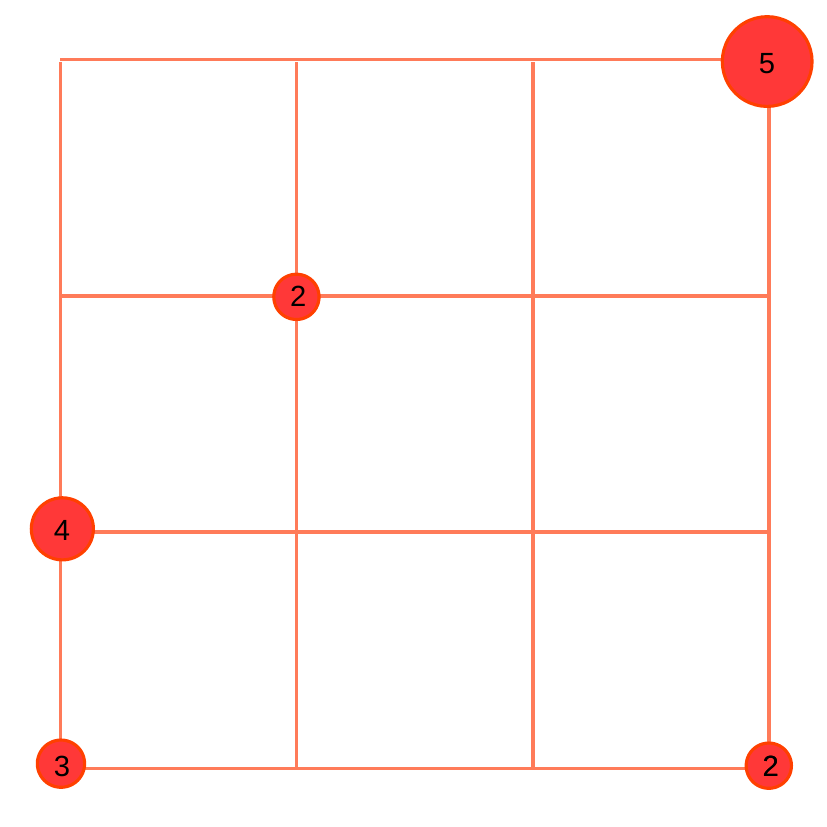}
	\caption{\small Schematic representation of the possible states of bosons with repulsive interactions on a square lattice. \underline{Left:} Insulating state. \underline{Right:} Superfluid state. The vertex labels represent the number of bosons at each lattice vertex.\normalsize
		\label{fig:hubbard}}
\end{figure}

A second lattice model realizing the Wilson-Fisher fixed point is the bosonic Hubbard model with the same number of bosons as lattice sites.  This is one of the simplest conformally invariant models. We consider spinless bosons hopping on the lattice of Fig.~\ref{fig:hubbard}, the operator $b^{\dagger}_i$ creating a boson at site $i$. The bosons can jump by tunneling betweeen nearest-neighbour sites, with a ``hopping'' amplitude $t$. We include short range repulsive interactions with coupling $U \geq 0$, and a chemical potential $\mu$. The Hamiltonian takes the following form,
\begin{equation}\label{hubbard}
	H=-t\sum_{\langle i j\rangle}(b^{\dagger}_i b_j+b^{\dagger}_j b_i) +
	U \sum_i n_i (n_i-1)+\mu \sum_i n_i\,,
\end{equation}
with $n_i=b^{\dagger}_i b_i$. This Hamiltonian has a $U(1)$ symmetry $b_i\to e^{i\phi}b_i$. 

By imposing a chemical potential that ensures an average integer number of bosons per site, we can infer that a zero-temperature quantum phase transition must occur at some critical value of $U/t$. For small values of $U$, the system will form a superfluid. However, when $U$ becomes too large, a bosonic Mott insulator will form. These two limits suggest the existence of a superfluid-insulator transition at an intermediate value of $U/t$.

Close to such continuous (second-order) phase transition, the correlation time and length scales become long compared with the UV scales, and hence the relativistic continuum field-theory description becomes appropriate in
the IR. The $U(1)$ symmetry is spontaneously broken on one side of the transition, and the critical theory is given in terms of the Ginzburg-Landau order parameter, which is now the condensate wavefunction represented by a complex number $\Phi$. This $\Phi$ realizes the dynamics of \eqref{eq:wf} with $N=2$ real components, for more details see \cite{cita13}.

These are just some of the models that have a quantum critical point and whose description must be made by means of a quantum field theory at strong interaction. They help us to understand the importance of the tools provided by the holographic duality, on understanding systems for which the perturbation theory approach fails.

\subsection{Criticality and scale invariance}

One of the main distinctive features of critical points is that the thermal and quantum correlation lengths for fluctuations diverge as a power of the distance to the critical values of the thermodynamic parameters. For example, for a quantum critical point at zero temperature $T_c=0$ and finite chemical potential $\mu_c$, we have that the thermal and quantum correlation lengths $\xi_q$ and $\xi_T$ satisfy
\begin{equation}
    \xi_T=(T-T_c)^{-\nu_T}\,,
    \qquad\qquad\quad
    \xi_q=(\mu-\mu_c)^{-\nu_q}\,,
\end{equation}
in terms of the {\em critical exponents} $\nu_T$ and $\nu_q$. This relation has the somewhat counter-intuitive consequence that the effects of a quantum critical point at zero temperature extend up to finite temperatures. Indeed, we can define the quantum dominated region as that where the quantum correlation length is larger than the thermal one $\xi_q/\xi_T>1$. From the above relation we see that it extends to finite temperatures as 
\begin{equation}
T>    (\mu-\mu_c)^{\nu_q/\nu_T}\,.%<T_c^{\nu_T}
\end{equation}
So, the system is critical in a wedge region centered on the critical chemical potential. 

Another interesting consequence of the emergence of scale invariance at the critical point is that the relations between physical observables become power laws. This can be explained as follows: assume that we have two observables $A$ and $B$ with length dimension $[A]=m$ and $[B]=n$ respectively. Whenever there is a non-vanishing correlation length $\xi$, we can relate them with equations of the form
\begin{equation}
    A=\xi^m f\left(B/\xi^n\right)\,,
\end{equation}
where $f$ is an arbitrary function. However, when the correlation length $\xi$ diverges, there is no other  possible relation but a power law  
\begin{equation}
    A=c\, B^{m-n}\qquad\qquad\quad{\text{with}}\qquad [c]=0\,.
\end{equation}
This is the reason why the appereance of a power law is taken as an indication of criticality, as we will do in the forthcoming chapters.

\begin{tcolorbox}[colback=red!5!white,colframe=red!75!black,title=Hint \arabic{boxcounter}] A quantum critical point at zero temperature has sensible effects at finite temperature, in a wedge-like region of the phase diagram when the chemical potential is close enough to its critical value. 
\end{tcolorbox}
\stepcounter{boxcounter}

\begin{tcolorbox}[colback=red!5!white,colframe=red!75!black,title=Hint \arabic{boxcounter}] One of the consequencies of criticality is the emergence of power law relations among physical observables. 
\end{tcolorbox}
\stepcounter{boxcounter}

\section{High $T_c$ superconductors}
\label{sec:highTc}

\begin{figure}[t]
	\centering
	\includegraphics[width=0.80\textwidth]{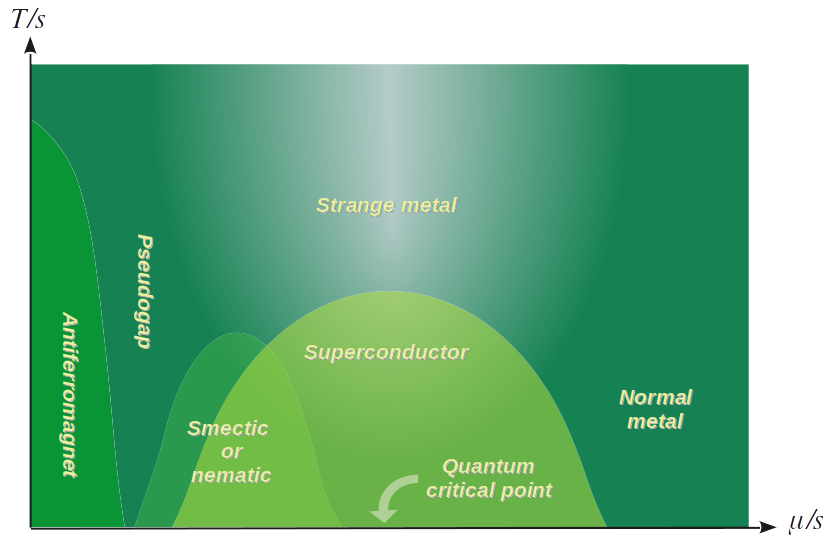}
	\caption{\small A schematic form of the phase diagram of High $T_c$ superconductors. At intermediate dopings we have a quantum critical point at zero temperature. It is surrounded by the superconducting dome, and its effects extend to higher temperatures in the strange metal region. At lower values of the doping we have an antiferromagnet while at higher values we see a normal Fermi liquid. Here $s$ is a dimensionful scale.\normalsize
		\label{fig:HihgTc}
	}
\end{figure}

The phase diagram of high $T_c$ cuprates, schematized in Fig.~\ref{fig:HihgTc}, is a complex and intriguing landscape that has been the focus of intense research for decades. One of its most fascinating features is the hypothesized quantum critical point, which is characterized by strong quantum fluctuations in the electronic degrees of freedom of the material. This critical point is located at zero temperature and intermediate values of doping, and its effects extend to a quantum critical region at higher temperatures as explained in the previous section.

The quantum critical point is hidden bellow the ``superconducting dome'', a phase in which the material can conduct electricity with zero resistance and exhibits Meissner effect. Above the superconducting dome lies the ``strange metal" phase, where the material displays some metallic properties but cannot be fully described by the Landau theory of the Fermi liquid. This phase is thought to be dominated by strong electronic correlations originated on the quantum critical point.

At low doping, the phase diagram is completed by an ``antiferromagnetic" state. It is separated from the superconducting dome by the ``pseudogap" region, within which the material exhibits a complex pattern of space-time symmetry breaking, including rotational or ``nematic'' and translational or ``smectic''  breaking. At higher doping, to the right of the superconducting dome, the material behaves as a Fermi liquid that can be accurately described by the Landau theory.

Overall, the richness of the phase diagram of high $T_c$ cuprates reflects the intricate interplay between temperature, doping, and quantum critical behaviour.

\begin{tcolorbox}[colback=red!5!white,colframe=red!75!black,title=Hint \arabic{boxcounter}] High $T_c$ superconductors are believed to have a quantum critical point at finite doping, which is responsible for a wedge-like critical region that extends up to finite temperature, onto the strange metal phase 
\end{tcolorbox}
\stepcounter{boxcounter}

\section{Why studying fermions?}
\label{sec:intro.fermions}
Throughout this thesis we will consider fermions as the only matter component on our gravitational solutions or ``stars''. From the dual point of view, this implies that we would be concentrating on the metallic degrees of freedom of the boundary quantum field theory. But why do we do this? 

Studying fermions at strong coupling is an open field of theoretical research and a hot topic in physics at the time this thesis is published. Its importance spans from the behavior of the quark-gluon plasma in high-energy experiments, to the description of the strange metal phase in high $T_c$ superconductors. 
In this last context, holography has been shown to be a very useful tool for understanding the properties of strongly coupled fermionic systems. Remarkably, the theoretical results of this approach share many features with phenomenological knowledge.

In the literature, charged fermionic excitations on non-backreactig backgrounds were explored in \cite{cita14,cita15}. The backreacting case was first investigated in \cite{cita18}, and further explored in \cite{cita19}-\cite{cita20}. The backreaction is considered by means of the energy momentum tensor of a perfect fluid, representing the fermionic background. The resulting Tolman-Oppenheimer-Volkoff equations are solved in an asymptotically AdS setup, in the Poincar\'e patch. Its finite temperature extension was investigated in \cite{cita21}-\cite{cita22}, in an approximation in which 
%Tolman-Oppenheimer-Volkoff
the equations are solved with a zero-temperature 
%fermionic
fluid, and temperature is introduced by means of a black hole horizon. 
This 
%The above described 
research results in a very rich set of features, which are qualitatively similar to those realized by the metallic degrees of freedom of High $T_c$ superconductors. 

\begin{figure}[ht]
	\centering
	\includegraphics[width=0.80\textwidth]{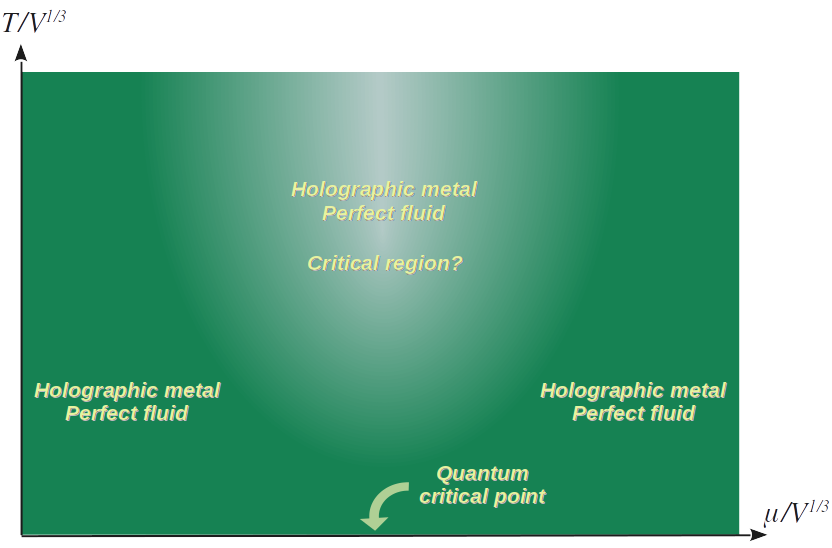}
	\caption{\small A schematic form of the expected  phase diagram for the holographic metal at finite volume. The effects of a quantum critical point at zero temperature extend to higher temperatures in the strange metal region.  %At lower values and higher values of the doping we see a normal Fermi liquid. 
		\normalsize \label{fig:phase.diagram.metal}
	}
\end{figure}

However, in the absence of an additional scale to play the role of $s$ in Fig.~\ref{fig:HihgTc}, all observables depend on the quotient $\mu/T$, what makes it impossible to locate the resulting phases on the cuprates phase diagram. 
Such limitation can be overcomed by introducing an additional chemical potential to play the role of the scale $s$, as was done in \cite{cita59}. A second possibility that we will explore on this thesis, is to confine the boundary theory to a finite volume vessel, whose scale is then given by the volume as $s\sim V^{1/3}$. This can be done by solving  Tolman-Oppenheimer-Volkoff equations with asymptotically {\em global} AdS boundary conditions, as was first done in \cite{cita16,cita17} for neutral fermions at zero temperature. We expect that the extension of these results to finite temperature would result on a phase diagram of the form shown in Fig.~\ref{fig:phase.diagram.metal}.
In this thesis we will present new results on this regard, published in \cite{cita54} and \cite{cita67}. 

 \begin{tcolorbox}[colback=red!5!white,colframe=red!75!black,title=Hint \arabic{boxcounter}] To dissentagle the temperature and chemical potential axes, the holographic description requires the introduction of an additional scale. This can be done by adding a second chemical potential or, as we will explore in this thesis, by confining the boundary theory to a finite volume vessel. 
\end{tcolorbox}
\stepcounter{boxcounter}
\newpage

%Neutron stars
\chapter{Some elements from Astrophysics}
\label{chap:astro}
\section{Neutron Stars}
\label{sec:neutronstars}
The so-called ``compact objects'' that are part of our universe are black holes, neutron stars and white dwarfs. They are born when a star dies, as occurs when most of its nuclear fuel has been consumed reaching the final stage of the stellar evolution.

An essential characteristic of these objects is that they cannot burn nuclear fuel, they are unable to support themselves against gravitational collapse by generating thermal pressure. Black holes are completely collapsed stars,  they don't have any means to hold back the inward pull of gravity and therefore collapse into singularities. White dwarfs are instead supported by the pressure of degenerated electrons. Finally, neutron stars are stabilized by the pressure of degenerated neutrons.

The primary factor in determining whether a star ends up as a white dwarf, a neutron star, or a black hole, is the mass of its progenitor. %\footnote{We are talking here about isolated objects since for binary or more complex systems things might become more complicated.}.
An important point to keep in mind, and which we will talk about later, is that there is a maximum mass for which the neutron star can no longer support itself, and collapses into a black hole.

Astrophysical neutron stars have a mass on the order of $1.4$ solar masses and a radius $R_{b}\sim 12km$, with a central density $\rho_{c}\sim 10^{17} kg/m^3$. This is on the order of nuclear density, situating them among the densest objects within the great galactic zoo. All fundamental interactions play an important role on their description, their extreme conditions being a perfect laboratory to help us understand quark matter, superfluidity, superconductivity and many other physical phenomena.

Understanding its physical aspects is relevant for this thesis, since we will introduce objects with similar characteristics into AdS space-time, and the equations we will find are very similar to those studied in this chapter. Here we work with the cosmological constant equal to zero so taking $L \rightarrow \infty$ in Eq.~\eqref{eq:action.Einstein.Hilbert}.

\subsection{Tolman-Oppenheimer-Volkoff equations}
% Esto lo saque del Hobson
Birkhoff's theorem states that the space-time geometry outside a general spherically symmetric matter distribution takes the form of a Schwarzschild geometry \eqref{eq:schwarzchild.metric}.
%
%\begin{equation}\label{sch-metric}
%ds^2=\ell^2\left(-\left(1-\frac{2 M}{r}\right)e^\chi dt^2+\left(1-\frac{2 M}{r}\right)^{-1}dr^2+r^2 d\Omega^2_2\right)
%\end{equation}
%where $d\Omega^2_2=d\vartheta^2+\sin^2\!\vartheta\, d\varphi^2$ is the metric on the two sphere, and we added two further ingredients for later convenience: a constant $\ell$ with units of length to have dimensionless time $t$ and radial $r$ variables, and a function $\chi$ which is constant in the vacuum. The Schwarzschild 
This  metric must then apply everywhere outside a spherical star, right up to its surface. To take the metric into the generic form \eqref{eq:generic.form} we  re-define 
\begin{eqnarray}
&&\left(1-\frac{2 M}{r}\right)e^\chi\equiv e^{\nu}\,,
%\label{eq:chi}\\&&
\qquad\quad\quad
\left(1-\frac{2 M}{r}\right)^{-1}\equiv e^{\lambda}\,,
\label{eq:lambda}
\end{eqnarray}
so the metric \eqref{eq:schwarzchild.metric} reads
\begin{equation}\label{sch-star}
ds^2=\ell^2
\left(-e^{\nu}dt^2+e^{\lambda}dr^2+r^2 d\Omega ^2_2\right).
\end{equation}
The metric \eqref{sch-star} also describes the gravitational field inside a spherical star, if we promote $\nu$ and $\lambda$ (or equivalently $M$ and $\chi$ in the Schwarzschild form \eqref{eq:schwarzchild.metric}) into independent arbitrary functions of the radius $r$. If we consider a star in hydrostatic equilibrium we can take $\lambda$ and $\nu$ independent of the time. We assume that the stellar material can be described as a perfect fluid in its rest frame,
\begin{equation}\label{eq:perfect-fluid}
T^{\mu\nu}=\frac{1}{G\ell^2}\left((\rho+P)u^\mu u^\nu+Pg^{\mu\nu}\right),
\end{equation}
where $u^\mu$ is the is the 4-velocity of the fluid, %Using the explicit form of metric tensor for Minkowski space-time we get
%\begin{equation}
%T^{\mu\nu}=\begin{pmatrix}
%-\rho & 0 & 0 & 0\\
%0 & p & 0 & 0\\
%0 & 0 & p & 0\\
%0 & 0 & 0 & p
%\end{pmatrix},
%\end{equation}
$\rho$ the rest frame energy density and $P$ the rest frame pressure, %so %at zero temperature we can define
that we can obtain from an equation of state $P(\rho, T)$. Again, a prefactor $1/G\ell^2$ was added for dimensional convenience. Einstein equations give
\begin{eqnarray}
    \frac{dM}{dr}&=&4\pi  r^2 \rho, \label{eq:mass}\\
    \frac{d\chi}{dr}  &=&
     8\pi r\left(\rho+P\right)e^{\lambda}
    \label{eq:Phi}
%    \frac{d\nu}{dr}&=&    \frac{M}{r^2}\left(1+\frac{4\pi P r^3}{M}\right)e^{\lambda}. \label{eq:Phi}
\end{eqnarray}
%
%\begin{eqnarray}
%    \frac{dM}{dr}&=&4\pi r^2 \rho, \label{eq:mass}\\
%    \frac{dP}{dr}&=&-\frac{\rho M}{r^2}\left(1+\frac{P}{\rho}\right)\left(1+\frac{4\pi P r^3}{M}\right)e^{\lambda}, \label{eq:ov}\\
%    \frac{d\nu}{dr}&=&-\frac{2}{\rho}\frac{dP}{dr}\left(1+\frac{P}{\rho}\right)^{-1}. \label{eq:Phi}
%    \frac{d\nu}{dr}&=&    \frac{M}{r^2}\left(1+\frac{4\pi P r^3}{M}\right)e^{\lambda}. \label{eq:Phi}
%\end{eqnarray}
%
%from which we can obtain
%
%\begin{eqnarray}
%    \frac{dP}{dr}&=&-\frac{1}{r^2}\left(P+{\rho}\right)\left({M}+{4\pi P r^3}\right)e^{\lambda}, \label{eq:ov}
%\end{eqnarray}
%
These are called the  Tolman-Oppenheimer-Volkoff (TOV) equations \cite{cita23}. %, the last one is the TOV equation of hydrostatic equilibrium \cite{cita23}.
The function $M(r)$ has the interpretation of the mass inside radius $r$, and we can find the total mass of a star of radius $r_{b}$ using Eq.~\eqref{eq:mass}.
\begin{equation}
\label{eq:totalmass}
M(r_{b})=\int_{0}^{r_{b}}4 \pi r^2 \rho \hspace{0.1cm}dr
\end{equation}

\begin{figure}[ht]
	\centering
	\includegraphics[width=0.48\textwidth]{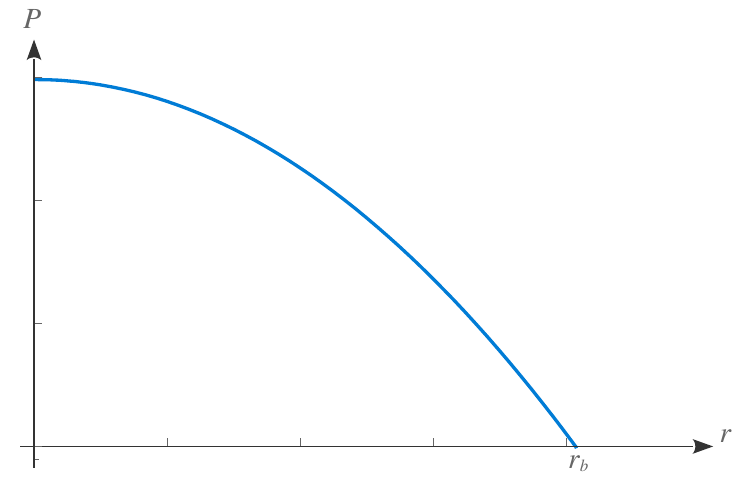}\hfill 
    \includegraphics[width=0.48\textwidth]{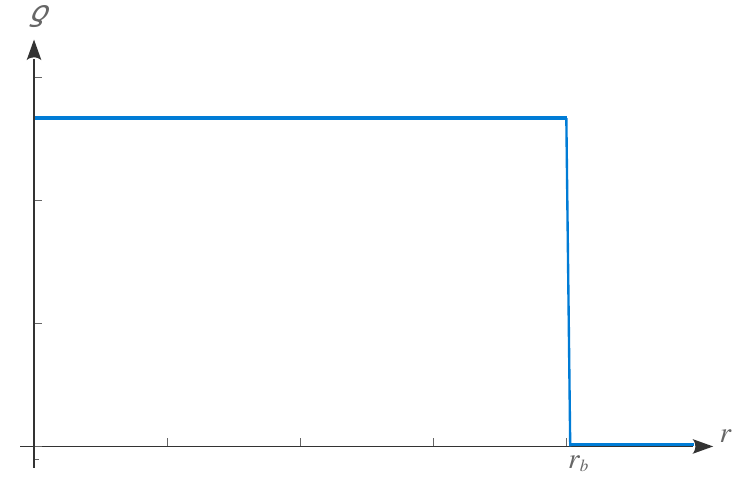}
	\caption{\small Radial profiles for a star with constant density. \underline{Left:} The fluid pressure. \underline{Right:} The energy density.  
		\normalsize\label{fig:Pconst}
	}
\end{figure}

When equations \eqref{eq:mass}, \eqref{eq:Phi} are supplemented with an equation of state $F(P,\rho)=0$, we can compute a numerical model for a general relativistic stellar configuration, as follows
\begin{itemize}
\label{num:neutron}
\item Choose a value for the central density $\rho(0)=\rho_c$ and temperature $T_c$, the equation of state then provides the central pressure $P_{c}$.
\item Integrate equations \eqref{eq:mass} and \eqref{eq:Phi} from $r=0$ using the boundary conditions $M(0)=0$ and $\nu(0)=0$. At each radial position obtain the value of $P$ and use the equation of state to find the corresponding value of $\rho$. Notice that to that end we need a way to obtain the local temperature, see bellow. %This can be done by solving the so-calle TOV equation for hydrostatic equilibrium, that can be obtained from \eqref{eq:mass} and \eqref{eq:Phi}
%
%\begin{eqnarray}
%    \frac{dP}{dr}&=&-\frac{1}{r^2}\left(P+{\rho}\right)\left({M}+{4\pi P r^3}\right)e^{\lambda}, \label{eq:ov}
%\end{eqnarray}
%
%As we will see below, in the case of a gas we can also obtain the local pressure and density from the distribution function.
\item Define the radius of the star as the value $r=r_{b}$ where $P(r_{b})=0$. Then the value of the function $M(r_{b})$ gives the total mass of the star. This condition ensures that we match with the exterior metric \eqref{eq:schwarzchild.metric}.
\item Assign the boundary condition $\nu(r_{b})=\chi(r_{b})+\frac{1}{2}\ln\left(1-\frac{2 M}{r_{b}}\right)$ to the metric function $\nu$. This ensures a smooth matching with the Schwarzchild metric \eqref{eq:schwarzchild.metric} at the surface. The constant $\chi(r_{b})$ can be set to zero by re-scaling time.
\end{itemize} 
As a simple example, %that will be useful to understand many of the studies carried out in this thesis,
we can consider a ``star'' with uniform density $\rho=$ constant. So we can solve Einstein equations \eqref{eq:mass}, \eqref{eq:Phi} %and \eqref{eq:ov}
using the package {\tt DSolve} of Mathematica 13.1 \cite{cita24} with the appropriates boundary conditions. We find
\begin{eqnarray}
    P&=&\frac{\sqrt{1- \frac{2 Mr^2}{r_{b}^3}}-\sqrt{1- \frac{2M}{r_{b}}}}{3\sqrt{1-\frac{2M}{r_{b}}}-\sqrt{1- \frac{2Mr^2}{r_{b}^3}}}\,\rho, \label{eq:Pconst} \\&~&\nonumber\\
    \nu&=&\log \left(\frac{3}{2} \sqrt{1-\frac{2 M}{r_{b}}}-\frac{1}{2} \sqrt{1-\frac{2 M
		r^2}{r_{b}^3}}\right).\label{eq:Phiconst}
\end{eqnarray}
We can see a plot of these solutions in Fig.~\ref{fig:Pconst}. The pressure becomes zero at $r=R_{b}$ (which in this case we chose as $r_{b}=1$), meaning that the edge of the star is located there. After that point, we have a Schwarzschild metric with $M$ being the total mass of the star.

\subsection{Thermodynamic stability analysis}\label{stabilityzerotemperature}
We will now analyze the general conditions for defining stability in these types of objects. We will use the so-called ``turning point" and Katz criteria, which are applicable to both flat space-time (as we will do in section \ref{sec:carlos}) and the AdS background (as we will discuss in chapter \ref{ch:holo_neutron_stars}).% that we will apply in our work, where we will consider similar systems but with different asymptotic limit. 
 
We will assume that the microscopic states of the star are described by a set of generalized variables $\eta_i$. If the value of the variables at thermodynamic equilibrium is $\eta_i$, then a fluctuation away from equilibrium can be characterized by $\eta_i+\delta \eta_i$. A simple example of such generalized variables can be given in the non-gravitational case, in which the equilibrium state is homogeneous. Any local thermodynamic quantity, such as the temperature or the chemical potential, can then be decomposed into Fourier modes. Thus, aside from the constant contribution, the amplitude of any higher mode parameterizes the deviation from homogeneity, and thus from equilibrium, and can be identified with one of the $\delta \eta_i$. 
In the self-gravitating case, the equilibrium state is given by the non-homogeneous solution obtained from Einstein equations. Any fluctuation of the corresponding local thermodynamic quantities moves the system out of equilibrium. By decomposing these fluctuations on a suitable basis, we get the variations $\delta \eta_i$. In any case, for the derivations below an explicit identification of the $\eta_i$ is not needed, being enough to acknowledge that  they exist.

In the following subsection we  explain two of the methods widely used to study stability of a macroscopic state under the fluctuations $\delta \eta_i$, the so-called {\em turning point} and {\em Katz} criteria. We study the first one in the microcanonical ensable, while for the second we work on the canonical ensamble (which is more suitable for holography). It is important to point out that, even if those approaches are completely equivalent in flat space, self-graviting systems often show the phenomenon of {\em ensamble inequivalence} \cite{cita47,cita53}.

\subsubsection{Turning point criterion}
\label{sec:turning}

In this subsection we will explain formally the turning point criterion, following \cite{cita39}.

Any state of the system is macroscopically characterized  by an entropy $S$ which is a function $S(M,N)$ of the total energy $M$ and the number of particles $N$. At equilibrium such entropy satisfies the first law of thermodynamics
\begin{equation}
    dM=T\,dS-\mu\, dN \,,
\end{equation}
which in particular implies that it is stationary $dS=0$ for fluctuations $\delta\eta_i$ such that $dM=dN=0$.
If we consider $S,M$ and $N$ as functions of the variables $\eta_i$, this implies
\begin{equation}
    \left(\partial_i M-T\,\partial_i S+\mu\, \partial_i N\right)\delta\eta_i =0 \,,
\end{equation}
where Einstein sumation convention on the index $i$ is intended. Expanding this equation one order further in the fluctuations, and then solving for the second order fluctuation of the entropy, we get
\begin{equation}
    \partial_i\partial_j S\,\delta\eta_i \delta\eta_j=\frac1T\left(\partial_i\partial_j M+\mu\, \partial_i\partial_j N-\partial_jT\,\partial_i S+\partial_j\mu\, \partial_i N\right)\delta\eta_i \delta\eta_j \,.
\end{equation}
We will concentrate in perturbations that keep the energy $M$ and the number of particles $N$ constant to first and second order, so the first two terms in the parenthesis vanish. 
%An important observation is that the
The resulting quadratic form is still symmetric 
% under the exchange of the indices. It reads
and reads
\begin{equation}
    \partial_i\partial_j S\,\delta\eta_i \delta\eta_j=\frac1T\left(\partial_j\mu\, \partial_i N-\partial_jT\,\partial_i S\right)\delta\eta_i \delta\eta_j \,.
\end{equation}
For a fluctuation to be stable, the entropy must be at its maximum, implying that the expression above must be negative. 

If we now have an uniparametric family of solutions written as $\eta_i(\rho_c)$ where $\rho_c$ is the central density, we can write an arbitrary fluctuation as
$%\begin{equation}
    \delta\eta_i=\partial_{\rho_c}\eta_i\,\delta\rho_c+\delta\eta_i^\perp \,,
$, %\end{equation}
where $\delta\eta_i^\perp$ guarantees that $M$ and $N$ are constant. To first order this implies
\begin{eqnarray}
    \partial_iM\,\delta\eta_i ^\perp=-\frac{dM}{d\rho_c}\delta\rho_c\,,
    \qquad\qquad\quad 
    \partial_iN\,\delta\eta_i ^\perp=-\frac{dN}{d\rho_c}\delta\rho_c
    \label{ecuuu}
\end{eqnarray}
Plugging back into the entropy variation we get
\begin{equation}
    \partial_i\partial_j S\,\delta\eta_i \delta\eta_j=
    -\frac1T\left(\frac{d\mu}{d\rho_c}\,\frac{dN}{d\rho_c}-\frac{dT}{d\rho_c}\,\frac{dS}{d\rho_c}\right)\delta\rho_c^2
    +
    \frac1T\left(\frac{dN}{d\rho_c}\, \partial_i \mu-\frac{dS}{d\rho_c}\,\partial_i T\right) \delta\rho_0\delta\eta_i^\perp \,.
    \label{ecucu}
\end{equation}
At a \emph{turning point} $\rho_c=\tilde\rho_c$ the following equations are satisfied by definition
\begin{eqnarray}
    \left.\frac{dM}{d\rho_c}\right|_{\rho_c=\tilde\rho_c}=
    \left.\frac{dN}{d\rho_c}\right|_{\rho_c=\tilde\rho_c}=0 \,.
\end{eqnarray}
Comparison with equation \eqref{ecuuu} shows that $\delta \eta_i^\perp$ has to be of order $\rho_c-\tilde\rho_c$, which in \eqref{ecucu} implies that the second term in smaller than the first as we approach the turning point, resulting in
\begin{equation}
    \partial_i\partial_j S\,\delta\eta_i \delta\eta_j=
    -\left.\frac{d}{d\rho_c}\left(\frac1T\left(\frac{d\mu}{d\rho_c}\,\frac{dN}{d\rho_c}-\frac{dT}{d\rho_c}\,\frac{dS}{d\rho_c}\right)\right)\right|_{\rho_c=\tilde\rho_c}\!\!\!
    (\rho_c-\tilde\rho_c\,)\delta\rho_c^2 \,.
    \label{ecurucucu}
\end{equation}
This implies that as we cross the turning point $\rho_c=\tilde\rho_c$, there is a change in the sign of the entropy flucutations. If it was stable at one side of the turning point, it becomes unstable at the other. 

Notice that turning points provide a sufficient condition for instability {\em i.e.} it could occur prior to a turning point or even without its presence at all. 
\begin{tcolorbox}[colback=red!5!white,colframe=red!75!black,title=Hint \arabic{boxcounter}] A turning point in the mass and particle number as a function of the central density provides a sufficient condition for thermodynamic instability. 
\end{tcolorbox}
\stepcounter{boxcounter}

\begin{figure}[ht]
	\centering
	\includegraphics[width=0.7\textwidth]{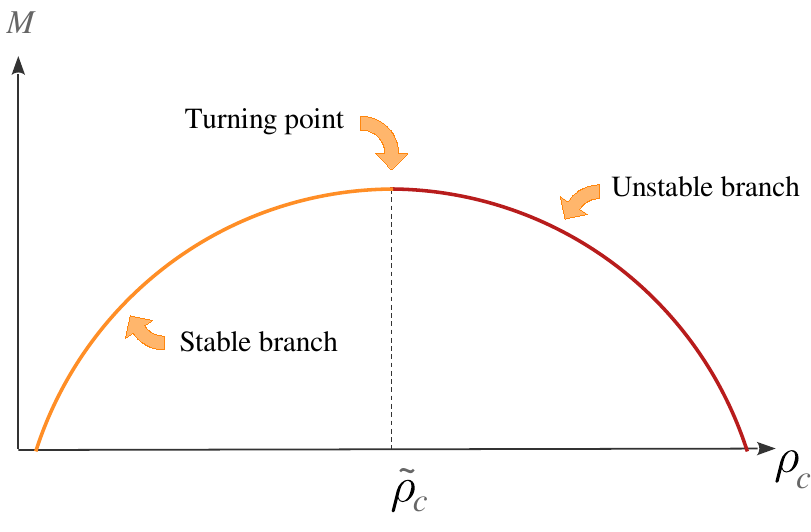}
 \vspace{-0.3cm}
 \caption{
 \small A schematic representation of the turning point criterion. On one side of the turning point, the system is stable, but after reaching $\tilde\rho_c$, the system becomes unstable.
		\normalsize \label{fig:turning_point}
	}
\end{figure}

\subsubsection{Katz criterion}\label{sec:katz}
In the next chapters we will find that the grand canonical potential might be multi-valued, so we need a criterion to determine which of its many branches corresponds to a stable phase of the boundary theory at a given temperature $T$ and chemical potential $\mu$. For this we use the so called
\emph{Katz criterion}, which was originally conceived for astrophysical systems \cite{cita26,cita27}. We sketch this criterion for our case of study.

Starting with the entropy of the system $S$ we define the grand canonical free entropy as
\begin{equation}
	\Phi(T,\mu)=
	S - \frac{1}{T}M-\frac{\mu}{T}N
	%- \frac{\Omega (\tilde{T},\tilde{\mu})}{\tilde{T}}
	=-\frac{\Omega}{T}
	\,,
	\label{eq:free.entropy}
\end{equation}
where $\Omega$ is the grand canonical potential. The derivative of this function with respect to $1/T$ gives minus the mass $-M$ of the configuration. 

Then lets assume that $\Phi( T,\mu)$ can be extended away from the equilibrium configuration into a function $\Phi(\eta_i)$ that depends on the generalized variables $\eta_i$ that parameterize the solutions of the system. 
The equilibrium states are stationary points of the extended grand canonical free entropy $\Phi(\eta_i)$, subject to the constraints of constant $T$ and $\mu$.
We can write the equilibrium solutions as $\eta_i=\eta_i( T,\mu)$ and then recover the equilibrium grand canonical free entropy as
\begin{equation}
	\Phi( T, \mu)=\Phi(\eta_i( T, \mu))\,.
	\label{eq:extended.free.entropy}
\end{equation}
We can then take a derivative of this expression 
with respect to the inverse temperature, to obtain
\begin{equation}
   	-M=\partial_i\Phi \,\partial_{1/T}\eta_i\,,
    \label{ocaa}
\end{equation}
where in the left hand side we recall that the derivative of $\Phi({T},{\mu})$ with respect to the inverse temperature $1/{T}$ at fixed ${\mu}/{T}$ gives minus the mass $-{M}$. A further derivative let us write
\begin{equation}
	-\partial_{1/ T}{M}=
	%-\partial_{1/ T}{M}^{\sf ext}
	-%\sum_i
	\partial_i{M} \,\partial_{1/ T}\eta_i(T,\mu)\,,
	\label{eq:mass.derivative}
\end{equation}
where the term proportional to $\partial_i\Phi$ vanishes at equilibrium.  If instead we differentiate the same equation with respect to the variable $\eta_i$ we get
\begin{equation}
   	-\partial_j M=\partial_j \partial_i\Phi \,\partial_{1/T}\eta_i \,.
\end{equation}
Which can be solved for $\partial_{1/T}\eta_i$ and plugged into \eqref{eq:mass.derivative} to finally obtain the form
\begin{equation}
	-\partial_{1/ T}{M}=
	-
	\partial_iM \,(\partial_i\partial_j\Phi)^{-1}\partial_j {M} \,.
	\label{eq:kats.criterion}
\end{equation}
Now we can parameterize the deformation away from equilibrium with coordinates $\eta_i$ such that the matrix $\partial_i\partial_j\Phi$ is diagonal, as
\begin{equation}
	-\partial_{1/ T}{M}= 
	-\sum_i
	\frac{(\partial_i {M})^2}{\lambda_i}\,.
	\label{eq:kats.criterion.diagonal}
\end{equation}
where $\eta_i$ is now the direction in the deformation space characterized by the eigenvalue $\lambda_i$ of the matrix $\partial_i\partial_j\Phi$. When any of such eigenvalues, say $\lambda$ associated to the coordinate $\eta$, is close enough to zero, it dominates the right hand side of equation \eqref{eq:kats.criterion.diagonal}, resulting in
\begin{equation}
	-\partial_{1/ T}{M}\approx
	-\frac{(\partial_{\eta}{M} )^2}{\lambda}\,,
\end{equation}
The crucial observation is that the sign of the eigenvalue is opposite to the sign of the expression $-\partial_{1/ T} M$. Since this is valid whenever $\lambda$ is approaching zero, the derivative is diverging at such points. In conclusion, whenever the plot of $-{M}$ versus $1/{T}$ at constant ${\mu}/{T}$ has a vertical asymptota,  the slope of the curve at each side of the asymptota is opposite to the sign of the eigenvalue of $\partial_i\partial_j\Phi$ that goes to zero there.

In a stable or meta-stable state, the  free entropy $\Phi$ is a maximum. This implies that all the eigenvalues of $\partial_i\partial_j\Phi$ are negative. As we move the inverse temperature at fixed ${\mu}/{T}$, the system evolves and the plot $-{M}$ versus $1/{T}$ eventually reaches a vertical asymptota. Since all the eigenvalues are negative, it approaches it with a positive slope. If at the other side of the asymptota the slope becomes negative, then one of the eigenvalues changed its sign, and the system reached an unstable region.

\begin{figure}[ht]
	\centering
	\includegraphics[width=0.8\textwidth]{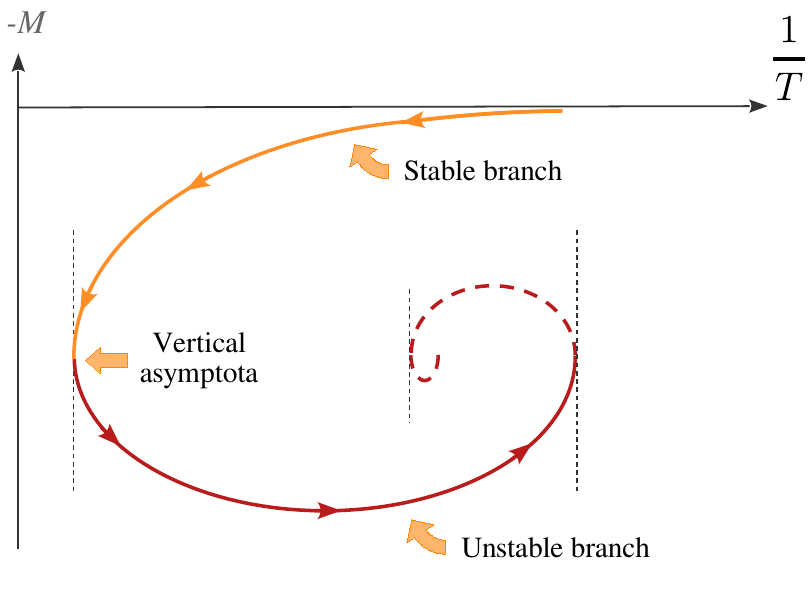}
	\caption{\small Parametric plot of minus the total mass $-{M}$ (right) as function of the inverse boundary temperature $1/{T}$, for some fixed value. The curve starts at low central densities where all the eigenvalues are assumed to be negative (orange line). As the plot on right hand side reaches its first vertical asymptota, an eigenvalue changes sign from negative to positive (red line). The process repeats on each vertical asymptota in which the slope changes from positive to negative.  
	\normalsize	\label{fig:katz}
	}
\end{figure}

In what follows we identify the stable equilibrium state in which all the eigenvalues are negative with the diluted configurations. Next, we follow the $-{M}$ versus $1/{T}$ curve until we reach an asymptota at which the slope changes its sign. For each change from positive to negative slope, we count a new positive eigenvalue. For each change from negative to positive slope, we count a new negative eigenvalue.  Any region with at least one positive eigenvalue, is unstable. A qualitative summary of the criterion is shown in Figure \ref{fig:katz}.

 %\begin{tcolorbox}[colback=blue!5!white,colframe=red!75!black,title=Hint 6] Both the turning point %criterion and the Katz criterion allow us to identify the values of $\theta_{0}$ for which the system %becomes unstable, locating the stable branch for configurations with $\theta_{0} < -20$. In addition, the %Katz criterion allows us to identify the stable branch of the grand canonical potential within this same %region.
%\end{tcolorbox}
 \begin{tcolorbox}[colback=red!5!white,colframe=red!75!black,title=Hint \arabic{boxcounter}] The Katz criterion allows us to identify the regions of the phase diagram where the system becomes unstable. We identify the stable branch as the one corresponding to configurations for with where the central density is low. %In addition, the Katz criterion allows us to identify the stable branch of the grand canonical potential within this same region.
\end{tcolorbox}
\stepcounter{boxcounter}

\section{Fermionic dark matter halos}
\label{sec:carlos}
The study of %the thermodynamics of
self-gravitating systems is extremely important in the astrophysical context. Among the many applications of this subject, we want to highlight %those that
the use self-gravitating fermions at finite temperature to describe the morphology of dark matter halos \cite{cita28}-%,cita29,cita30,cita31,cita32,cita33,cita34,cita35,
\cite{cita36}. In these references, the stability of these systems is studied in a similar manner to what we will do in this thesis \cite{cita1} for their AdS counterpart. In the following sections, we will focus on the main ideas and equations that describe this type of system, following the work of \cite{cita1}.
\subsection{Self-gravitating  gas at finite temperature}
The main assumptions made to describe these systems in the referred literature are
\begin{itemize}
	\item Dark matter halos consist on a spherically symmetric configuration of a neutral fluid in hydrostatic equilibrium. Then the metric has the form \eqref{sch-star} where the functions $\nu$ and $\lambda$ satisfy the TOV equations \eqref{eq:mass} and \eqref{eq:Phi}.
	\item The equation of state is that of a gas of non-interacting massive fermions, with an energy distribution which considers relativistic effects and the existence of a escape velocity above which particles are not gravitationally bound.
\end{itemize}
To fulfill the second requirement, we consider a variant of the Fermi-Dirac distribution function, given by
\begin{equation}\label{F-D}
f(p)=H\left(\epsilon_m-\sqrt{p^2+m^2}\right)\,
\frac{1-e^{\beta\left(\sqrt{p^2+m^2}-\epsilon_{m}\right)}}{1+e^{\beta\left(\sqrt{p^2+m^2}-\mu\right)}} \,.
\end{equation}
Here $H$ is the Heaviside function satisfying $H(u)=1$ for $u>0$ and $H(u)=0$ for $u<0$, the constant $\mu$ is the chemical potential, $\beta=1/T$ is the inverse temperature, $\vec p$ is the particle momentum, and $m$ is the particle mass. The constant $\epsilon_{m}$ is a cut-off on the energy above which particles escape the gravitational pull of the configuration.
%
%The temperature $T$, chemical potential $\mu$, and cutoff $\epsilon_c$ are $r$-depending parameters.

With this distribution function, the matter source for the corresponding Einstein equations are given in terms of the parametric equation of state defined by
\begin{align}
\rho &= \frac{g}{8\pi^3%G
\ell^2}\int f(p)\sqrt{p^2+m^2 }\,d^3p\,,
\label{eq:rho_flat}\\
P &= \frac{g}{24\pi^3%G
\ell^2}\int
f(p)\frac{p^2}{\sqrt{p^2+m^2 }} \,d^3p,
\label{eq:p_flat}
\end{align}
here, due to the Heaviside function in $f(p)$, the integration in momentum space is bounded by the escape energy $\epsilon_{m}$. The constant $g$ is the number of particle types, for the present astrophysical context of a spin $1/2$ fermion we have $g=2$.

In this setup, the local temperature $T$ and chemical potential $\mu$ are radial functions, %. They are 
defined by the %thermodynamic equilibrium conditions of
Tolman and Klein local equilibrium conditions, respectively
\begin{eqnarray}\label{eq:tolman.klein}
e^{\frac{\nu}2} T = T_c,\qquad\qquad\qquad
e^{\frac{\nu}2} \mu = \mu_c\equiv\ \Theta_c T_c+m\,.
\end{eqnarray}
where $T_c$ and $\mu_c$ are the central values of temperature and chemical potentia. The parameter $\Theta_c$ is called the {\em central degeneracy}. Tolman relation can be understood as follows: in order to put the system in a thermal bath, a Wick rotation $t_E=it$ is performed, and the resulting Euclidean time direction is compactified in a circle whose physical length is the inverse temperature. If we assume $\nu(0)=0$, then the period of the Euclidean time $t_E$ is the inverse of $T_c$, and Tolman relation follows.

%of length $1/T$.
%In order for the fluid in the bulk to be in equilibrium with the thermal bath, we need to impose $\beta=e^{\frac{\nu}2}/T_c$. On the aforementioned cylinder, the holographic dual is defined as a conformal field theory.
%Since the Euclidean time direction of the conformal theory is given by $t_E$, its temperature corresponds to $T=e^{-\frac{\nu_\infty}2}/T_c$ where ${\nu_\infty}$ is the value of $\nu$ at $r\to\infty$.

Conditions \eqref{eq:tolman.klein} together with TOV equations \eqref{eq:mass} and \eqref{eq:Phi} with density and pressure given by \eqref{eq:rho_flat} and \eqref{eq:p_flat} lead to a system of non-linear differential equations that must be solved by numerical means.
%
%Notice that the fermion mass $m$ plays no role other that setting the scale.
The boundary conditions are% at the center of the star $r=0$ are chosen as
\vspace{-.5cm}
\begin{eqnarray}\label{eq:NumericalConditions}
{M}(0)&=&0\,,\qquad\qquad\mbox{(or in other words }\lambda(0)=0\mbox{)}\,,
\\
\vspace{-.5cm}
\chi(0)&=&0\,,\qquad\qquad\mbox{(or in other words }\nu(0)=0\mbox{)}\,.
\label{eq:NumericalConditions2}
\end{eqnarray}
The resulting solutions are indexed by the parameters $T_c, \Theta_c$ and $\gamma^2\equiv gG\ell^2m^4/2\pi^2$.

The first consistent solutions %with regular boundary conditions at the center of the galaxy 
were found in \cite{cita37}. For suitable choices of the dark matter particle mass $m$ and the central values of the temperature $T$ and chemical potential $\mu$, they match the dark matter halo observables of the Milky Way. For certain regions of parameters, the dark matter profiles develop a ``dense core - diluted halo'' morphology, see Fig.~\ref{fig:carlos3} (top-right). The central core is governed by Fermi-degeneracy pressure, while the outer halo holds against gravity by thermal pressure. When the central degeneracy is low, the model is assumed to be in the diluted-Fermi regime. There are many advantages on using such core-halo model to represent dark matter in galaxies, which are explained in more detail in \cite{cita1}.

\subsection{Stability of dark matter halos}
Studying the stability of these models is important in the astrophysical context, as it can change the massive black hole in the center of some galaxies into a degenerate compact core made of neutral dark matter fermions with mass $m\sim1$KeV$-10$KeV . This core is surrounded by a diluted halo composed of the same particles, which is responsible of the observed rotation curves.

\begin{figure}[ht]
	\qquad \quad \includegraphics[width=.9\textwidth]{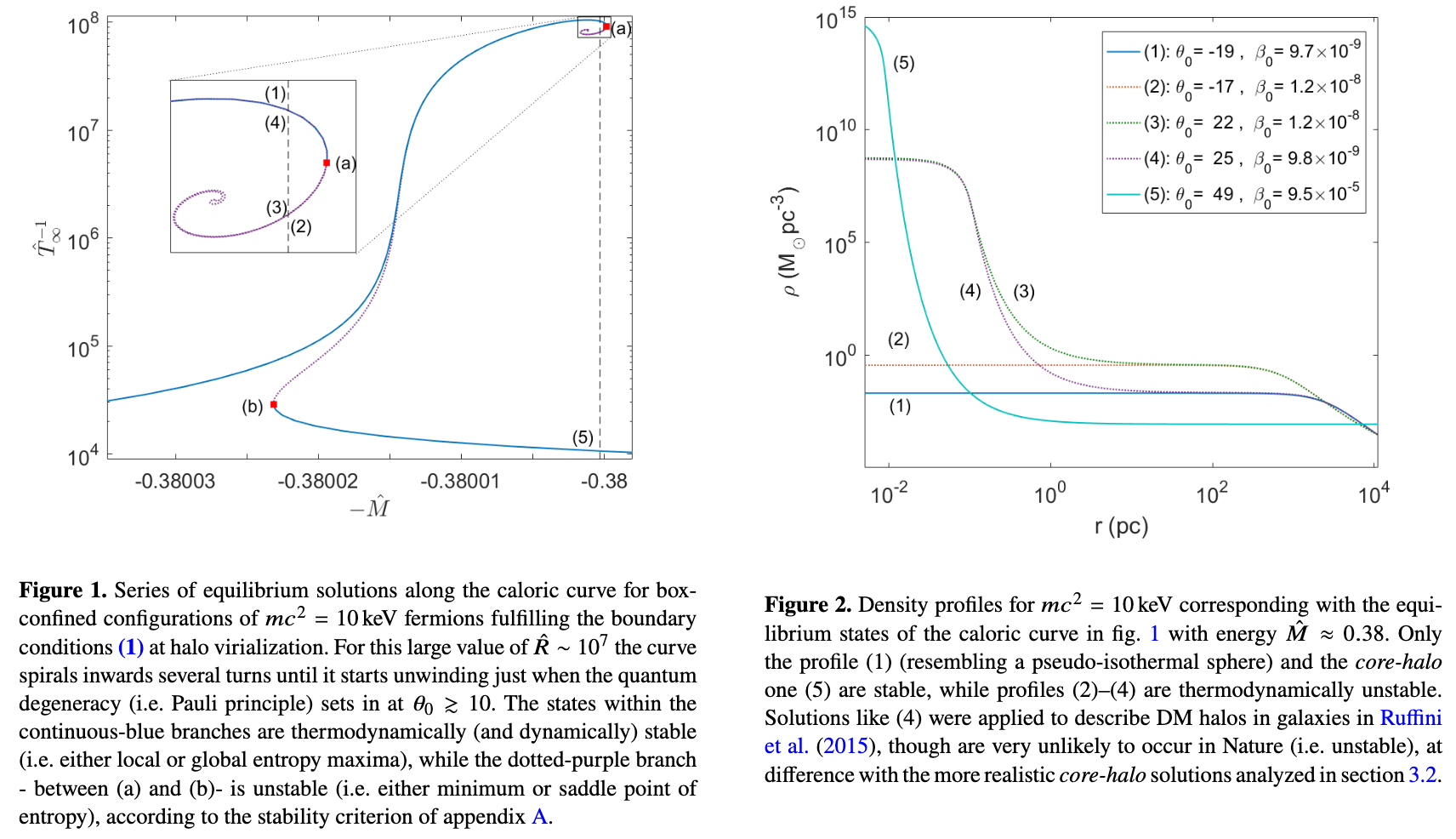}
	 \\~\\
	\includegraphics[width=1.07\textwidth]{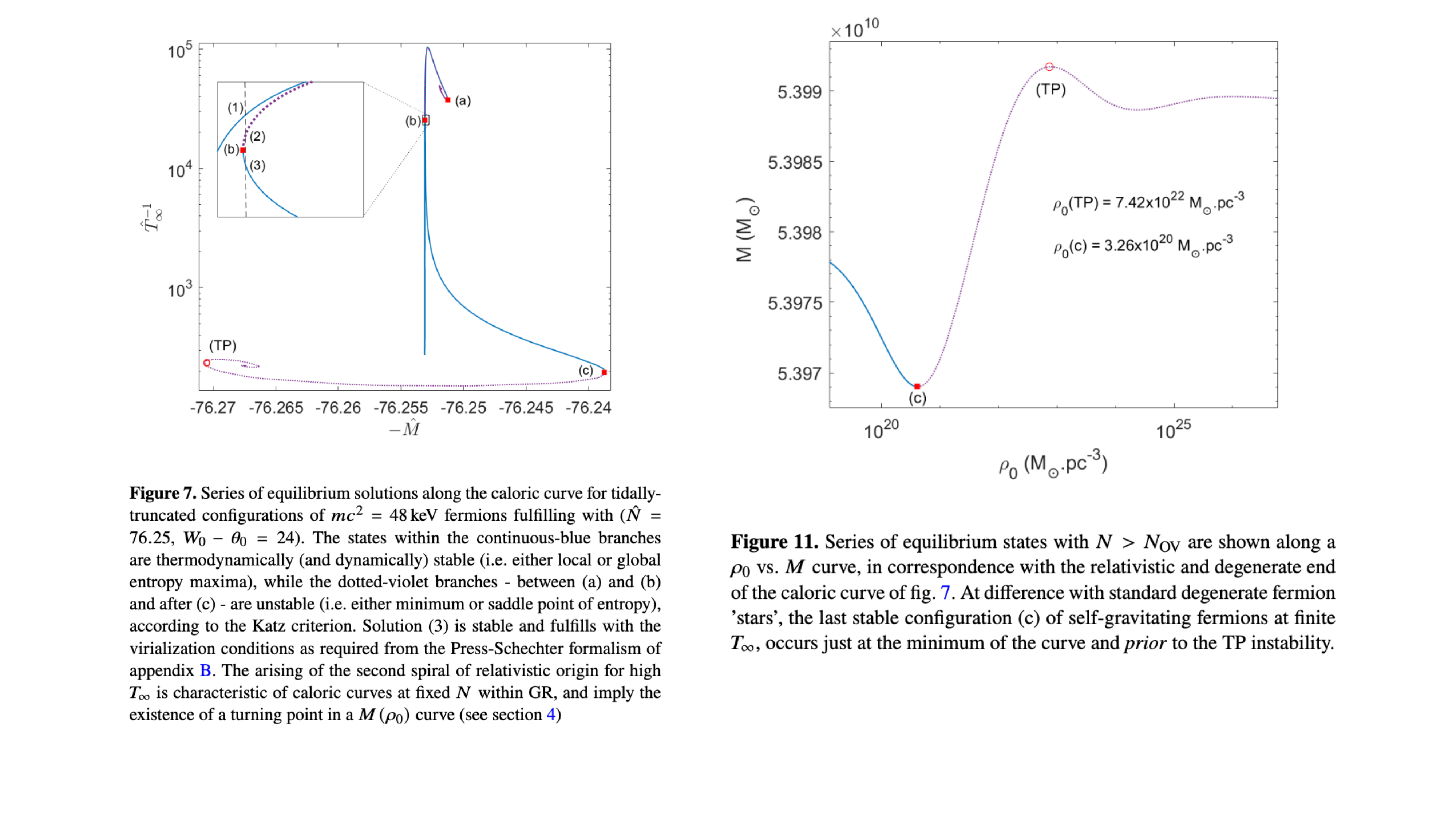}
	\caption{\small Figures taken from the work of \cite{cita1}.
		\normalsize\label{fig:carlos3}
	}
\end{figure}

To determine the stability of the solutions along the series of hydrostatic equilibrium, the Katz criterion  was used. The analisys was performed in the microcanonical ensamble, which required to bind the system in a box of radius $R$ in order for the entropy to reach a maximum\footnote{As we will see in the forthcoming chapters, this somewhat artificial construction is not needed in our work, since we have a natural ``box'' given by the AdS boundary.}. So extra equations are needed, fixing the total number of particles $N$ 
\begin{equation}\label{nofparticles}
{N}=
\frac{g}{2\pi^2}
\int_{0}^{{R }}dr\,{r}^2\, e^{\lambda /2}
\int_{0}^{\epsilon_{c}}f\left(p\right)\,d^3 p,
\end{equation}
and ensuring the continuity of the metric at the boundary $R$ of the box
\begin{equation}\label{schcondition}
	e^{\nu(R)}=1-\frac{2{M}(R )}{{R}}\,.
\end{equation}
Adding these components, the problem is solved for a wide range of parameters, for fixed $m=10$KeV and for different values of ${N}$. We can see in Fig.~\ref{fig:carlos3} some of the solutions for the density present in \cite{cita1} along with an example of a caloric curve.

~ 

~

%\section{Gravitational collapse and turning point}
As we explained in previous sections, the turning point is defined as the point where the total mass is a maximum respect to the central density  ${dM}/{d\rho_{c}}=0$. In \cite{cita38} was shown that the existence of a turning point yields a sufficient condition for the existence of a thermodynamic instability along a family of thermodynamic equilibrium solutions. However, turning points do not provide a necessary condition for thermodynamic instability, {\em i.e.}, the onset of a thermodynamic instability could occur prior to a turning point or without the presence of any turning point at all.
 
A proof that instability can occur before the turning point can be observed by analyzing the figures in \ref{fig:carlos3}. In the left lower end of the caloric curve, where the spiral of relativistic origin rotates clockwise (empty circle in Fig.~\ref{fig:carlos3} right), the turning point occurs at a different energy respect to the last stable configuration (c) along the unstable branch of the curve.

%Thus is show that for self-gravitating system at finite $T_{\infty}$ (temperature as seen by an observer at infinity) in General Relativity, the turning point instability does not coincide with the last stable configuration at (c). 

 \begin{tcolorbox}[colback=red!5!white,colframe=red!75!black,title=Hint \arabic{boxcounter}] For flat space-time, astrophysical solutions for dark matter droplets are found, where we can highlight profiles with a diluted part in the region of low central densities and a core-halo structure for high central densities.
\end{tcolorbox}
\stepcounter{boxcounter}

\part{Contents}
%Holographic Neutron stars
\chapter{Holographic neutron stars}
\label{ch:carlos_nico}

%This chapter provides a review of \cite{cita16} and \cite{cita2}. These works present the basic ideas and equations that will serve as the cornerstone of our thesis.

%\section{Neutron stars at zero temperature}
The work of \cite{cita16, cita17}, and its finite temperature generalization \cite{cita2}, serve as the starting point for our main work on holographic neutron stars. In those papers, the authors numerically solve the Tolman-Oppenheimer-Volkoff equations in an asymptotically anti-de Sitter space. Similarly to \cite{cita23,cita40}, they find a maximum value for the mass as a function of the central density. That such a ``turning point'' behaviour results in an instability has already been explained in the previous chapters. %The authors assume that the star collapses and forms a black hole. %, after which the fermionic operator begins to mix with generic states and thermalizes.  
%They suggest that the collapse is associated with a phase transition on the boundary theory, that turns a high density (baryonic) state into a thermal state (quark gluon plasma).

In the present chapter we present the equations that describe a holographic neutron star at finite temperature \cite{cita2}, and the main results regarding its solutions \cite{cita54}.

\section{Neutron star in AdS spacetime}
\label{sec:bulk3} 
The system  consist in a very large number of neutral self-gravitating fermions in thermodynamic equilibrium, treated within a $3+1$ asymptotically  AdS space-time. The Einstein equations for such system and the fermion energy-momentum tensor are \eqref{eq:Einstein.equations} and \eqref{eq:perfect-fluid} respectively. 
%
% \begin{equation}
% G_{\mu\nu}+\Lambda g_{\mu\nu}= 8\pi G\, T_{\mu\nu}\,,
% \label{eq:einstein-crudas}
% \end{equation}
% %
% being $\Lambda$ a negative cosmological constant, that can be written in terms of the AdS length $L$ as $\Lambda=-3/L^2$, and $G$ the gravitational constant. Regarding the fermion energy-momentum tensor, they approximate it as a perfect fluid
% %
% \begin{equation}
% T_{\mu\nu}=P g_{\mu\nu}+(\rho+P)u_\mu u_\nu\,.
% \label{eq:energy-momentum}
% \end{equation}
%
In the limit $mL\gg 1$ in which there is a huge amount of particles within one AdS radius \cite{cita16}, the density and pressure are given by 
equations %
%\begin{align}
%\rho &= \frac{g}{8\pi^3}\int f(p)\sqrt{p^2+m^2 }\,d^3p\,,
%\label{eq:rho}\\
%P &= \frac{g}{24\pi^3}\int
%f(p)\frac{p^2}{\sqrt{p^2+m^2 }} \,d^3p,
%\label{eq:p}
%\end{align}
%
%that are the same that 
\eqref{eq:rho_flat} and \eqref{eq:p_flat} but here $g$ is the number of fermionic species (that in the holographic limit is taken to be very large),
$\ell$ is replaced by the AdS length $L$,
and the integration runs over all momentum space, with the distribution function $f(p)$ taking the form
\begin{equation}
f(p)=\frac1{e^{\beta\left(\sqrt{p^2+ m^2}-\mu\right)}+1} \,.
\label{eq:fermi-dirac-distribution}
\end{equation}
This is the Fermi-Dirac distribution function for a fermion of mass $m$. This expressions set a double parametric dependence of the density $\rho$ and pressure $P$ on the temperature $T$ and chemical potential $\mu$, which in turn depend on the metric as explained below. Notice that it is not necessary to put a cut off as in %the previous chapter 
Eq.~\eqref{F-D} because AdS has a repulsive wall that prevents the particles from being lost.

To deal with equilibrium configurations of neutral fermions in global AdS, %they consider an 
a stationary spherically symmetric metric with the form \eqref{sch-star} is needed, where again $\ell$ is replaced by the AdS length $L$. The functions in this Ansatz allow  to write the local temperature and chemical potential in terms of the Tolman and Klein relations \eqref{eq:tolman.klein}, which are consistent with the choice $\nu(0)=0$.
%
%
%
%\begin{equation}
%ds^2=L^2\left(-e^{\nu(r)}\,dt^2+e^{\lambda(r)}\,dr^2+r^2\,d\Omega^2_2\right)\, ,
%\label{eq:metricAdS}
%\end{equation}
%
%with $d\Omega^2_2=d\vartheta^2+\sin^2{\vartheta}d\varphi^2$ the same as \eqref{sch-star} but with the multiplicative constant $L$.

%The authors found a convenient re-scale the temperature and chemical potential as a dimensionless combinations $\tilde \mu=\mu/m$ and $\tilde T=T/m={1}/{\tilde \beta}$. Using the re-scaled temperature and chemical potential in Tolman and Klein equations \eqref{tolman_klein}
%
%\begin{equation}
%\tilde T=\tilde T_0 e^{\frac{\nu_0-\nu(r)}{2}}\, ,
%\label{eq:tolman3}
%\end{equation}
%
%
%\begin{equation}
%\tilde \mu=\tilde \mu_0 e^{\frac{\nu_0-\nu(r)}{2}}\, ,
%\label{eq:klein3}
%\end{equation}
%
%where the re-scaled temperature $\tilde T_0$, the re-scaled chemical potential $\tilde \mu_0$ and the metric component $\nu_0$ are measured at a reference point. 
%
%Further re-scalings of the energy density and pressure as $\tilde \rho=GL^2\rho $ and $\tilde P= GL^2P$ gives the result
%
%\begin{align}
%\tilde\rho &= \gamma^2\int_1^\infty \frac{\epsilon^2\sqrt{\epsilon^2-1}}{e^{\tilde \beta\left(\epsilon-\tilde \mu\right)}+1}\,d\epsilon\,,
%\label{eq:rhoe}\\
%\tilde P &= \frac{\gamma^2}{3}\int_1^\infty
%\frac{(\sqrt{\epsilon^2-1})^3}{e^{\tilde \beta\left(\epsilon-\tilde \mu\right)}+1}\,d\epsilon\,.
%\label{eq:pe}
%\end{align}
%
%Where they have re-written the integrals in terms of the variable $\epsilon = \sqrt{1+p^2/m^2}$, and they defined the dimensionless coupling $\gamma^2=gGL^2m^4/2\pi^2$.

A convenient re-parametrization of $\lambda$ and $\nu$ in terms of new functions ${M}$ and $\chi$ is
\begin{eqnarray}
&&e^{\lambda}=\left(1-\frac{2 {M} }{r}+{r^2}\right)^{-1},	
%\label{eq:redefinition-lambda}\\&&
\qquad\qquad\quad
e^{\nu}=e^\chi\left(1-\frac{2 {M} }{r}+{r^2}\right)\,.
\label{eq:redefinition-nu}
\end{eqnarray}
These differ from \eqref{eq:lambda} in the inclusion of the AdS asymptotics through the $r^2$ term.
The resulting metric takes the form of \eqref{eq:AdSBlackHole} but now $M$ and $\chi$ are functions of $r$. The Einstein equations are again given by \eqref{eq:mass}-\eqref{eq:Phi}, which must be solved numerically with the energy density and pressure given by \eqref{eq:rho_flat}-\eqref{eq:p_flat} where the distribution function is \eqref{eq:fermi-dirac-distribution} in terms of the temperature and chemical potential obtained from \eqref{eq:tolman.klein}.

%read
%
%\begin{align}
%&\frac{d \tilde{M}}{d r}=4\pi\,r^2 \tilde\rho \,,
%\label{eq:eqs2a}
%\\
%&\frac{d\nu}{d r}=\frac{2}{r^2}\left(M+r^3\left(4\pi\,\tilde P-\frac32%+P_{vac}\right)\right)e^{\lambda(r)}, \\
%&\frac{d\chi}{d r}
%=8\pi r\left(\tilde P+\tilde\rho\right)e^{\lambda},
%\\
%\label{eq:eqs2b}
%\end{align}
%
%The equations \eqref{eq:eqs2a}-\eqref{eq:eqs2b}, together with the definitions \eqref{eq:rhoe}-\eqref{eq:pe} in terms of the spatially varying temperature and chemical potentials given by \eqref{eq:tolman3}-\eqref{eq:klein3} must be solved numerically for the variables $\tilde{M}, \chi,\tilde{\rho},\tilde{P}$. Notice that the fermion mass $m$ plays no role other that setting the scale. They chose boundary conditions at the center of the star $r=0$ as
%
%\begin{eqnarray}\label{eq:NumericalConditions}
%\tilde{M}(0)&=&0\,,\qquad\qquad\mbox{(or in other words }\lambda(0)=0\mbox{)}\,,\nonumber\\\chi(0)&=&0\,,\nonumber\\
%&&\nonumber\\
%\tilde T(0)&=&\tilde T_0\,,\nonumber\\\tilde \mu(0)&=&\tilde \mu_0\ \equiv\ \Theta_0\Tilde T_0+1\,.
%\label{mu0}
%\end{eqnarray}
%
%Here $\Theta_0$ can be regarded as the central degeneracy. The resulting solutions are indexed by the parameters $\tilde T_0, \Theta_0$ and $\gamma$.
%

\section{Holographic dual of the neutron star}
%
%\subsection{A highly degenerate fermionic state}
In order for holography to work, the conformal field theory must have a large central charge. This can be achieved in the present approximation, as explained in \cite{cita16}, by taking a proportionally large mass $mL$.

The presence of a self-gravitating fermionic gas in the AdS interior is interpreted from the dual point of view as a highly degenerate fermionic state of the strongly coupled field theory defined on the boundary. 
Since the geometry asymptotes to global AdS \eqref{eq:globalAdS}, its conformal boundary is a cylinder $\mathbb{R}\times S_2$. The $\mathbb{R}$ direction is coordenatized by the variable $t$ on our metric Ansatz, while the $S_2$ represents the spatial directions. This implies  
\begin{itemize}
    \item The time in the boundary is measured by $e^{\frac{\nu_\infty}2}t$ where ${\nu_\infty}$ is the value of $\nu$ at $r\to\infty$. The corresponding Wick rotation gives a temperature $T=e^{-\frac{\nu_\infty}2}T_c$ for the boundary field theory.
    \item The boundary theory is contained in a spherical vessel $S_2$. Such finite volume provides a scale which disentangles the temperature and chemical potential axes in the resulting phase diagram, as explained in section \ref{sec:intro.fermions}.
\end{itemize}
Finally, the local chemical potential's boundary value $\mu=e^{-\frac{\nu_\infty}2}\mu_c$  acts as a source for the particle number operator in the boundary theory and can thus be identified with the boundary chemical potential.
 \begin{tcolorbox}[colback=red!5!white,colframe=red!75!black,title=Hint \arabic{boxcounter}] The holographic dual of the neutron star in AdS corresponds to a highly degenerate fermionic state on a conformal field theory at finite temperature and chemical potential, which is contained into a finite volume spherical vessel. 
\end{tcolorbox}
\stepcounter{boxcounter}

\section{Results}
\begin{figure}[H]
	\centering
    \includegraphics[width=0.45\textwidth]{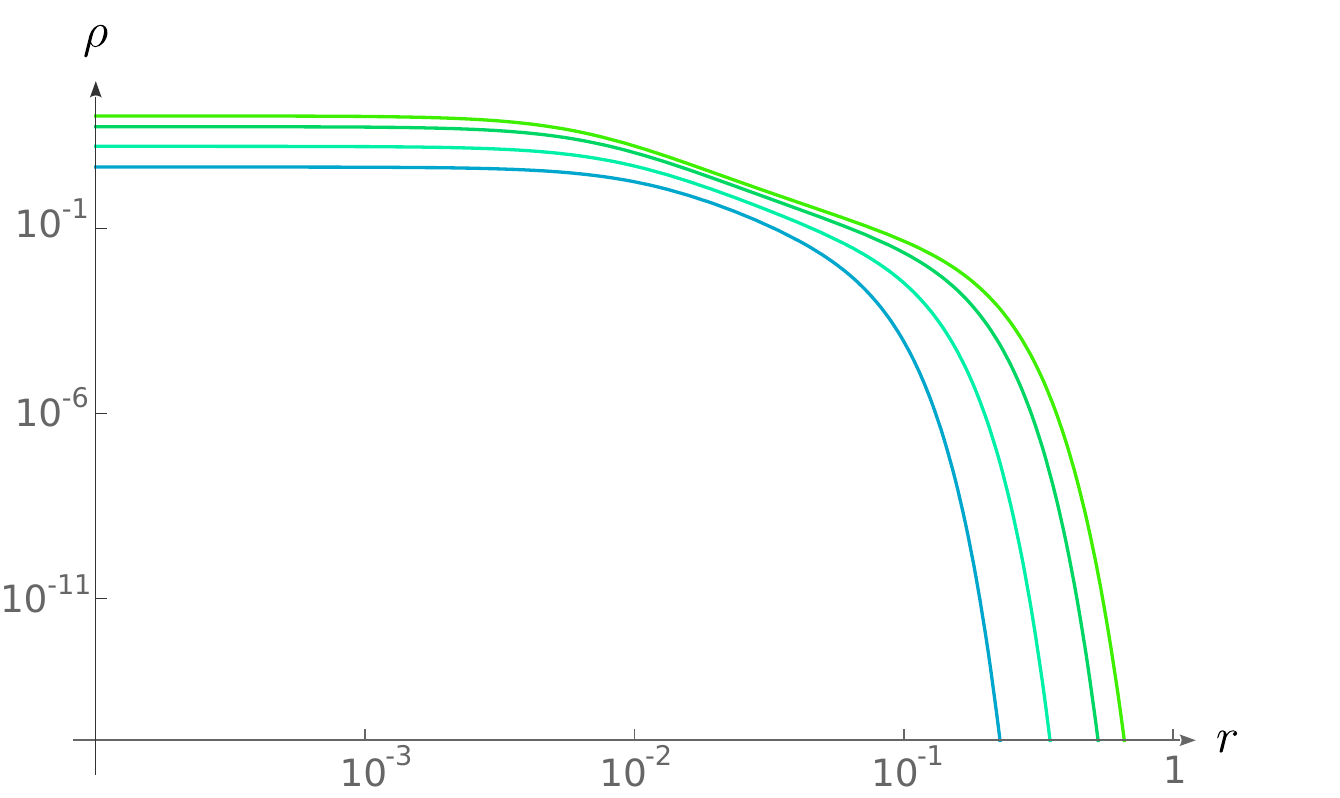}
    \hfill
%    \includegraphics[width=0.45\textwidth]{Figures/turning.point.-15.pdf}
%    \\
    \includegraphics[width=0.45\textwidth]{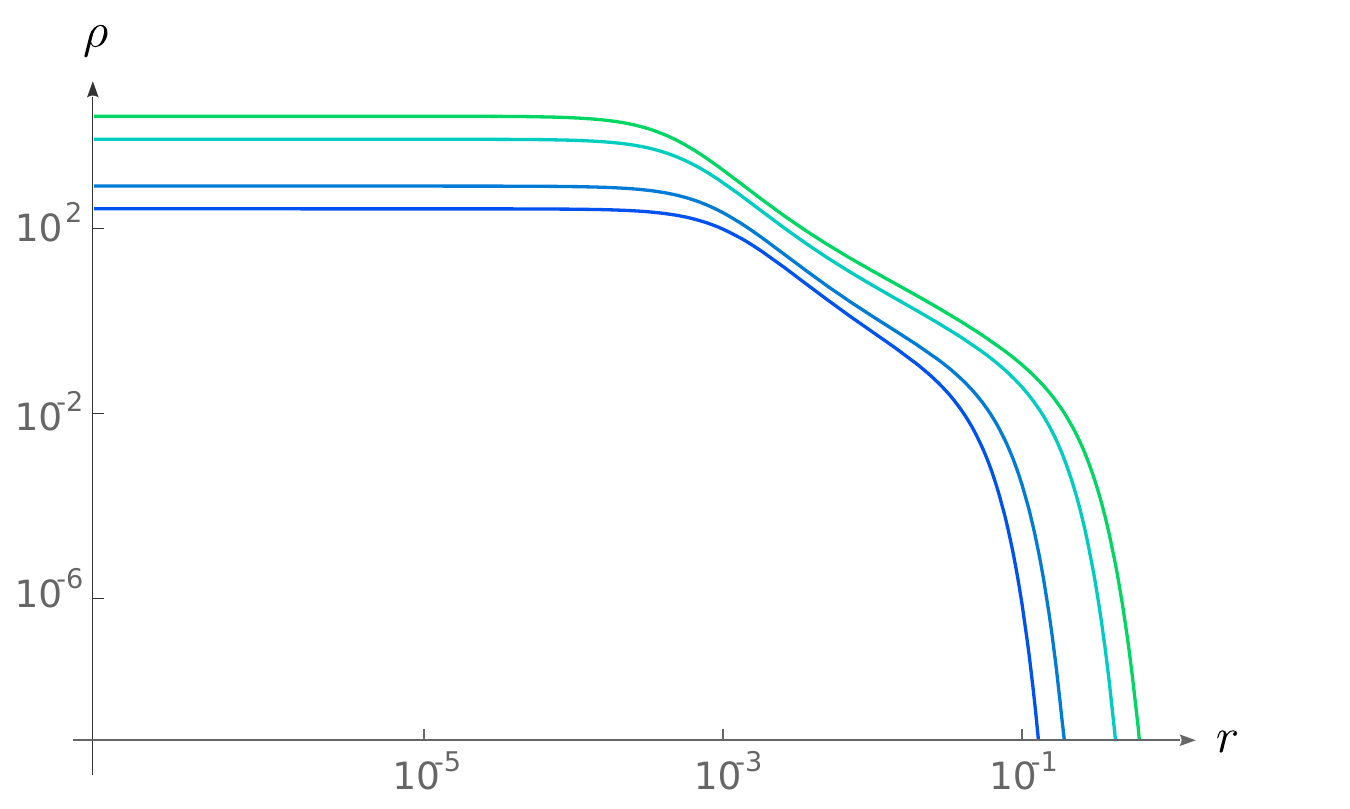}
\\
%\hfill
%    \includegraphics[width=0.45\textwidth]{Figures/turning.point.-10.pdf}
%    \\    
	\includegraphics[width=0.45\textwidth]{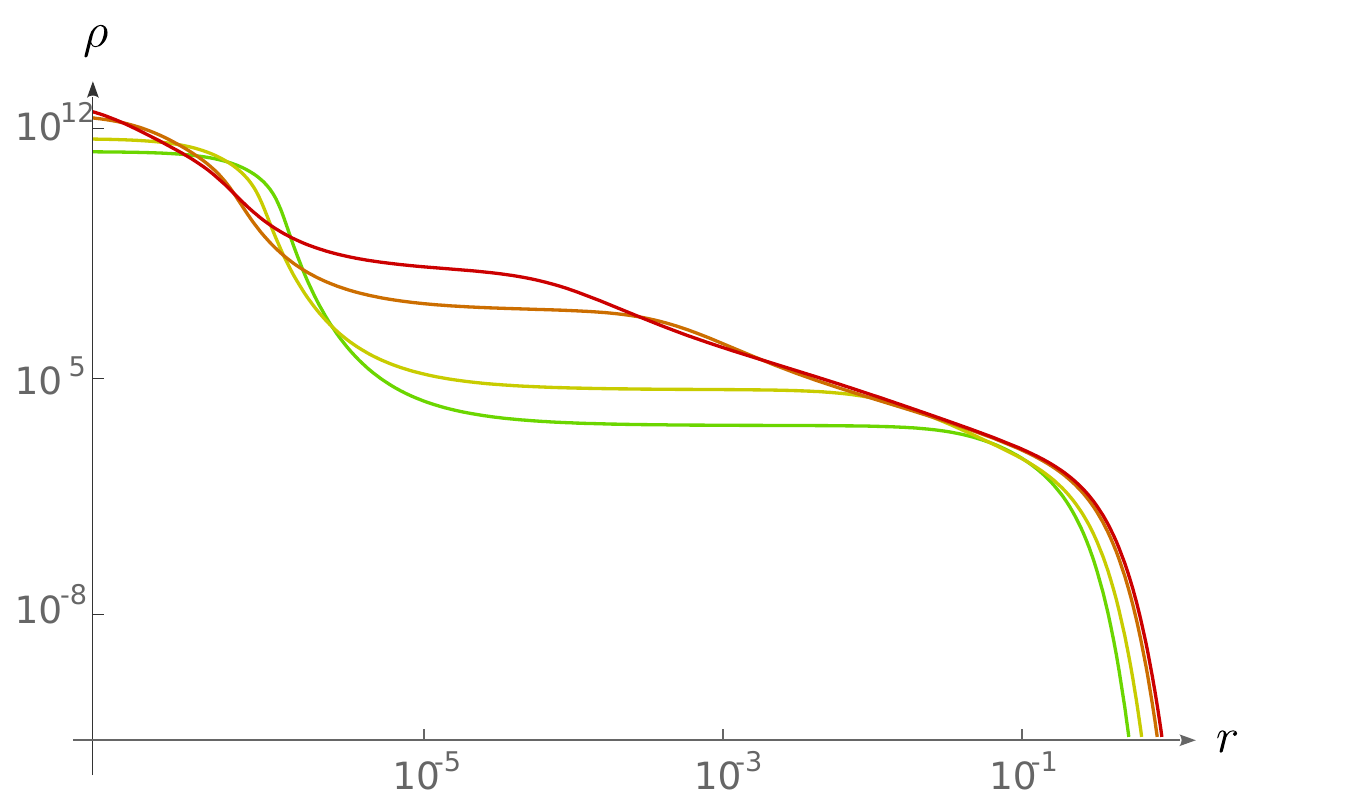} 
    \hfill
%	\includegraphics[width=0.45\textwidth]{Figures/turning.point.30.pdf}
%    \\
    \includegraphics[width=0.45\textwidth]{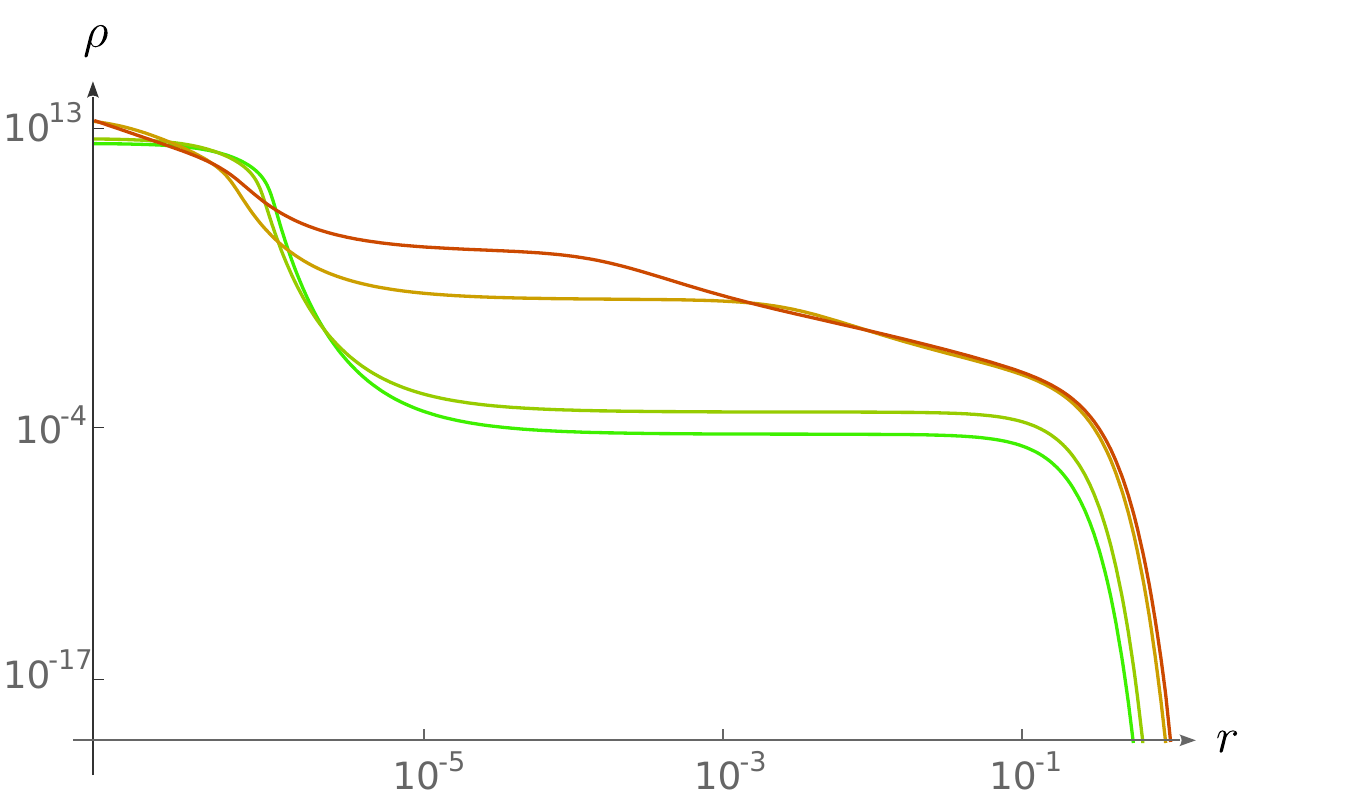}
%	\hfill
%	\includegraphics[width=0.45\textwidth]{Figures/turning.point.50.pdf}
\\
\vspace{0.7cm}
\flushleft
\includegraphics[trim=-5mm 0mm 0 10mm,width=0.16\textwidth]{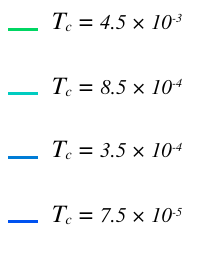}	
\qquad\quad\,
\includegraphics[trim=-5mm 0mm 0 10mm,width=0.16\textwidth]{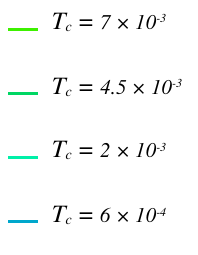}	
\qquad\qquad\;
\includegraphics[trim=-5mm 0mm 0 10mm,width=0.16\textwidth]{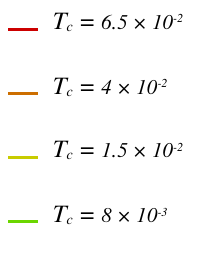}	
\qquad\quad\,
\includegraphics[trim=-5mm 0mm 0 10mm,width=0.16\textwidth]{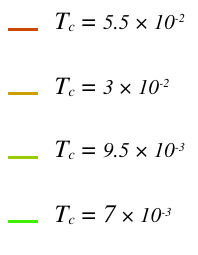}
\hfill 
%\put(-428,-15){\scriptsize$\Theta_c=-15, -10, 30, 50$   from top to bottom.\normalsize}
%\vspace{0.3cm}
\hfill\\
	\caption{\label{fig:positivetheta1} 
    \small Logarithmic plots of the density profiles. At high central degeneracies, the density develops a dense core and a diluted halo that decreases sharply at the boundary of the star for low temperatures. For higher temperatures the edge of the star takes a power law form.  \underline{Top:} Central degeneracies $\Theta_c=-15$ and $\Theta_c= -10$. \underline{Bottom:} Central degeneracies $\Theta_c=30$ and  $\Theta_c= 50$.
   \normalsize\normalsize }
\end{figure}

The system of equations defining the background was solved numerically using routines in {Fortran} and {Mathematica} for different values of the parameters $T_c$, $\Theta_c$,  $\gamma$. 
The resulting density profiles are depicted in Fig.~\ref{fig:positivetheta1}, % and \ref{fig:negativetheta3}, 
and can be sumarized as follows:

\begin{itemize}
\item 
At negative central degeneracy $\Theta_c < 0$, fermions follow the Boltzmannian regime of the Fermi-Dirac distribution, the star being fully supported against gravity by thermal effects. 

In this case,  for small enough central temperature $T_c$ the solutions have a diluted density profile with a plateau followed by a smooth transition towards a sharp edge. Increasing the central temperature causes the star to become more extended, with the density exhibiting a power law edge. % Interestingly, this coincides with the critical temperature  $\tilde T_c$ at which the turning point in the mass shows up. 
\item 
For positive and large enough central degeneracies, quantum effects become important in holding the star against gravity. 

At  small central temperatures $T_c$, the density profiles exhibit a well-defined core-halo structure. This is analogous to the flat space case \cite{cita44, cita45, cita46, cita47, cita37} and coincides with the results of \cite{cita2}. The density develops a central plateau that extends up to a well-defined radius $r=r_c$, which is identified as the ``core'', and a lower exterior plateau identified as the ``halo'' which ends at a sharp edge at $r=r_b$ where the density drops to zero.  
The highly-dense core is supported by degeneracy pressure, while the outer halo is held  by thermal pressure

As the central temperature increases, the core becomes more compact.
%, until at a critical temperature $\tilde T_c$ a turning point on the mass is reached. Above this critical value (such as $\tilde T_c \approx 10^{-2}$ for $\Theta_c = 30$)
At a certain temperature the outer halo region as well as the inner dense core develop a power-law morphology.  This is again analogous to the asymptotically flat case \cite{cita48}, and it can be taken as an indication of a critical phenomenon taking place. We will further explore such hypothesis in the forthcoming chapters. 
\end{itemize}

\begin{figure}[ht]
	\centering
	\includegraphics[width=1.\textwidth]{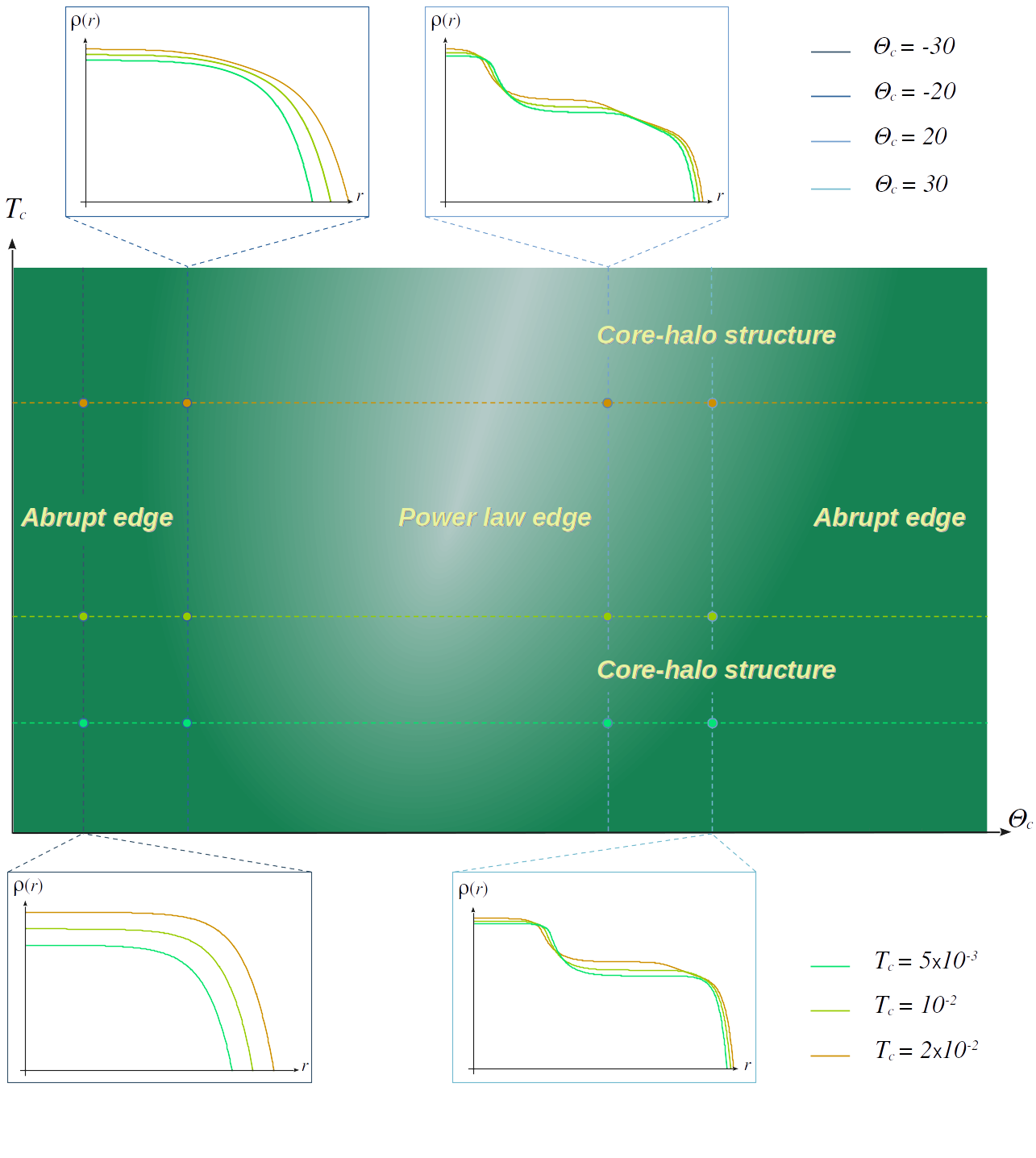}
	\caption{ \small
		Central temperature ${T}_c$ {\em vs.} central degeneracy $\Theta_c$ phase diagram. The dotted horizontal lines represent constant temperatures, the vertical ones represent constant values of $\Theta_c$. The density profiles plotted inside each box are located in the phase diagram by a dot with the same color sitting along the corresponding vertical line. 
		Solutions have an abrupt edge in the darker region, while in the lighter zone they develop a power law edge. The comparison with Fig.~\ref{fig:phase.diagram.metal} is suggestive.
		\normalsize}
	\label{fig:diagram.background}
\end{figure}

The results obtained so far can be summarized in a phase diagram in the central temperature $T_c$ versus central degeneracy ${\Theta_c}$ plane, as shown in Fig.~\ref{fig:diagram.background}. The power law behavior at the boundary of the density profile, appear at intermediate degeneracies and high enough temperatures. Notice the resemblance of the ``power law edge'' zone with the critical region of  Fig.~\ref{fig:phase.diagram.metal}.

 \begin{tcolorbox}[colback=red!5!white,colframe=red!75!black,title=Hint \arabic{boxcounter}]  The density profile of the solutions shows a core-halo structure at large central degeneracies, that disappears as the degeneracy is decreased. 
\end{tcolorbox}
\stepcounter{boxcounter}

\newcounter{boxnew}
\setcounter{boxnew}{1} 
 \begin{tcolorbox}[colback=blue!5!white,colframe=blue!75!black,title=Hint \arabic{boxnew}] A power law behavior   occurs at the boundary of the density profile, which can be taken as an indication of a critical phenomenon. This happens at intermediate values of the central degeneracy in a wedge-like region, which results very suggestive when compared to the critical region of the metallic degrees of freedom of a High $T_c$ superconductor. 
\end{tcolorbox}
\stepcounter{boxnew}
\newpage

%Correlators
\chapter{Bosonic two-point correlator}
\label{ch:nmodes}

In the previous chapter, we examined the physics of a holographic neutron star at finite temperature. Our investigation revealed a diverse solution space, defined by the central temperature $T_c$, the central degeneracy $\Theta_c$, and the effective number of degrees of freedom $\gamma$. This space includes configurations with a dense core and a diluted halo, as well as more regular non-cored solutions. Furthermore, we generated a phase diagram for bulk solutions.  

In the present chapter we probe the backgrounds we have found with a bosonic scalar two-point correlator and find the normal modes of a scalar operator. Additionally, we investigate the relationship between these observables and the different regions of the phase diagram \ref{fig:diagram.background}.

\section{Scalar correlator in the worldline limit}
\label{sec:correlators3}
\subsection{Massive Euclidean geodesics}
As explained in sections \ref{sec:worldline} and \ref{sec:global}, the Matsubara two point correlator of a boundary scalar operator with large conformal dimension $\Delta={\sf m} L$ can be obtained in terms of Euclidean bulk geodesics of a particle with large mass ${\sf m}$ joining the corresponding points. It is given as a function of the angular span of the points on the boundary $\vartheta, \varphi$ and the elapsed Euclidean time $t_E$ in terms of the  on-shell form of the Euclidean world line action \eqref{eq:worldline.action}.  In our  coordinates \eqref{sch-star} it reads
\begin{equation}
S_{\sf EWL} ={\sf m} L\int d\tau_E\sqrt{e^{\nu(r)}\, {t'_E}^2+e^{\lambda(r)}\, {r'}^2+r^2({\vartheta'}^2+\sin^2\!\vartheta\,{\varphi'}^2)}\, ,
\label{eq:action-particle}
\end{equation}
where $\tau_E$ is the an Euclidean affine parameter and $(~')=\partial_{\tau_E}(~)$. 
Without loss of generality, we can concentrate on trajectories completely contained in the equatorial plane $\vartheta=\pi/2$. Furthermore, this action is invariant under arbitrary re-parametrizations of $\tau_E$, which means we can fix $\tau_E = \varphi$. We then obtain
\begin{equation}
S_{\sf EWL} ={\sf m} L\int d\varphi\sqrt{r^2+e^{\lambda(r)}\,r'^2}\, .
\label{eq:action-particle-fixed}
\end{equation}
Where we restricted to trajectories with constant $t_E$.
The resulting equations for the single dynamical variable $r$ are invariant under $\varphi$ translations, implying the conservation of the quantity
\begin{equation}
r_*=\frac{\,r^2}{\sqrt{r^2+e^{\lambda(r)}\,r'^2}}\,.
\label{eq:p-phi}
\end{equation}
By evaluating the right hand side at the tip of the trajectory where $r'=0$, we see that $r_*$ corresponds to the radial position of the tip $r|_{r'=0}=r_*$. Solving the above equation for $r'$ we get
\begin{equation}
r'=\frac r{r_*}e^{-\frac{\lambda(r)}2}\sqrt{r^2-r_*^2}\,.
\label{eq:velocity}
\end{equation}
This allows to relate the integration constant $r_*$ with the angular separation at the boundary $\varphi$ of the initial and final points of the trajectory
\begin{eqnarray}
\varphi &=&  \int \frac{dr}{r'}
=2r_*\int_{r_*}^{r_\epsilon}\!\!dr \frac{e^{\frac{\lambda(r)}2}}{r\sqrt{r^2-r_*^2}}\,,
\label{eq:anmgular-span}
\end{eqnarray}
where the cutoff $r_\epsilon$  must be taken to infinity at the end of the calculations.
On the other hand, plugging eq. \eqref{eq:velocity} into the gauge-fixed action  \eqref{eq:action-particle-fixed} results in %the on shell form
\begin{equation}
S_{\sf EWL}^{\sf on\,shell}%(\Delta\varphi) %=\frac{{\sf m}  L}{r_*}\int\!\!d\varphi r^2
=2{\sf m}  L\int_{r_*}^{r_\epsilon}\!\!dr  \frac{re^{\frac{\lambda(r)}2}}{\sqrt{r^2-r_*^2}} \,,
\label{eq:action-particle-on-shell}
\end{equation}
where the same cutoff was included. Notice that, when re-inserted into the correlator \eqref{eq:correlator}, the %logarithmic
divergence of this integral on the cutoff is canceled by the pre-factor%, as anticipated.

\begin{figure}[t]
	\centering
	\vspace{-1.8cm}
	\includegraphics[width=1.\textwidth]{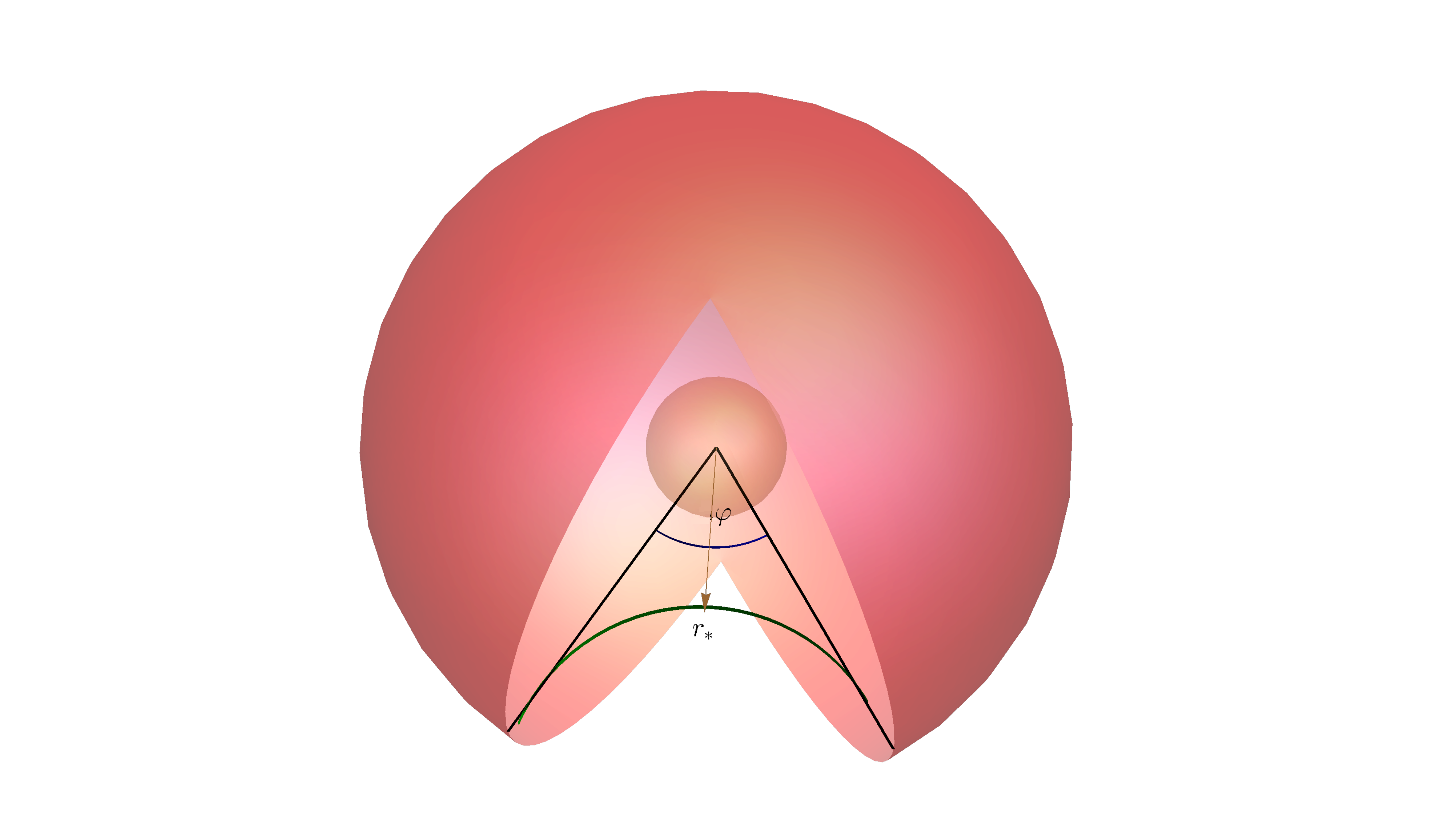}
	%\vspace{-1.7cm}
	\caption{\label{fig:sphere}\small We study the scattering problem of a massive Euclidean particle entering the geometry from infinity, approaching the neutron star up to a tip radius $r_*$, and then moving again to the asymptotic region, spanning an angle $\varphi$.	\normalsize}
\end{figure}

\begin{figure}[ht]
	\centering
	\begin{subfigure}[b]{\textwidth}
		\includegraphics[width=.45\textwidth]{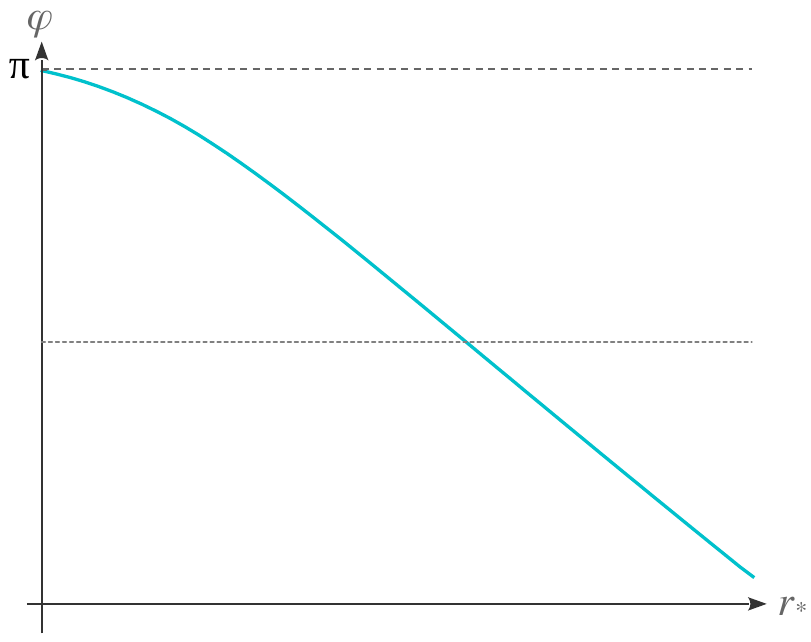}
		\hfill
		\includegraphics[width=.45\textwidth]{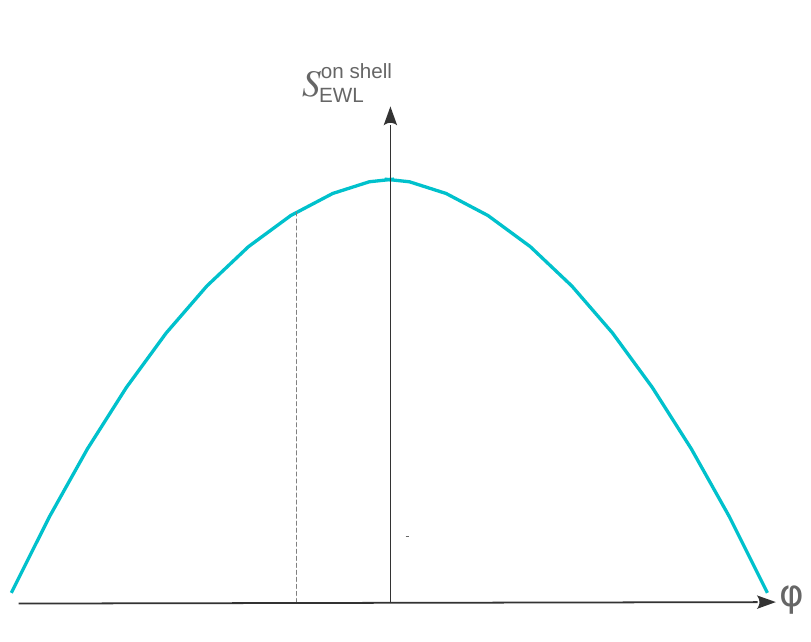}
	\end{subfigure} 
	\\~\\
	\begin{subfigure}[a]{\textwidth}
		\includegraphics[width=.45\textwidth]{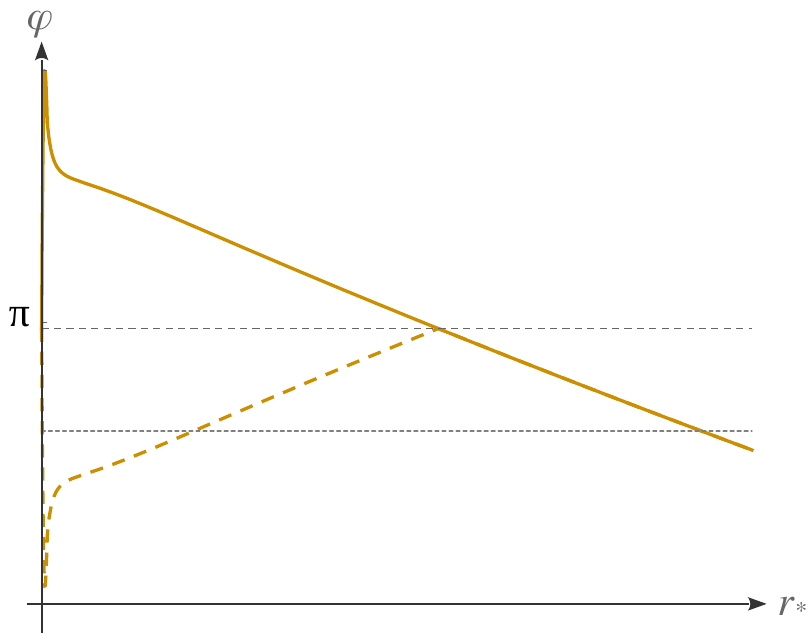}
		\hfill
		\includegraphics[width=.45\textwidth]{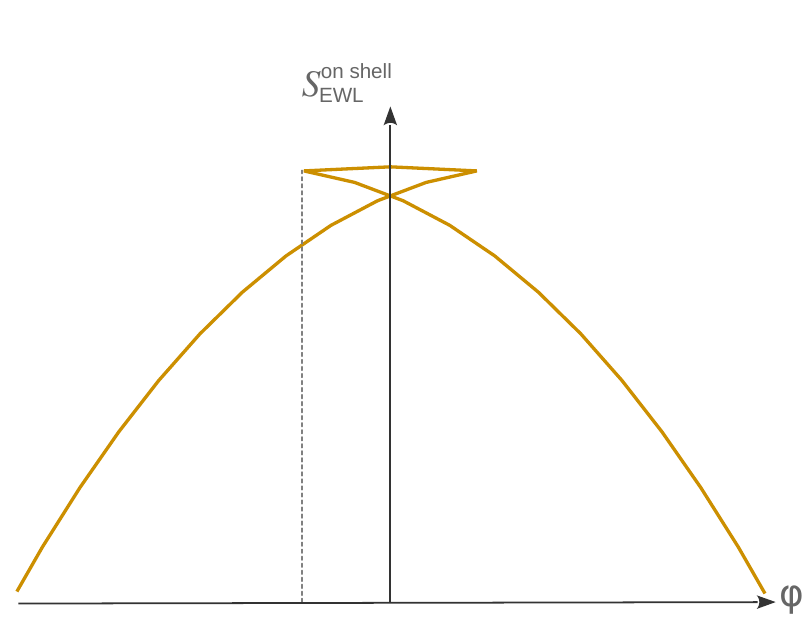}
	\end{subfigure}
	\caption{\small 
		\underline{Left:} Plots of the angular span $\varphi$ as a function of the apsidal position $r_*$. \underline{Right:} The corresponding on shell Euclidean action. \underline{Top:} if the angle $\varphi$ is a monotonic function of $r_*$ we get an single-valued action (the dotted line at $\varphi=3$ intersects the curves only once).
    	\underline{Bottom:} if the angle oscillates as a function of $r_*$, it be reflected at $\varphi=\pi$ into ist natural range $(0,\pi]$ (dashed line),  resulting in a multivalued action (the dotted line sits at $\varphi=3$ intersects the curves three times). \normalsize
		\label{fig:angular}}
\end{figure}

Formulas \eqref{eq:anmgular-span} and \eqref{eq:action-particle-on-shell} give us a parametric description of $S_{\sf EWL}^{\sf on\,shell}(\varphi)$ with parameter $r_*$.
For $r_*=0$ we have a head on collision, the particle being not scattered by the neutron star $\varphi=\pi$. On the other hand in the limit of very large $r_*$ the particle does not sink into the bulk at all, and we get $\varphi=0$. In the intermediate region, the behavior of $\varphi$ as a function of $r_*$ can be either monotonic or non-monotonic. In the last case, the same angle $\varphi$ is spanned by geodesics with different values of the apsidal radius $r_*$. Since they correspond to different values of the on-shell action, a multivalued relation appears, see Fig. \ref{fig:angular}. The scalar correlator being given by the shorter branch, it develops a non-vanishing  derivative at $\varphi=\pi$.

%
%Fig6

\subsection{Results}
\label{sec:results3}
 
\begin{figure}[H]
	\centering
	\includegraphics[width=0.45\textwidth, height=0.21\textheight]{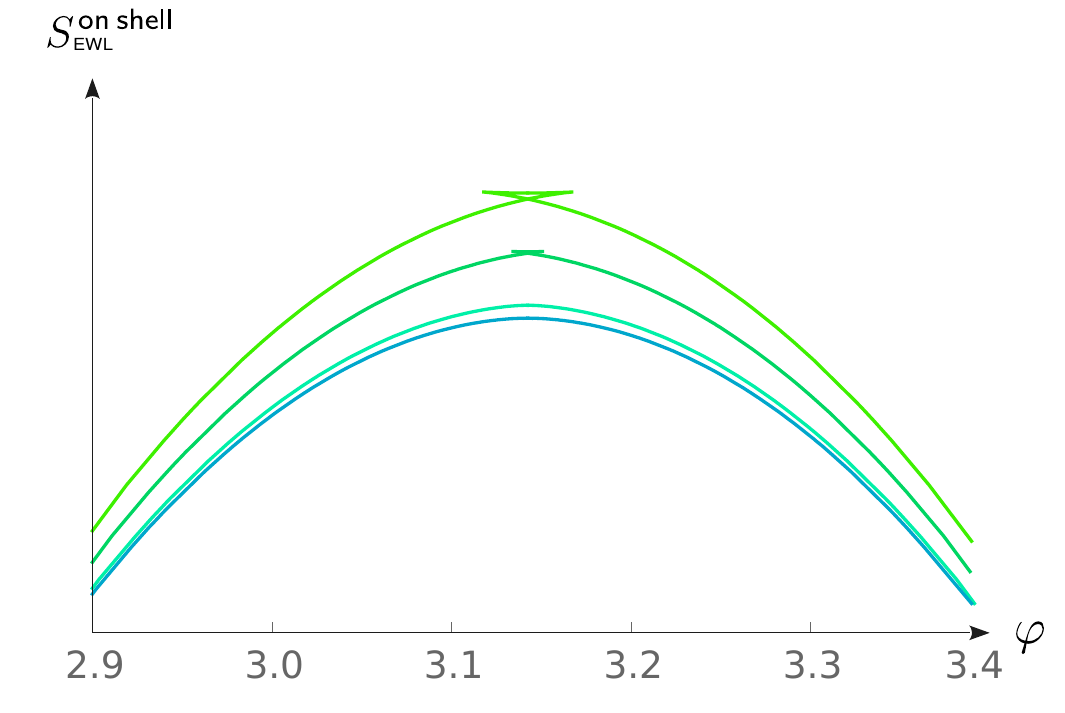}
	\hfill
	\includegraphics[width=0.45\textwidth]{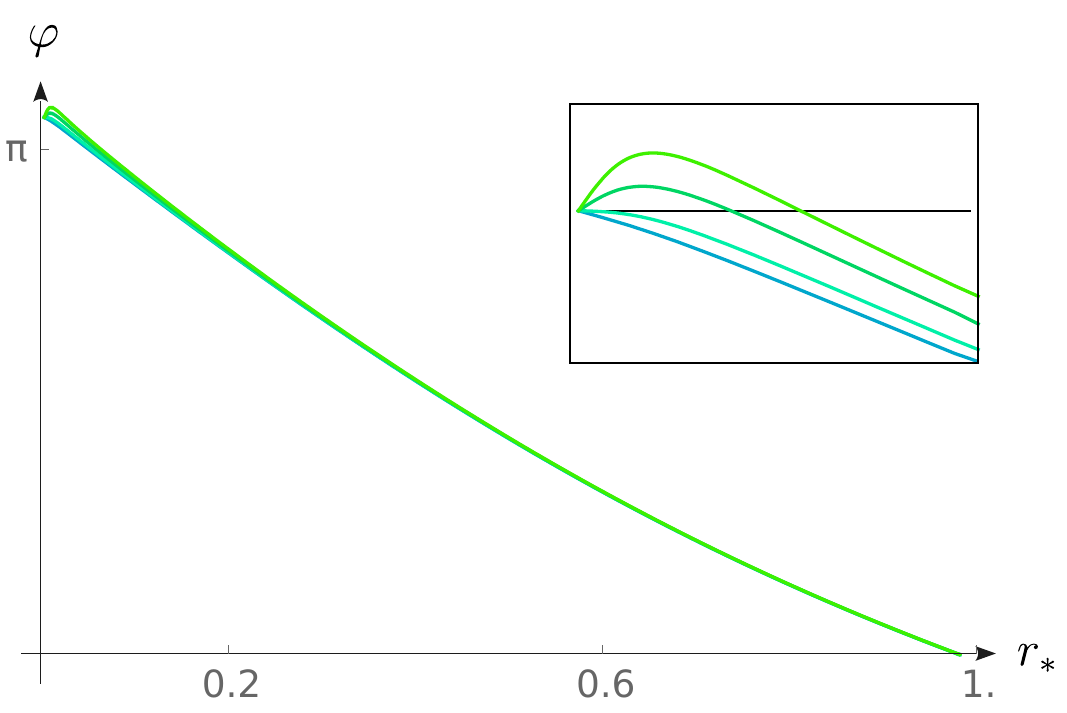}
    \\
    	\includegraphics[width=0.45\textwidth, height=0.20\textheight]{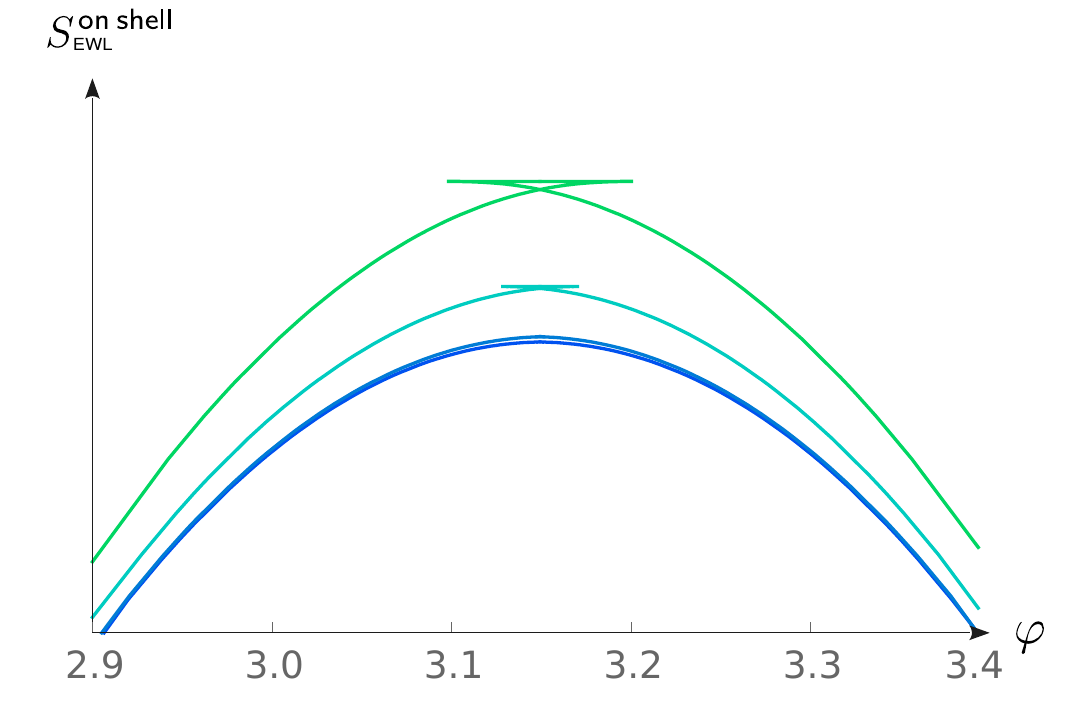}
	\hfill
	\includegraphics[width=0.45\textwidth]{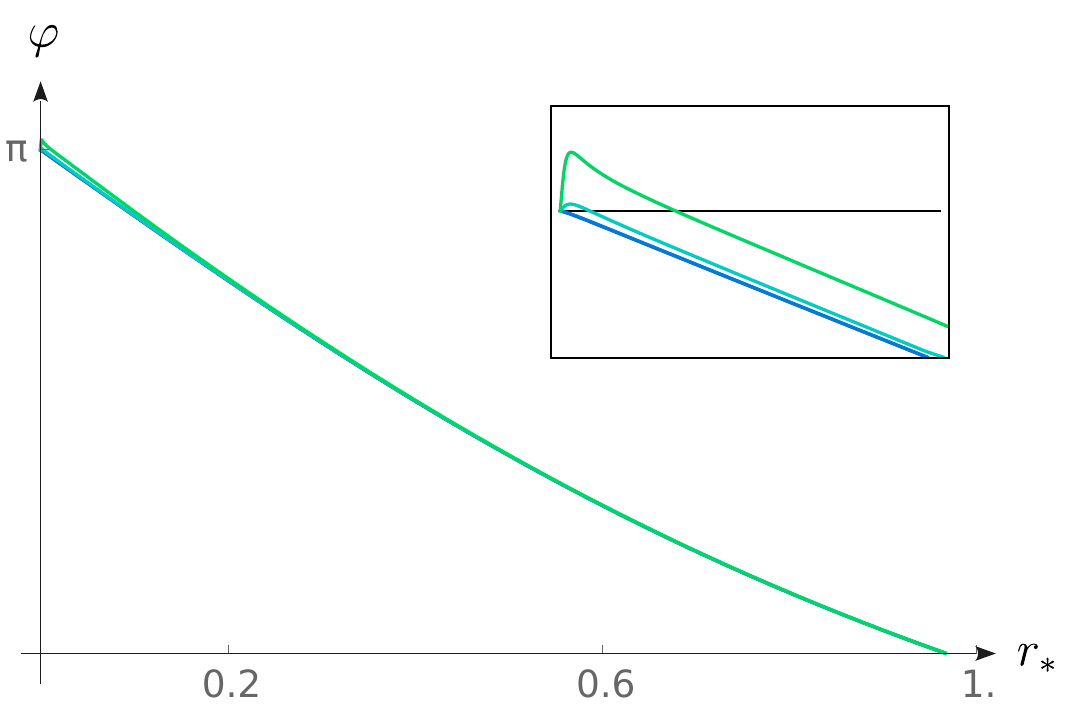}
    \\
	\includegraphics[width=0.45\textwidth]{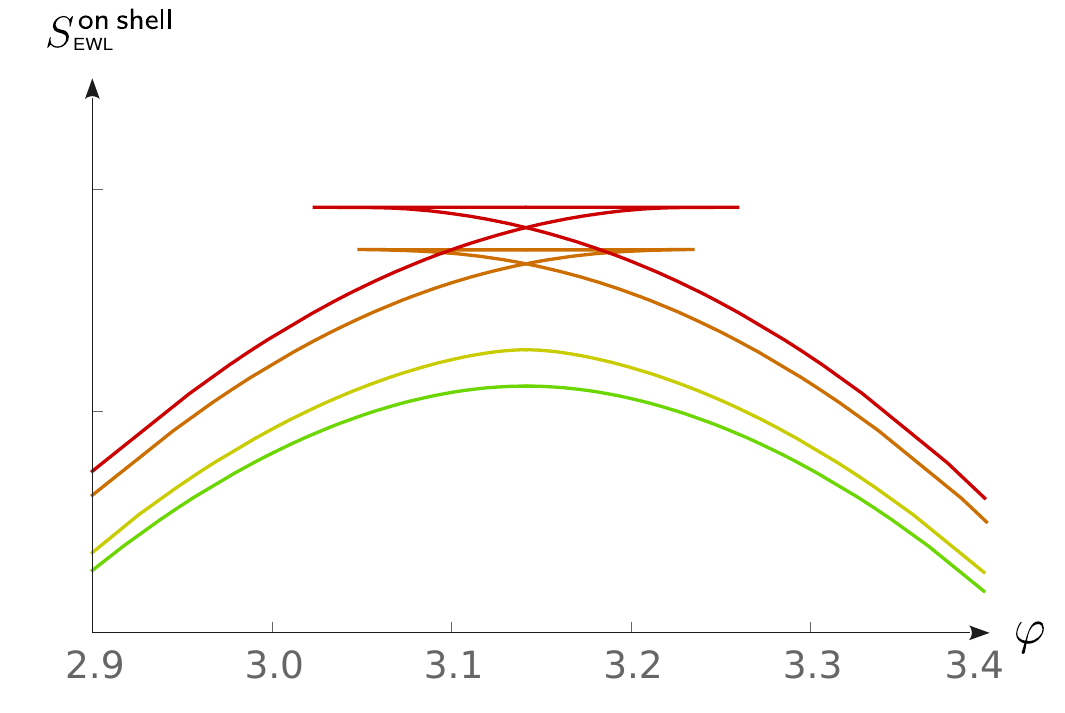}
	\hfill
	\includegraphics[width=0.45\textwidth]{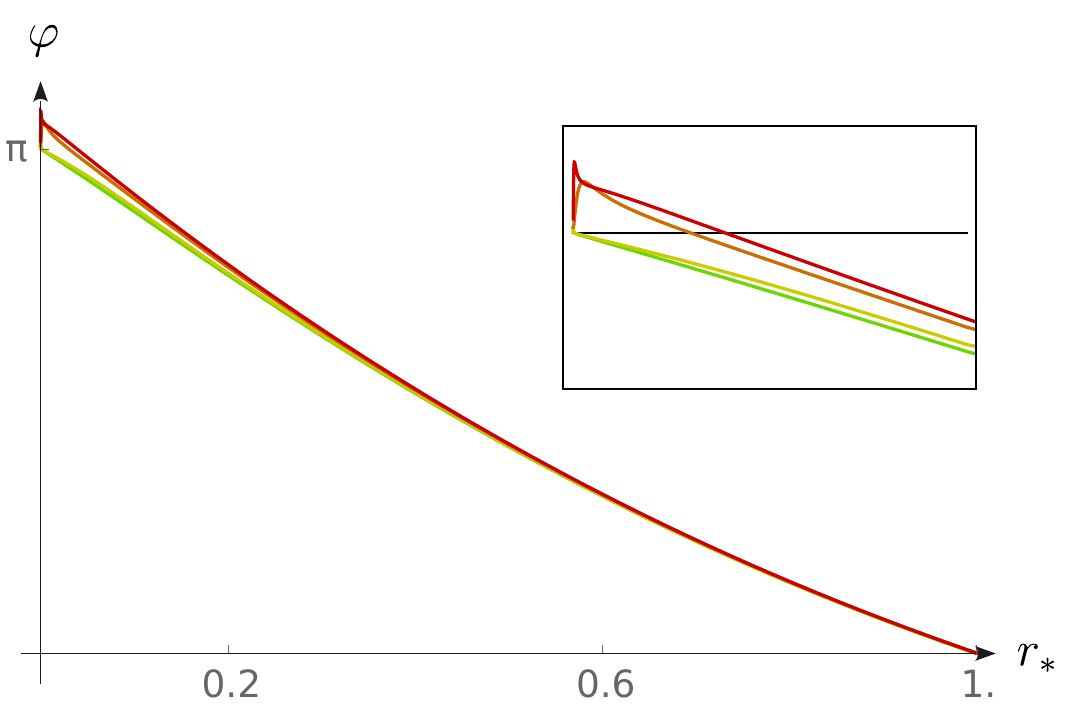}
    \\
    
	\includegraphics[width=0.45\textwidth]{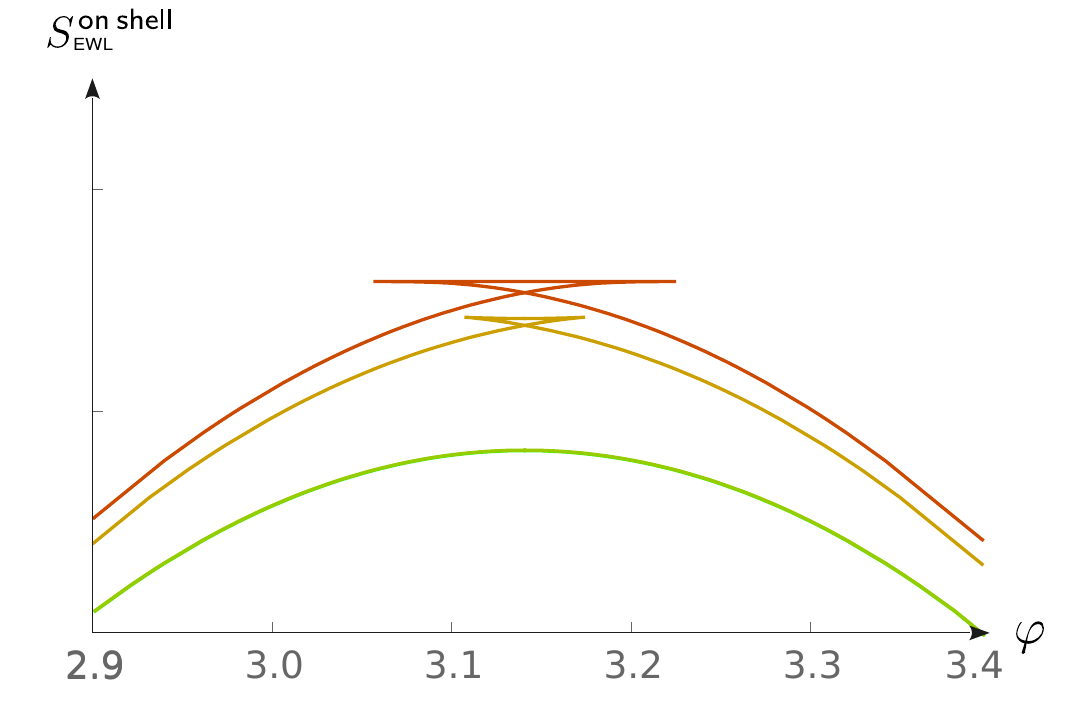}
	\hfill
	\includegraphics[width=0.45\textwidth]{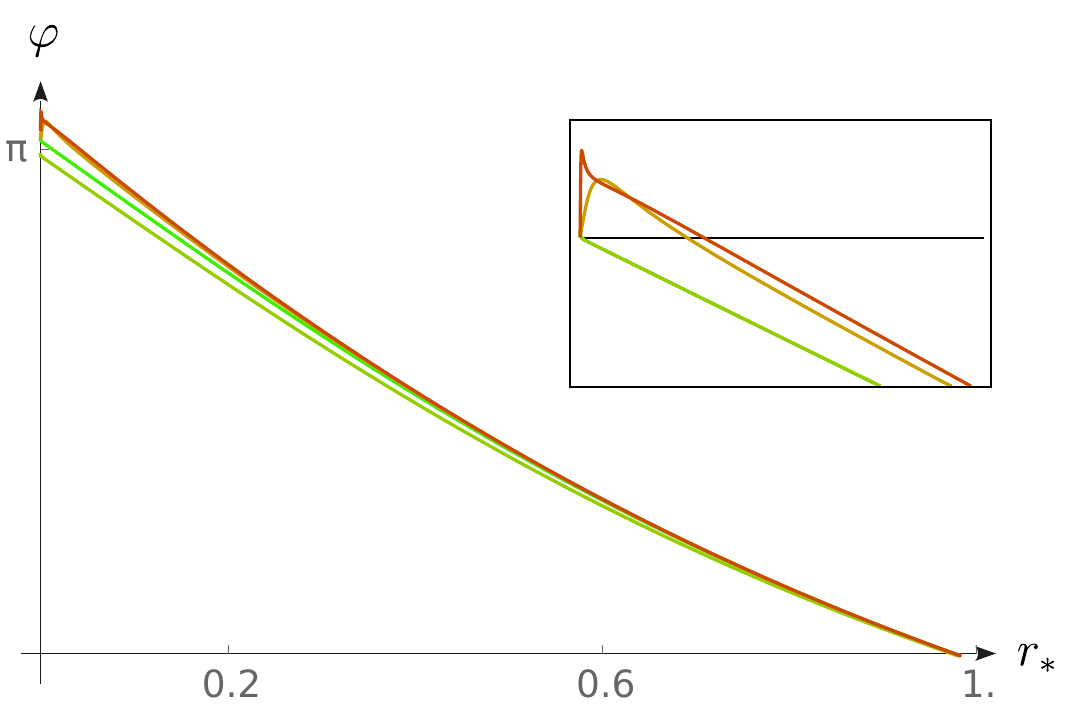}
     \caption{\small\label{fig:correlator} Plots for different values of $\Theta_c$. \underline{Left:} the on-shell Euclidean particle action as a function of the angular span $\varphi$; a swallow tail structure appears as the central temperature is increased. \underline{Right:} the angular span as a function of the geodesic tip $r_*$; the non-monotonicity is what originates the swallow tail on the left. Temperatures and degeneracies are as in Fig.~\ref{fig:positivetheta1}.\normalsize}
\end{figure}

We solve for the massive geodesics in the backgrounds of the previous chapter, using a Mathematica language script that we developed from scratch \cite{cita69}. The results are shown in Fig.~\ref{fig:correlator},  where the on-shell action (which corresponds to minus the logarithm of the correlator) is plotted for different values of the central degeneracy and the central temperature. In these figures, the natural range $(0,\pi)$ of the polar angle $\varphi$ is extended along the opposite meridian up to $2\pi$. 
 
\begin{itemize}
	\item  Within the diluted regime $\Theta_c\sim -20$ where density profiles have an abrupt edge, the angular span $\varphi$ is monotonic as a function of the tip position $r_*$, and the correlator behaves smoothly at $\varphi=\pi$.
	\item For larger but still negative $\Theta_c$ the angle $\varphi$ becomes a non-monotonic function of the tip position $r_*$. This results in a multivalued on-shell action with a ``swallow tail'' structure, and consequently in a scalar correlator with a non-vanishing derivative at $\varphi=\pi$. These features persist up to positive values of $\Theta_c$. %Interestingly, this second transition occurs for $\tilde T_c\ll\tilde T_c^{cr}$, and therefore always prior to the turning point in the total mass as a function of central density, this behaviour is similar to the one shown for flat space-time \ref{sec:carlos}.
	
	\item There always exists a positive $\Theta_c$ such that the the angular span $\varphi$ as a function of the geodesic tip $r_*$ becomes monotonic again, and the swallow tail behavior in the on-shell action disappears, resuting in a smooth correlator at $\varphi=\pi$. 
\end{itemize}

Interestingly, the non-smooth behaviour of the correlator occurs in the same region of phase space at which the boundary of the star has a power law form. This suggest that it may be related with the criticality of the dual theory. 

This ``multivalued'' form for correlators has been  reported for quenched states in thermalization studies, where the correlator is plotted as a function of gauge theory time \cite{cita41},\cite{cita42}, \cite{cita70} and \cite{cita71}. In this case, the swallow tail structure appears for near-critical equilibrium states, as a function of spatial separation. On the other hand, when a swallow tail appears in the free energy as a function of temperature, it is typically considered a signal of a phase transition. However, the present case is different and further investigation is necessary to support this claim, as explained in detail in \cite{cita2}.
\begin{tcolorbox}
[colback=blue!5!white,colframe=blue!75!black,title=Hint \arabic{boxnew}]
In the same region of parameters in which the power law edge shows up on the density profiles, the on-shell action for massive particle exhibits a swallow tail structure, resulting in a two-point correlator for a scalar operator which is non-smooth at the pole of the sphere $\varphi=\pi$.
\end{tcolorbox}
\stepcounter{boxnew}

\section{Scalar field perturbations}
\label{sec:scalar.perturbations}
\subsection{Normal modes}
We would like to extend the above exploration to finite conformal dimension $\Delta$. To that end, we calculate the perturbations of a probe scalar field in the neutron star background. 

The dynamics of a scalar probe $\Phi$ is described by the Klein-Gordon equation in the presence of the background metric \eqref{sch-star}.
\begin{equation}
	\left(
	{\Box-{\sf m}^2}\right)
	{\Phi}=0\ .\label{eq}%
\end{equation}
This equation can be separated by writing ${\Phi}$ as a Fourier decomposition in time, and a superposition of spherical harmonics in the angular variables,
\begin{eqnarray}
	{\Phi}\left(  t,r,\vartheta,\varphi\right)  &=&\sum_{\ell m }\int d\omega\ e^{-i\omega
		t}R_{\omega \ell m}\left(  r\right)  Y_{\ell m}\left(  \vartheta,\varphi\right)
	\\ &&\nonumber \phantom{\frac{\frac12}1}
	\qquad\mbox{with}\quad
	Y_{\ell m}(\vartheta,\varphi)=
	(-1)^m
	\mbox{\scriptsize$
		\sqrt{\frac{(2\ell+1)}{4\pi}\frac{(\ell-m)!}{(\ell+m)!}}
		$\normalsize}
	\,e^{i m \vartheta}P_{\ell}^m(\cos\varphi)
	\ .\label{sep}
\end{eqnarray}
Here $(\vartheta,\varphi)$ denotes the spherical angles and
$P_\ell^m(\cdot)$ an associated Legendre polynomial. 
The spherical harmonics fulfill 
$\nabla_{S^{2}}^{2}Y_{\ell m}\left(  \vartheta,\varphi\right)  =$ $-\ell\left( \ell+1\right)$ $
Y_{\ell m}\left(  \vartheta,\varphi\right)$.

In our spherically symmetric background only the eigenvalue $\ell$ appears in the wave equation. The radial dependence then satisfies
\begin{equation}
	R^{\prime\prime}_{\omega \ell}
	+
	\frac{1}{2}
	\left(
	\nu^\prime-\lambda^\prime
	+\frac {4}{r}
	\right) R^{\prime}_{\omega \ell}
	+
	e^{\lambda}\left(  \omega^2 e^{-\nu}-\frac{\ell\left(
		\ell\!+\!1\right)}{r^2} -L^2 {\sf m}^{2}\right)  R_{\omega \ell}=0\ ,\label{eq:R}%
\end{equation}
where we omitted the $m$ index in $R_{\omega \ell m}=:R_{\omega \ell}$ since it is evident from the equation that there will be no dependence on it.

The boundary conditions are obtained by expanding the equation at the extremes of the radial interval. First we go to large $r$ where the equation takes the form
\begin{equation}
	R^{\prime\prime}_{\omega \ell}
	+
	\frac {{4}}{r}
	R^{\prime}_{\omega \ell}
	-
	\frac{L^2{\sf m}^{2}}{r^2}
	R_{\omega \ell}=0\ ,\label{eq:Rlarger}%
\end{equation}
whose solution is
\begin{equation}
	R_{\omega \ell}
	=
	a_{\omega\ell}(1+\dots) \,r^{-\Delta_-}
	+
	b_{\omega\ell} (1+\dots)\,r^{-\Delta_+}\ ,
	\label{eq:conditionsInf}
\end{equation}
with $\Delta_\pm={\frac{3}{2}\pm\sqrt{\frac{9}{4}+\,{\sf m}^2L^2}}$ and the dots represent subleading negative integer powers of $r$. For our numerical calculations we choose ${\sf m}L=2$ which implies $\Delta_+=4$ and $\Delta_-=-1$. Notice that a normalizable mode would then require $a_{\omega\ell}=0$.
On the other hand, when we expand the equation at small values of the radius, we get
\begin{equation}
	R^{\prime\prime}_{\omega \ell}
	+
	\frac 2r
	R^{\prime}_{\omega \ell}
	-
	\frac{\ell\left(
		\ell\!+\!1\right)}{r^2}
	R_{\omega \ell}=0\ .\label{eq:Rsmallr}%
\end{equation}
The solution of this is immediately
\begin{equation}
	R_{\omega \ell}= A_{\omega\ell}\, r^\ell + B_{\omega\ell}\, r^{-\ell-1}\,.
	\label{eq:asymp}
\end{equation}
A regular solution corresponds to $B_{\omega\ell}=0$.
This results in the non-independence of the coefficients of the leading and subleading parts $a_{\omega\ell}$ and $b_{\omega\ell}$ at infinity, what would in turn quantize the values of the frequency $\omega$ for which a normalizable mode $a_{\omega\ell}=0$ is obtained. They correspond to the {\em normal modes} of the scalar field in the bulk.

\begin{figure}[ht]
	\centering
	\includegraphics[width=0.45\textwidth]{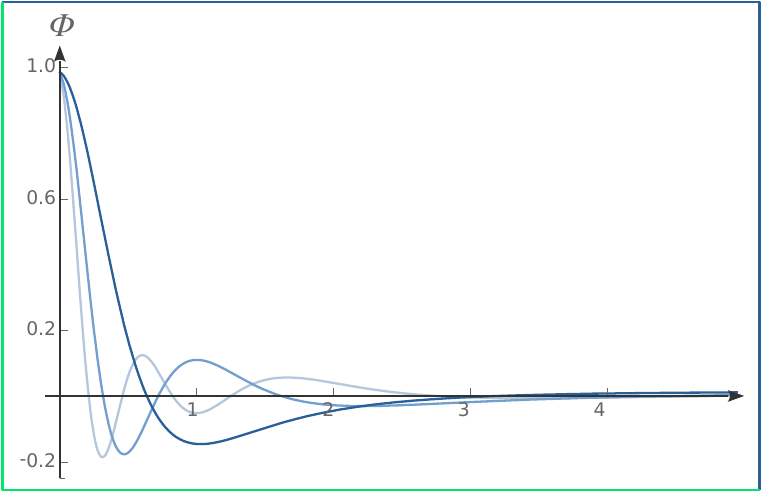}
 \hfill 
	\includegraphics[width=0.45\textwidth]{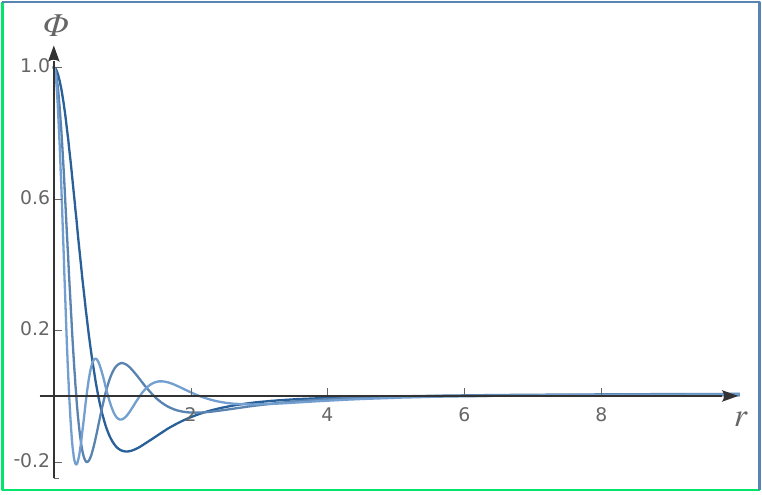}
	\caption{\small \label{fig:profiles} Profiles of the normal modes of a probe scalar field in the neutron star background. \underline{Left:} solutions for ${T}_c=10^{-2}$, $\Theta_c=-20$. \underline{Right:} for ${T}_c=5\times10^{-3}$, $\Theta_c=20$.
 \normalsize}
\end{figure}

Typical profiles of the radial function are shown in Fig.~\ref{fig:profiles}.
\newpage

\vspace{-1.3cm}
\subsection{Scalar two-point correlator}\label{findings}
The correlator of the dual scalar operator can be written as
\begin{equation}
	\langle
	{\cal O}(\vartheta,\varphi)
	{\cal O}(0)
	\rangle
	=\sum_{\ell m} g_{\omega\ell}\,Y_{\ell m}(\vartheta,\varphi)\ ,
	\label{eq:correlator2}
\end{equation}
where use of the symmetries has been made to discard any dependence of $g_{\omega\ell}$ on $m$. 
According to the holographic prescription \ref{intro:correlator}, the coefficients $g_{\omega\ell}$ are given by the quotient of the subleading to the leading part of the dual bulk scalar field in \eqref{eq:conditionsInf}
\begin{equation}\label{gdeomegaele}
	g_{\omega\ell} = \frac{b_{\omega\ell}}{a_{\omega\ell}}\ ,
\end{equation}

On general grounds, it is natural to expect that, as a function of the frequency $\omega$, the correlator $g_{\omega\ell}$ would have a set of simple poles plus an analytic contribution,
\begin{equation}
	g_{\omega\ell}=\sum_{n}\frac{\rho_{n\ell}}{\omega-\omega_{n\ell}}+h_{\ell}(\omega)\, ,
	\label{eq:fit}
\end{equation}
where $h_{\ell}(\omega)$ is analytic in $\omega$. The poles show up at the values of $\omega$ for which $a_{\omega\ell}=0$ in \eqref{eq:conditionsInf}, {\em i.e.} to the normal modes. Their positions $\omega_{n\ell}$ are identified with the energy of the corresponding boundary excitations. On the other hand, the residua $\rho_{n\ell}$ correspond to their decay amplitudes.
\newpage 

\subsection{Results}

\begin{figure}[t]
	\centering
	\includegraphics[width=0.48\textwidth]{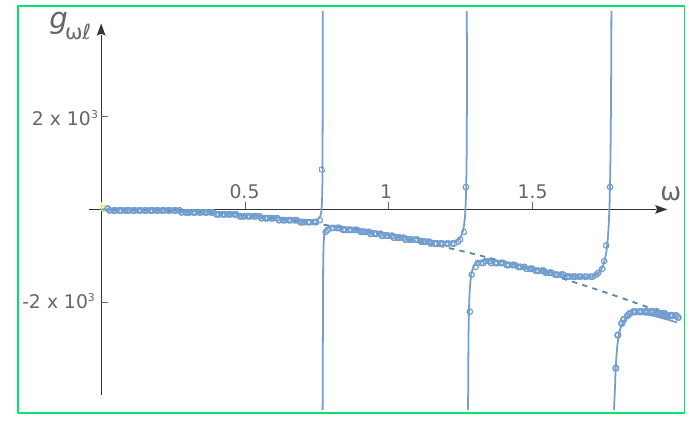}
 \hfill 
 	\includegraphics[width=0.48\textwidth]{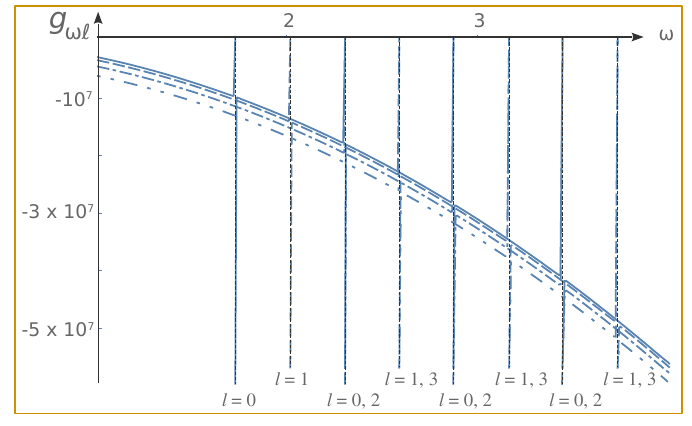}
	\caption{\small \label{fig:ajuste} \underline{Left:} Numerical form of the scalar correlators at $\Theta_c=30$ and $T_c=5\times10^{-3}$ for $\ell=0$, compared with the fit defined in equation \eqref{eq:fit} with a quadratic form for the analytic component $h_{\ell}(\omega)$. \underline{Right:} Plots of the scalar correlators at $\Theta_c=20$ and $T_c=10^{-2}$ for different values of $\ell$. To our numerical precision, the shift on the position of the poles as $\ell$ is increased is well approximated by the AdS expression $\omega_{n\,\ell\!+\!\Delta\ell}-\omega_{n\ell}=\Delta\ell$.\normalsize}
\end{figure}

In our numerical explorations of the parameter space, the expansion of the smooth part $h_{\ell}(\omega)$ up to the quadratic order in $\omega$ resulted in a good fit of the numerical data, see Fig.~\ref{fig:ajuste}.

%\begin{figure}[h]
%	\centering
%	\includegraphics[width=0.5\textwidth]{Chapters/Thesis4/diffl.pdf}
%	\caption{\label{fig:correlators2} Plots of the scalar correlators at $\Theta_c=20$ and $T_c=1/100$ for different values of $\ell$. To our numerical precision, the shift on the position of the poles as $\ell$ is increased is well approximated by the AdS expression $\omega_{n\,\ell\!+\!\Delta\ell}-\omega_{n\ell}=\Delta\ell$ 	.}
%\end{figure}

As can be seen in Fig.~\ref{fig:ajuste}, as we increase $\ell$ the energies of the normal modes jump by an amount that, to our numerical precision, is well approximated by the AdS expression $\omega_{n\,\ell\!+\!\Delta\ell}-\omega_{n\ell}=\Delta\ell$ (see Appendix \ref{AppendixE}). However, as it could have been expected, the energy distance between different normal modes with the same angular momentum do not match the AdS formula $\omega_{n\!+\!\Delta n\,\ell}-\omega_{n\ell}\neq 2\Delta n$.

\begin{figure}[ht]
	\vspace{.15cm}
	\centering
	\includegraphics[width=0.98\textwidth]{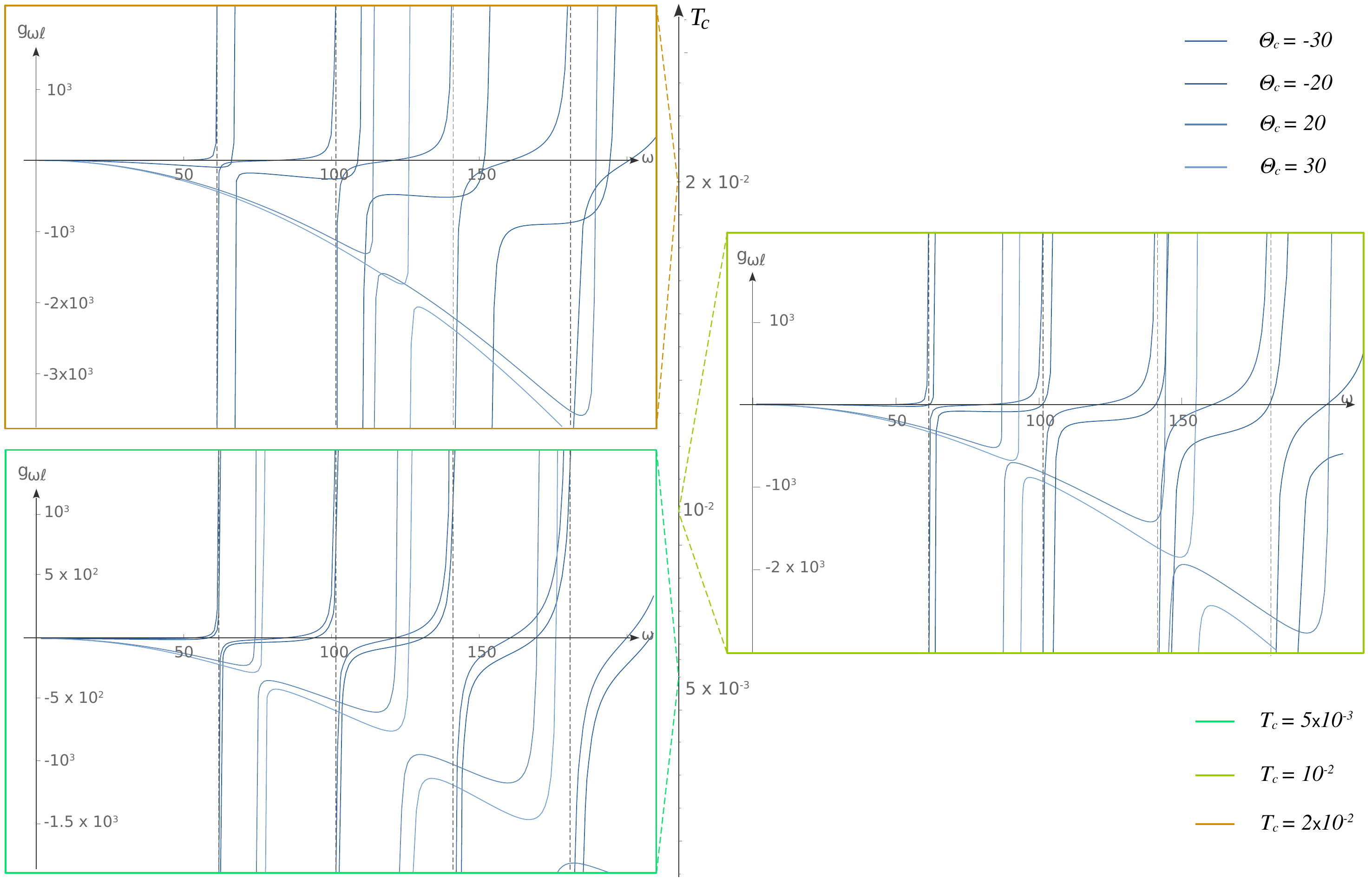}
	\caption{\label{fig:correlators}\small
		Plots of the scalar correlators. The frame color corresponds to the central temperatures $T_c$, while that of the curves correspond to the different values of the central degeneracy $\Theta_c$.\normalsize
	}
\end{figure}

The behaviour of the correlators as we change the central temperature $T_c$ and degeneracy $\Theta_c$ are shown in Fig.~\ref{fig:correlators}. In the stable region of the phase diagram, the correlators look very much like those of the pure AdS case (see Appendix \ref{AppendixB}). As we increase $\Theta_c$ at fixed $T_c$, moving into the unstable region, the analytic part becomes more important, eventually dominating the plot. This behaviour is again interpreted as an indication that the system is getting more and more critical, developing a power law correlator.

\begin{figure}[ht]
	\vspace{.01cm}
	\centering
	\includegraphics[width=0.43\textwidth]{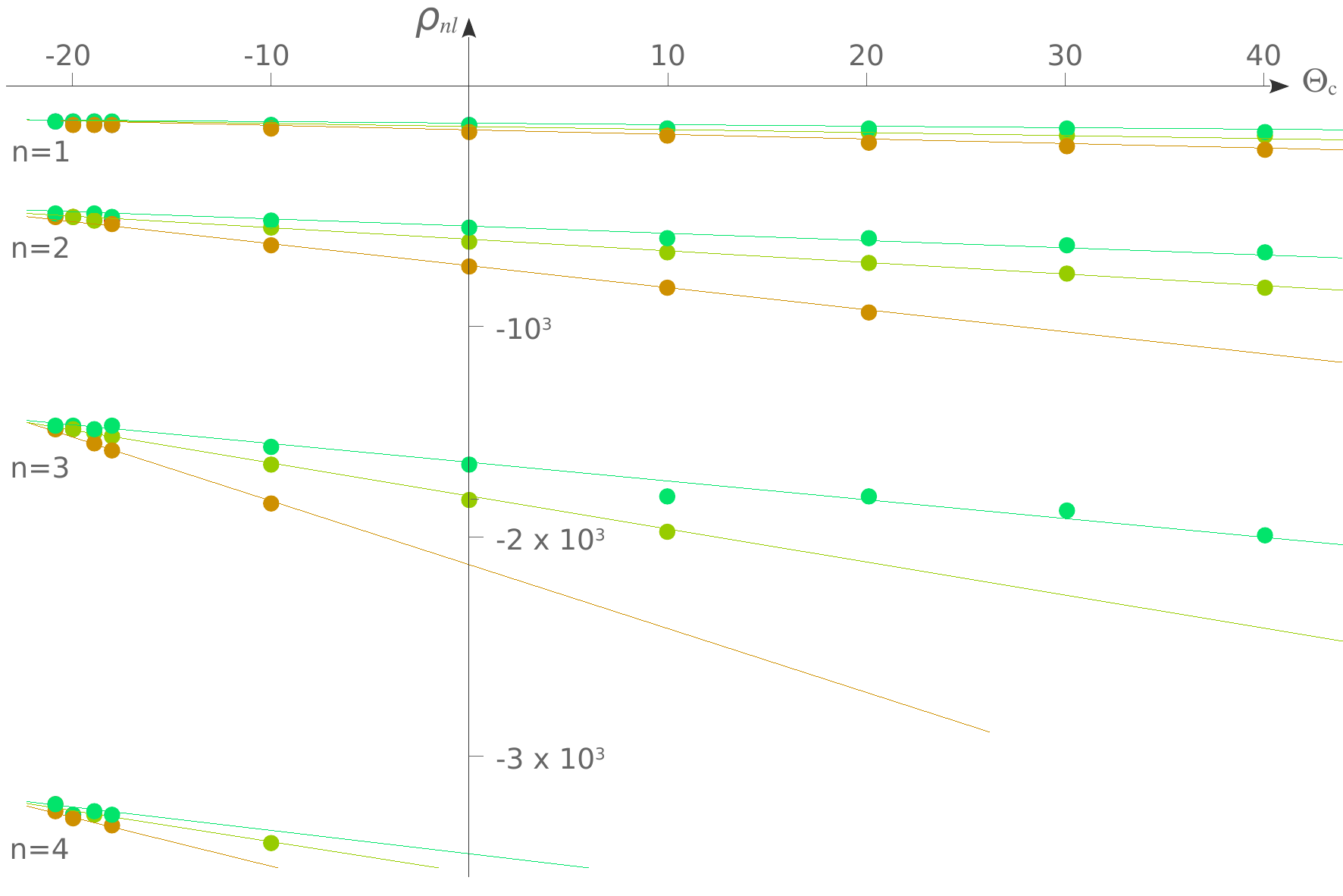}
	\hspace{0.05\textwidth}
	\includegraphics[width=0.43\textwidth]{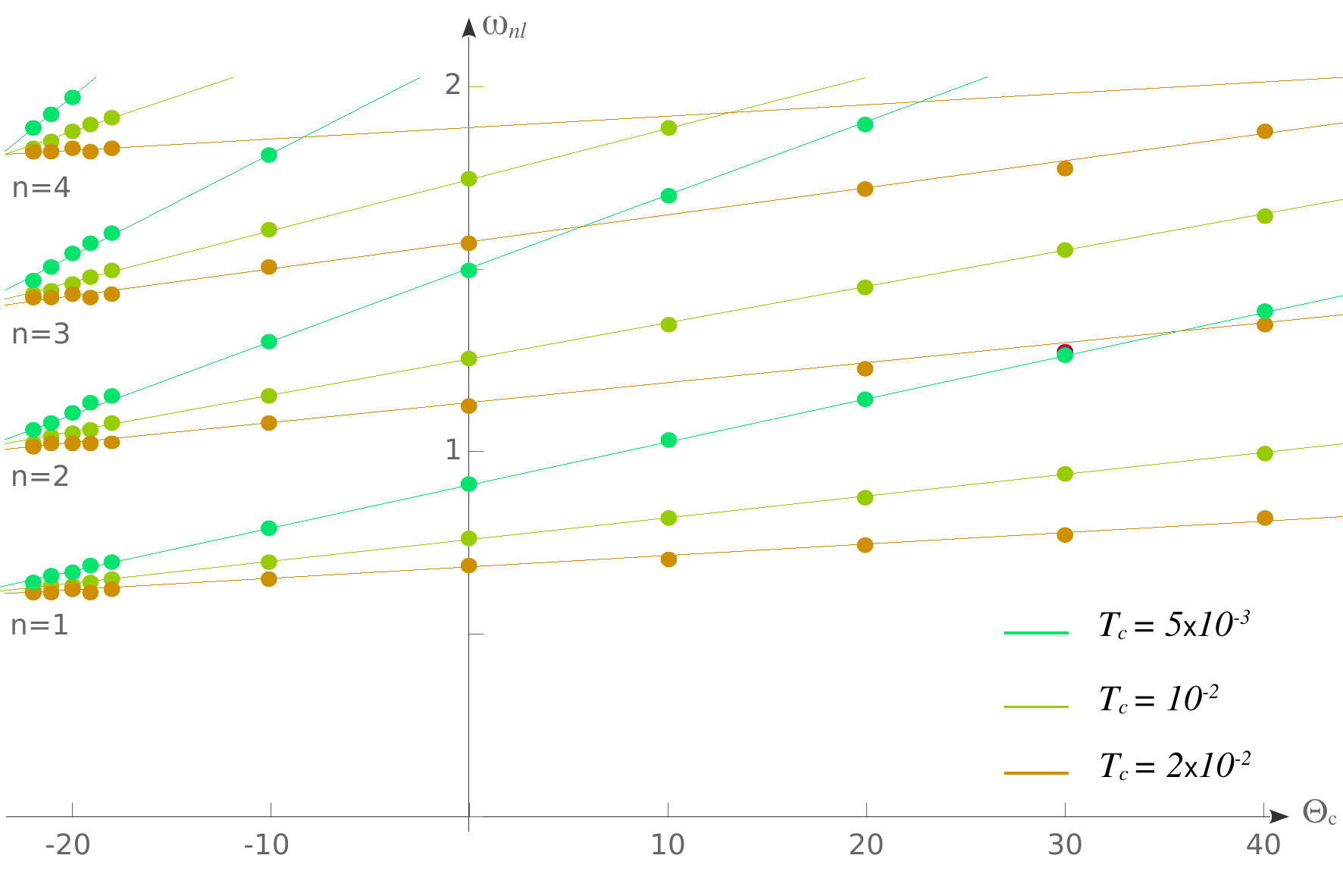}
	\caption{\label{fig:ajustes2}\small \underline{Left:} Normal mode' decay constants $\rho_{n\ell}$. \underline{Right:} Normal mode' energies $\omega_{n\ell}$. Both are plotted as functions of the central degeneracy $\Theta_c$ for different central temperatures $T_c$. \normalsize}
\end{figure}

In the expression \eqref{eq:fit} the normal mode energy $\omega_{n\ell}$ and the absolute value of its decay constant $|\rho_{n\ell}|$ grow linearly both with the central temperature $T_c$ at fixed central degeneracy $\Theta_c$, and with  the central degeneracy $\Theta_c$ at fixed central temperature $T_c$, see Fig.~\ref{fig:ajustes2}.

 \begin{tcolorbox}[colback=blue!5!white,colframe=blue!75!black,title=Hint \arabic{boxnew}]  In the region of the phase diagram in which the star present a sharp boundary, the scalar correlator resembles that of the pure AdS case, being dominated by the pole contribution. As we increase $\Theta_c$ at a fixed $T_c$, moving into the region in which the edge of the star becomes power law, the analytic part becomes more significant and eventually dominates the plot. This indicates that the system is becoming critical, developing a power law correlator.
\end{tcolorbox}
\stepcounter{boxnew}

\newpage

\section{Discussion  }\label{sec:main_results_6}

At this stage, an intriguing question remains open: how does the swallow tail structure on the correlator, which was found in Section \ref{sec:correlators3} in the geodesic limit of large conformal dimension $\Delta$, manifest itself in the finite $\Delta$ context of Section \ref{sec:scalar.perturbations}? The scalar correlators, as defined in \eqref{eq:correlator2} and \eqref{gdeomegaele}, from the real frequency stationary states of a scalar field,  do not exhibit any new distinctive features when entering the unstable region, but instead undergo a gradual transformation from a pole-dominated curve into a power-law form. Nonetheless, we would expect that analyzing the classical solutions of the scalar field would provide a means to test whether the swallowtail also arises at finite conformal dimensions.

In general, the correlator of a quantum field diverges at coinciding points (in our case, when the angular separation $\varphi$ vanishes), but is expected to be smooth at antipodal points (i.e., when $\varphi=\pi$).
For finite $\Delta$, a non-vanishing derivative at $\varphi=\pi$ can be inferred from the coefficients $g_{\omega\ell}$ of the partial wave expansion of the correlator. %Due to the spherical symmetry of the bulk background, the correlator admits a Fourier expansion in terms of Legendre polynomials $P_\ell(\cos{\vartheta})$ with certain coefficients $C_\ell$.
 
We can study the expansion of $g_{\omega\ell}$ in powers of $\ell$ numerically. We verified that the leading behavior for large $\ell$ is $g_{\omega\ell}\sim \ell^{2\Delta_+-1}$, as expected from the singularity at $\varphi=0$. Positive powers of $\ell$ in the asymptotics of $g_{\omega\ell}$ would lead to divergent series, but upon Borel summation, they give finite contributions to the correlator. In particular, they are singular at $\varphi=0$, but have a vanishing derivative and are thus smooth at $\varphi=\pi$.

We can obtain the derivative of the correlator at $\varphi=\pi$ by examining the term in the asymptotics of $g_{\omega\ell}$ that decreases as $\ell^a$ with $-3\leqslant a<-1$. A term with this type of falloff at large $\ell$, if it has no alternating sign, would contribute to the derivative. However, even for small conformal dimensions, we need to explore the large-$\ell$ form of the field with higher numerical precision to determine such subleading behavior.

%\begin{tcolorbox}[colback=blue!5!white,colframe=blue!75!black,title=Hint \arabic{boxnew}]  To study the flow from the power law form of the finite $\Delta$ correlator into the non-vanishing derivative at $\vartheta=\pi$ of its large $\Delta$ limit, we would need to analyze the behaviour of the correlator with $\ell^a$ with $-2\leq a<-1$.
%\end{tcolorbox}
%\stepcounter{boxnew}

\section{Results}

\begin{figure}[ht]
	\centering
	\vspace{-.6cm}
	\includegraphics[width=1\textwidth]{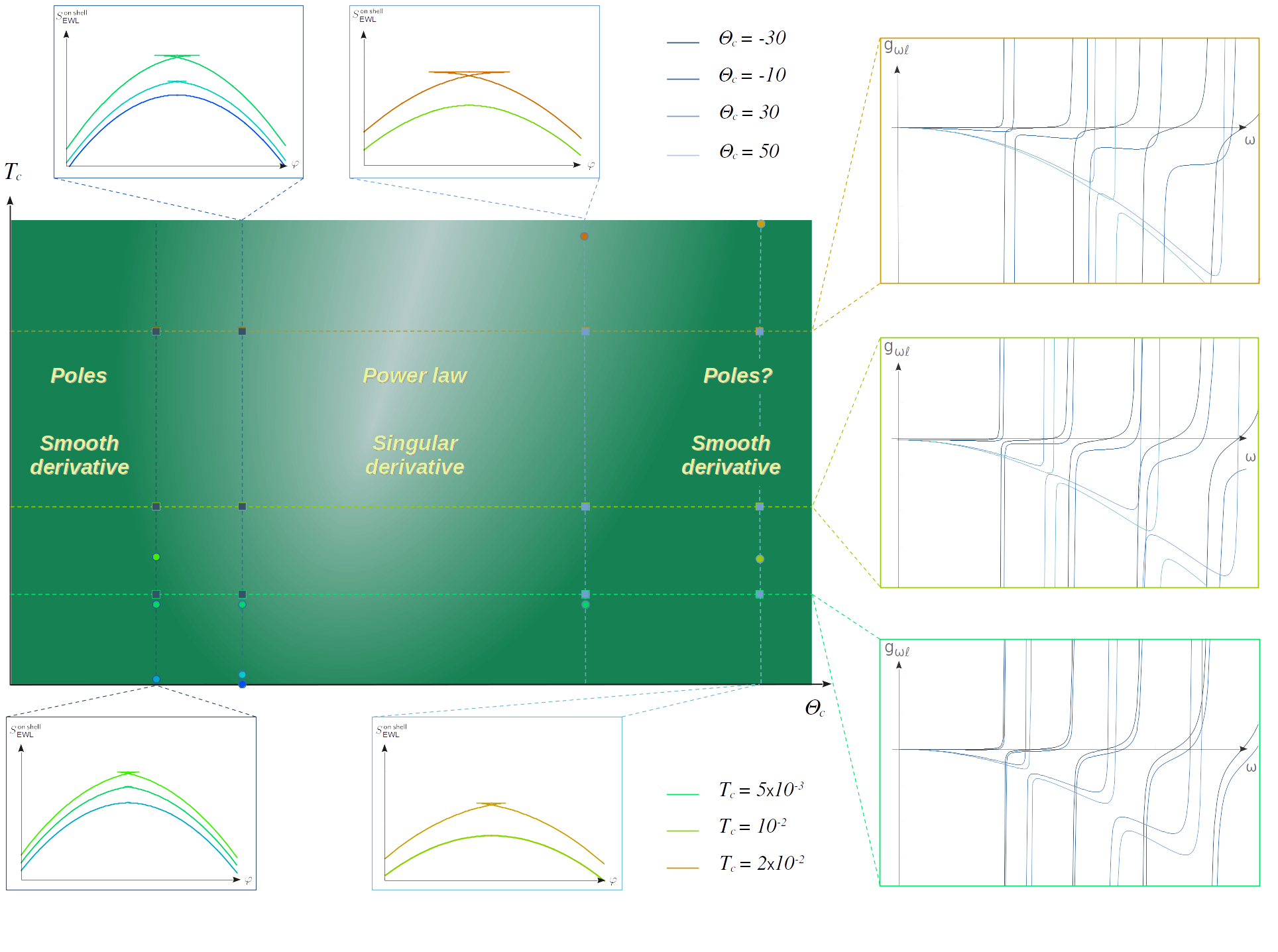}
	\vspace{-1cm}
	\caption{\label{fig:phasediagram}
\small		Central temperature ${T}_c$ {\em vs.} central degeneracy $\Theta_c$ phase diagram. The dotted horizontal lines represent constant temperatures, the vertical ones represent constant values of $\Theta_c$. The on shell actions plotted inside each box at the top and and the bottom of the figure are located in the phase diagram by a dot with the same color sitting along the corresponding vertical line. The correlators plotted on the boxes at the right of the figure are located by a square with the same color sitting along the corresponding horizontal line.  Again, the comparison with Fig.~\ref{fig:phase.diagram.metal} is suggestive.\normalsize}
\end{figure}

With the results of the present chapter, we can upgrade our phase diagram, as shown in Fig. \ref{fig:phasediagram}. We see a clear picture emerging: there is a region at intermediate values of the central degeneracy, at which the system manifest many features that are characteristic of a critical phenomenon. This is to be compared with the plot \ref{fig:phase.diagram.metal}.

%Fermions
\chapter{Fermionic two-point correlator}\label{ch:fermions}

As mentioned in Section \ref{sec:intro.fermions}, strongly correlated electron systems have been extensively studied in condensed matter theory. Of particular interest in such investigations is the two-point correlator of the electronic degrees of freedom, since it provides valuable information on the  structure of the Fermi surface and the existence of long-lived excitations. 

In this chapter, we will use the previously studied neutron star background to propagate a Dirac spinor. The goal will be to solve the Dirac equation for the perturbation, in order to use its solutions to calculate the fermionic two-point correlator of the holographic field theory. 

Even if not necessary, we will consider the spinor as an excitation of the perfect fluid sourcing the background. A consequence of such a choice is that the mass of the spinor is $m$, which has been taken to be very large in order to have a well-defined holographic setup. As we will see, this allow us to make a WKB approximation on the Dirac equation.

\newpage
\section{Spinor field perturbations}
\subsection{Equivalent Schrödinger problem}
Let us consider a spinorial excitation moving in a metric of the form \eqref{sch-star} with $\ell$ replaced by the AdS length. 
To write the corresponding Dirac equation we need to introduce a vierbein basis $\omega_\mu{}^a$ and dual vector $e^\mu{}_a$  as
\be
\begin{array}{lllll}
\omega^0 =  L\,e^{\frac{\nu}2}\,dt\,,
&\omega^1 = L\,r\,d\vartheta\,,
&~
&~e_0 = \frac1L\,e^{-\frac{\nu}2}\,\partial_t\,,
&e_1 = \frac{1}{L\,r}\,\partial_\vartheta\,,
\\
\omega^2 = L\, r\,\sin\vartheta\, d\varphi\,, 
&\omega^3 = L\,e^{\frac{\lambda}2}\,dr\,, 
&~
&~e_2 = \frac1{L\,r\,\sin\vartheta}\,\partial_\varphi\,,
&e_3 = \frac1L\,e^{-\frac{\lambda}2}\;\partial_r\,.
\end{array}
\ee
This results in the components $\omega_{\mu ab}$ of the spin connection
\be
\begin{array}{lll}
\omega_{12}= - \cos\vartheta\,d\varphi 
&\quad \omega_{23}= e^{-\frac{\lambda}{2}}\,\sin\vartheta\,d\varphi
&\quad \omega_{01}=0
\\
\omega_{13}= e^{-\frac{\lambda}{2}}\,d\vartheta &\quad \omega_{03} = -\frac{1}{2}\,e^{\frac{\nu-\lambda}2}\,\nu' \,dt
&\quad \omega_{02}=0
\end{array}
\ee
We define tangent space gamma matrices obeying the conmutation relation $\{\Gamma^a,\Gamma^b\} = 2\,\eta^{ab}$ as $\Gamma^0\equiv i\,\sigma_1\otimes 1_{2\times 2}$, $\Gamma^I\equiv\sigma_2\otimes\gamma^I$, $\Gamma^3\equiv\sigma_3\otimes 1_{2\times 2}$ 
where $\{\sigma_1,\sigma_2,\sigma_3\}$ are the Pauli matrices and 
$\gamma^I$ correspond to the gamma matrices on the sphere, see Appendix \ref{app:conventions}. 
 
With the above, we can write the Dirac equation for a fermionic perturbation of our background in the form
\be\label{eq:eom-fermionic}
(\slashed\nabla -i\,\slashed A - m)\,\Psi(t,r,\vartheta,\varphi) =0
\ee
where $\slashed\nabla\equiv \Gamma^\mu\nabla_\mu= 
\Gamma^\mu\left(e_\mu +{\omega_\mu{}^{bc}}[\Gamma_b;\Gamma_c]/8\right)$ is the curved  space covariant derivative, and the coupling to a gauge field  $\slashed A =\Gamma^\mu \,A_\mu=e^\mu{}_a\Gamma^a A_\mu$ was introduced in order to include the effects of the local chemical potential by writing $A_t=\mu$, which results in the term $\slashed A={\mu}\, e^{-\frac {\nu}2}\,\Gamma^0 /L$.

Using a basis $\psi_{jm\epsilon}$ of eigenspinors on the sphere with eigenvalues $\epsilon(j+1/2)$ where the index $j\in\mathbb{Z}$  is an integer, the index $m$ fullfills $-j\leq m\leq j$ and $\epsilon=\pm1$ is a sign (see Appendix \ref{app:conventions}), we can decompose the spinor perturbation as
\be
\label{eq:Psiexpan}
\Psi(t,r,\vartheta,\varphi) = 
\frac{e^{-\frac{\nu}{4}}}{r}\,
\sum_{jm\epsilon} \int\frac{d\omega}{2\pi}\;
\varphi_{\omega jm\epsilon}(r)\otimes \psi_{jm\epsilon}(\vartheta,\varphi)\,e^{-i\,\omega t}
\ee 
By plugging into the Dirac equation we obtain
 the pair of equations
\bea\label{eq:fieq2}
\varphi^{+'}_{\omega  jm\epsilon}(r) 
- mL\,
e^{\frac{\lambda}{2}}
\,\varphi^+_{\omega jm\epsilon}(r)  
+
e^{\frac{\lambda}2} 
\left(({\omega+\mu})\,e^{-\frac{\nu}{2}} +\frac{{\epsilon}}{ r}\left(j+\frac12\right)\right) 
\varphi_{\omega  jm\epsilon}^-(r)&=&0
\cr
\varphi^{-'}_{\omega  jm\epsilon}(r) 
+mL\, e^{\frac{\lambda}{2}}\varphi^-_{\omega \,jm\epsilon}(r)  
-e^{\frac{\lambda}2}\left(({\omega+\mu})\,e^{-\frac{\nu}{2}}-\frac{{\epsilon}}{ r}\left(j+\frac12\right)\right) 
\varphi_{\omega  jm\epsilon}^+(r)&=&0
\phantom{\frac{1^2}{1}}~~~~~~
%\cr&&
\eea
For positive $\epsilon$ we can solve the first equation for the function without derivatives, and the same can be done for negative $\epsilon$ solving the second equation. We get
\be
\varphi^{-\epsilon}_{\omega  jm\epsilon}(r) = \frac{1}{f }
\left(-\epsilon\,\varphi'^{\epsilon}_{\omega  jm\epsilon}(r) +  
m L\;e^{\frac{\lambda}{2}}\;\varphi^\epsilon_{\omega jm\epsilon}(r)\right)
\ee
where we have defined 
$f =e^{\frac{\lambda}{2}} \left({(\omega+\mu)}e^{-\frac{\nu}{2}} + {\left(2j+1\right)}/{ 2r}\right)$. By plugging in the remaining equation, and after the further rescaling 
%s $\varphi_{\omega jm\epsilon}(r) = \sqrt{f }\;\Phi_{\omega  jm\epsilon}(r)$ and 
$\varphi^{\epsilon}_{\omega jm\epsilon}(r) =\sqrt{f }\;\phi_{\omega  jm\epsilon}(r)
$
the bi-spinor $\varphi_{\omega jm\epsilon}$ can be written as
\be \label{eq:veremos}
\varphi_{\omega  jm\epsilon}(r) =  
\frac{1}{2\sqrt f }
\small
\left(\!\!
\begin{array}{c}
      1\!-\!\epsilon \\
      1\!+\!\epsilon
\end{array}\!\!
\right)
\normalsize
\left(\phi'_{\omega jm\epsilon}(r) + \left(\frac{f' }{2 f }
-\epsilon\, m L\;e^{\frac{\lambda}{2}}\right) \phi_{\omega  jm\epsilon}(r)\right)
%+\nonumber\\&&\quad
+\frac{\sqrt f}2 
\small
\left(\!\!
\begin{array}{c}
     1\!+\!\epsilon  \\
     1\!-\!\epsilon
\end{array}\!\!
\right)
\normalsize
\phi_{\omega  jm\epsilon}(r)
\ee
where the functions $\phi_{\omega jm\epsilon}(r)$ satisfy a Schr\"odinger-like  equation 
\be\label{eq:Schroeq}
\phi''_{\omega  jm\epsilon }(r) - U_{\omega  jm\epsilon }(r)\;\phi_{\omega  jm\epsilon }(r)=0
\ee
with potential
\be \label{eq:Upot}
U_{\omega jm\epsilon }(r) =
\sqrt{f  } \left(\frac{1}{\sqrt{f }}\right)'' 
+\epsilon\, mL\,f  \left(\frac{e^{\frac\lambda2}}{f }\right)'
%\cr&&
+e^\lambda 
\left(-{(\omega+\mu)^2}e^{-\nu} + \frac{1}{r^2}\left(j+\frac12\right)^2 + m^2 L^2\right)
\ee 
\subsection{WKB approximation}
By taking a large spinor mass $m$ or equivalently a large conformal dimension $\Delta=mL$ of the dual fermionic operator, we can rewrite the potential as
\be \label{eq:Upot2}
U_{\omega jm\epsilon }(r) \approx
m^2L^2\;e^\lambda 
\left(1-{E^2}e^{-\nu} + \frac{J^2}{r^2}  \right)
\equiv
m^2L^2\,V(r)\,,
\ee 
where we defined $E=(\omega+\mu)/mL$ and $J=(2j+1)/2mL$ as finite quantities in the limit. This potential multiplied for a large constant sets precisely the conditions for a WKB approximation to work.

By looking to the expressions \eqref{eq:redefinition-nu}, we see that $e^{-\lambda}\sim e^\nu\sim r^2$ at large radius, implying that the potential decays as $1/r^2>0$. On the other hand close to the origin we have $e^{-\lambda}\sim e^\nu\sim 1$ but there is a centrifugal barrier. This implies that the number of turning points is even. Numerical exploration of parameter space shows that, depending on the values of the constants $E$ and $J$ and the background parameters, the potential has either two turning points or none.  We will concentrate  in what follows on the case of two turning points, at $r=r_L$ and $r=r_R$ with $r_L<r_R$. 

Since we are in the conditions in which the WKB approximation is valid, it is immediate to derive the semiclassical form of the wave function
\begin{equation}
    \phi_{\omega  jm\epsilon }(r)=
    \frac{A}{\sqrt[4]{|V|}}
    \left\{
    \begin{array}{ll}
    %\left(
    %A_L e^{-mL\int_{r_L}^r dr\sqrt{V}}
    %+
    e^{mL\int_{r_L}^r dr\sqrt{V}}
    %\right)
    &\qquad \mbox{for}\ r<r_L
    \\
    %\left(
    \cos\left(mL\int_{r_L}^r dr\sqrt{-V}+\frac\pi4\right)
    %+
    %B \sin\left(mL\int_{r_L}^r dr\sqrt{-V}+\frac\pi4\right)
    %\right)
    &\qquad \mbox{for} \ r_L<r<r_R
    %\\
    %\left(
    %\cos\left(mL\int_{r_R}^r dr\sqrt{-V}
    %+\frac\pi4 \right)
    %\cos\left(mL\int_{r_L}^{r_R} dr\sqrt{-V}
    %\right)
    %-
    %\sin\left(mL\int_{r_R}^r dr\sqrt{-V}
    %+\frac\pi4 \right)
    %\sin\left(mL\int_{r_L}^{r_R} dr\sqrt{-V}
    %\right)
    %+
    %B \sin\left(mL\int_{r_L}^r dr\sqrt{-V}+\frac\pi4\right)
    %\right)
    %&\qquad \mbox{for} \ r_L<r<r_R
    \\
    %\left(
    \frac12\alpha  e^{-mL\int_{r_R}^r dr\sqrt{V}}
    +
    \beta e^{mL\int_{r_R}^r dr\sqrt{V}}
    %\right)
    &\qquad \mbox{for}\ r_r<r
    \end{array}
    \right.
    \label{eq:WKBsolution}
\end{equation}
where $A$ is a normalization constant, and we imposed regular boundary conditions at the origin. The quantities $\alpha$ and $\beta$ are fixed by the standard WKB connection formulas as
\begin{equation}
    \alpha=\sin\!\left(mL\int_{r_L}^{r_R} dr\sqrt{-V}
    \right)\qquad\qquad\qquad
    \beta=\cos\!\left(mL\int_{r_L}^{r_R} dr\sqrt{-V}
    \right)
\end{equation}

\section{Fermionic correlator and normal modes}
Using the  WKB solution we are in conditions to find the normal modes of the system and its two-poin correlator, as follows. 

First, since the integral in the exponent in the third line of \eqref{eq:WKBsolution} is positive and diverges at large radius, normalizable states require $\beta=0$. This results immediately in 
the Bohr-Sommerfeld quantization condition 
\begin{equation}
    mL\int_{r_L}^{r_R} dr\,\sqrt{-V}=\left(n+\frac12\right)\pi\qquad\qquad\mbox{with~} n\in\mathbb{Z}
\end{equation}
which provides the dispersion relation $E_n(J)$ for the normal modes of the system. 

Next, replacing the WKB solution in \eqref{eq:veremos} and approaching the AdS boundary, we obtain the asymptotic form
\be 
\varphi_{\omega  jm\epsilon}(r) =  
-B\sqrt{\frac{mL}{J+E}}
\small
\left(\!\!
\begin{array}{c}
    {(1-\epsilon)\beta}\normalsize   \left(\frac r{r_\epsilon}\right)^{mL}e^{\int_{r_R}^{r_\epsilon}\sqrt{V}} \\
      \frac{1}{2}(1+\epsilon)\alpha\left(\frac r{r_\epsilon}\right)^{-mL }e^{-\int_{r_R}^{r_\epsilon}\sqrt{V}}
\end{array}\!\!
\right)
\normalsize
\ee 
where $r_\epsilon$ is  a radial cutoff and $B$ is an overall constant proportional to $A$.  With this we can write
\begin{equation}
    \langle {\cal O}_{-mj}(\omega){\cal O}_{m j}(\omega) \rangle
    =
    \frac{\varphi_{\omega  jm+}^-(r_\epsilon)}{\varphi_{\omega  jm-}^+(r_\epsilon)}=\frac1{2 } e^{-2\int_{r_R}^{\infty}\sqrt{V}}\tan\!\left(mL\int_{r_L}^{r_R} dr\sqrt{-V}
    \right) 
    \label{eq:femionic.correlator}
\end{equation}
This result is analytic, even if to evaluate it explicitly we have to make use of the numerical background. Notice that, as expected, the correlator has simple poles at the normal modes. 
\begin{tcolorbox}[colback=blue!5!white,colframe=blue!75!black,title=Hint \arabic{boxnew}]   
The fermionic two-point correlator can be obtained in analytic form in the limit of large conformal dimension, by making use of the WKB approximation on the bulk spinor. As expected, it has simple poles at the normal modes. 
\end{tcolorbox}
\stepcounter{boxnew}

\section{Results}
\label{sec:main_results_7}
\begin{figure}[ht]
	\vspace{.01cm}
	\centering
\includegraphics[width=0.47\textwidth]{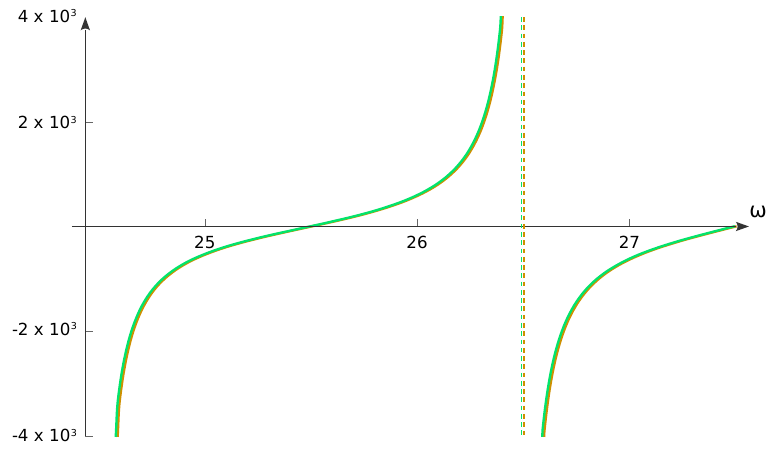}
	\hfill
	\includegraphics[width=0.47\textwidth]{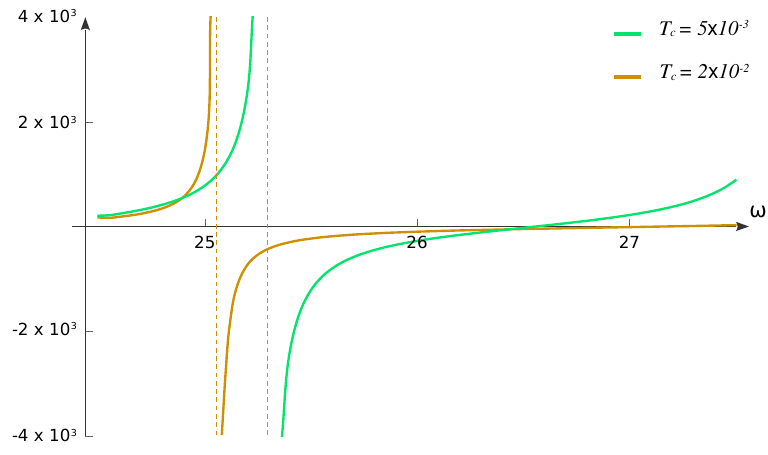} 
    \small
 	\caption{\label{fig:dependence.temperature} Plots of the two-point fermionic correlator as a function of the frequency $\omega$ for two different temperatures. \underline{Left:} $J=6.5$. \underline{Right:} $J=8$. As we can see, for larger $J$ the dependence on the temperature of the pole energies becomes stronger.
  \normalsize
 }
\end{figure}

\begin{figure}[ht]
	\vspace{.01cm}
	\centering
\includegraphics[width=0.47\textwidth]{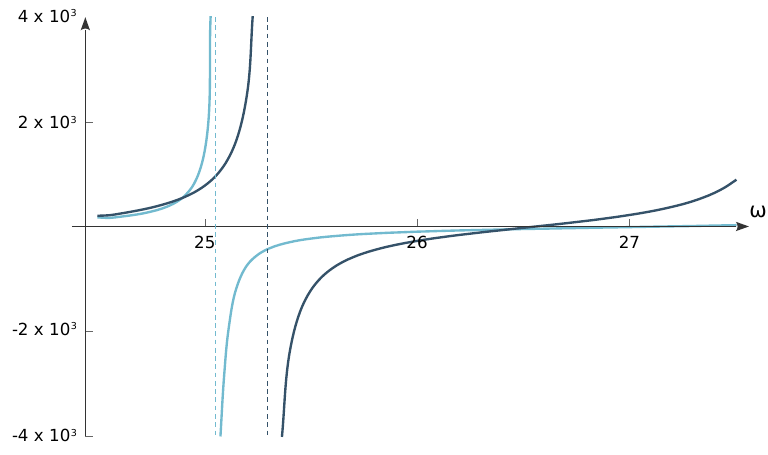}
	\hfill
	\includegraphics[width=0.47\textwidth]{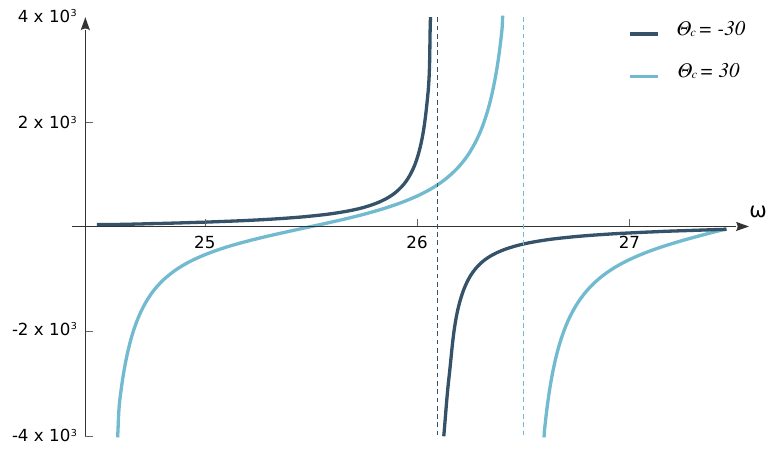} 
 	\caption{\label{fig:dependence.theta}
    \small Plots of the two-point fermionic correlator as a function of the frequency $\omega$ for two different values of $\Theta_c$. \underline{Left:} $J=6.5$. \underline{Right:} $J=8$. As we can see, for larger $J$ the dependence on the central degeneracy of the pole energies becomes stronger.\normalsize
 }
\end{figure}

\begin{figure}[ht]
	\vspace{.01cm}
	\centering
    \vspace{0.5cm}
    \includegraphics[width=0.45\textwidth]{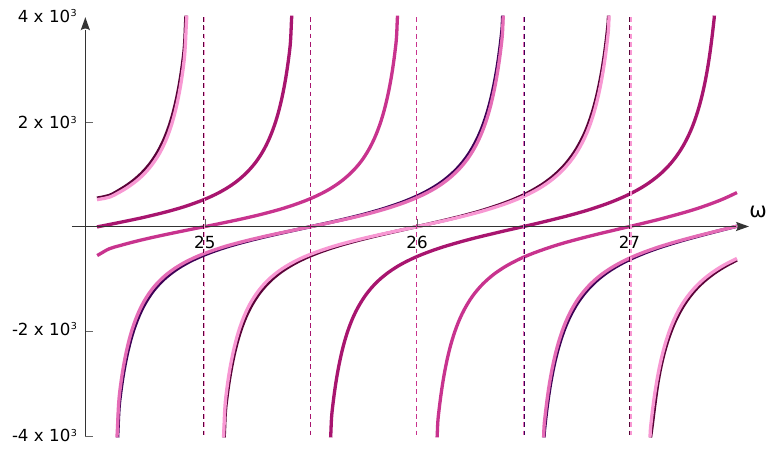}
	\hfill
	\includegraphics[width=0.54\textwidth]{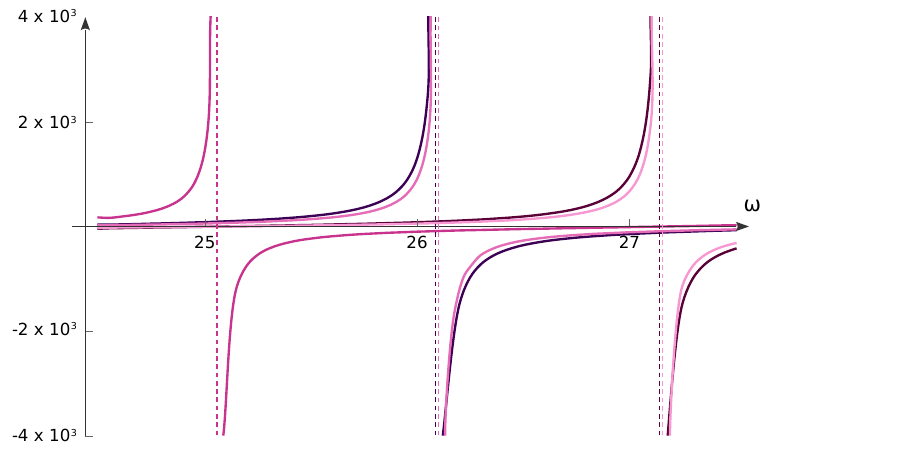}
\\~\\
    \includegraphics[width=0.45\textwidth]{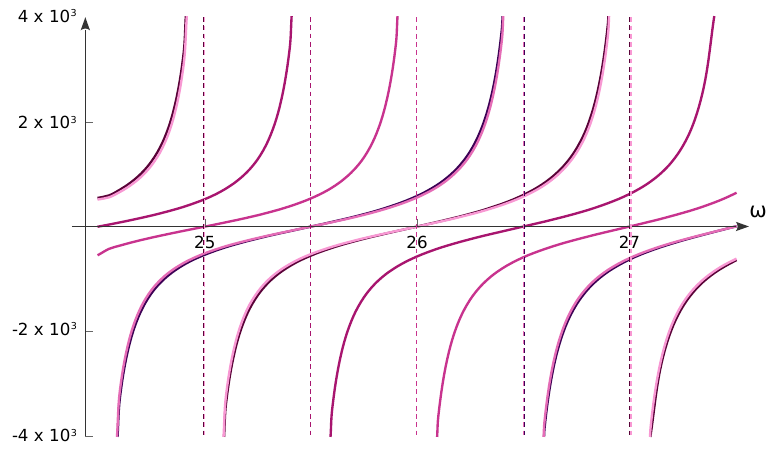}
	\hfill
	\includegraphics[width=0.54\textwidth]{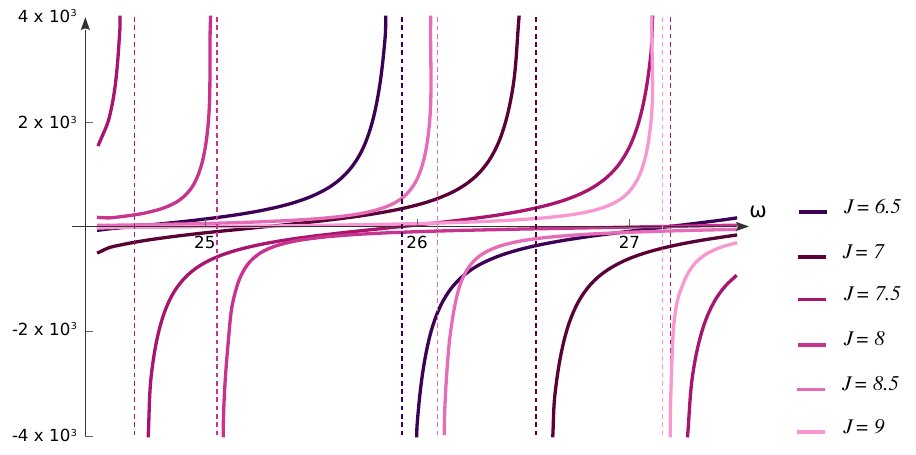}
 	\caption{\label{fig:evsImG_30} \small Plots of the two-point fermionic correlator as a function of the frequency $\omega$ for temperature $T_c=2\times 10^{-2}$ (top line) and  temperature $T_c=5\times 10^{-3}$ (bottom line) for different values of the angular momentum $J$. \underline{Left:} $\Theta_c=-30$. \underline{Right:} $\Theta_c=30$. The dependence on temperature is almost undetectable at large negative central degeneracy. Moreover, there is an approximate periodicity in the position of the poles, those with $J=6.5, 7$ almost coincide with those corresponding to  $J=8.5, 9$, which breaks down at low temperatures.\normalsize
 }
\end{figure}

Using the numerical background we obtained in the previous chapters, we evaluated the expression \eqref{eq:femionic.correlator} using a Mathematica script \cite{cita24}, to have the two-point correlator for fermionic operators on the dual field theory. The results are presented in the plots of Figs.~\ref{fig:dependence.temperature}, \ref{fig:dependence.theta} and \ref{fig:evsImG_30}.

As we can see in Fig.~\ref{fig:dependence.temperature} and \ref{fig:dependence.theta}, the dependence of the normal mode position on the central temperature and central degeneracy is more evident for modes with higher $J$. In Fig.~\ref{fig:evsImG_30}, it can be checked that, as we move from the region of large negative $\Theta_c$ into large positive values, the separation between the normal modes grows. Moreover, at fixed $\Theta_c$ this separation is more sensitive to temperature for large positive values of $\Theta_c$. On the other hand, the dependence on $J$ seems to have an approximate periodicity, which breaks down as we move to larger $\Theta_c$ at lower temperatures. This last feature concides with the entrance on the zone where critical properties were manifest in the previous chapters.

\begin{figure}[ht]
	\centering
    \includegraphics[width=1\textwidth]{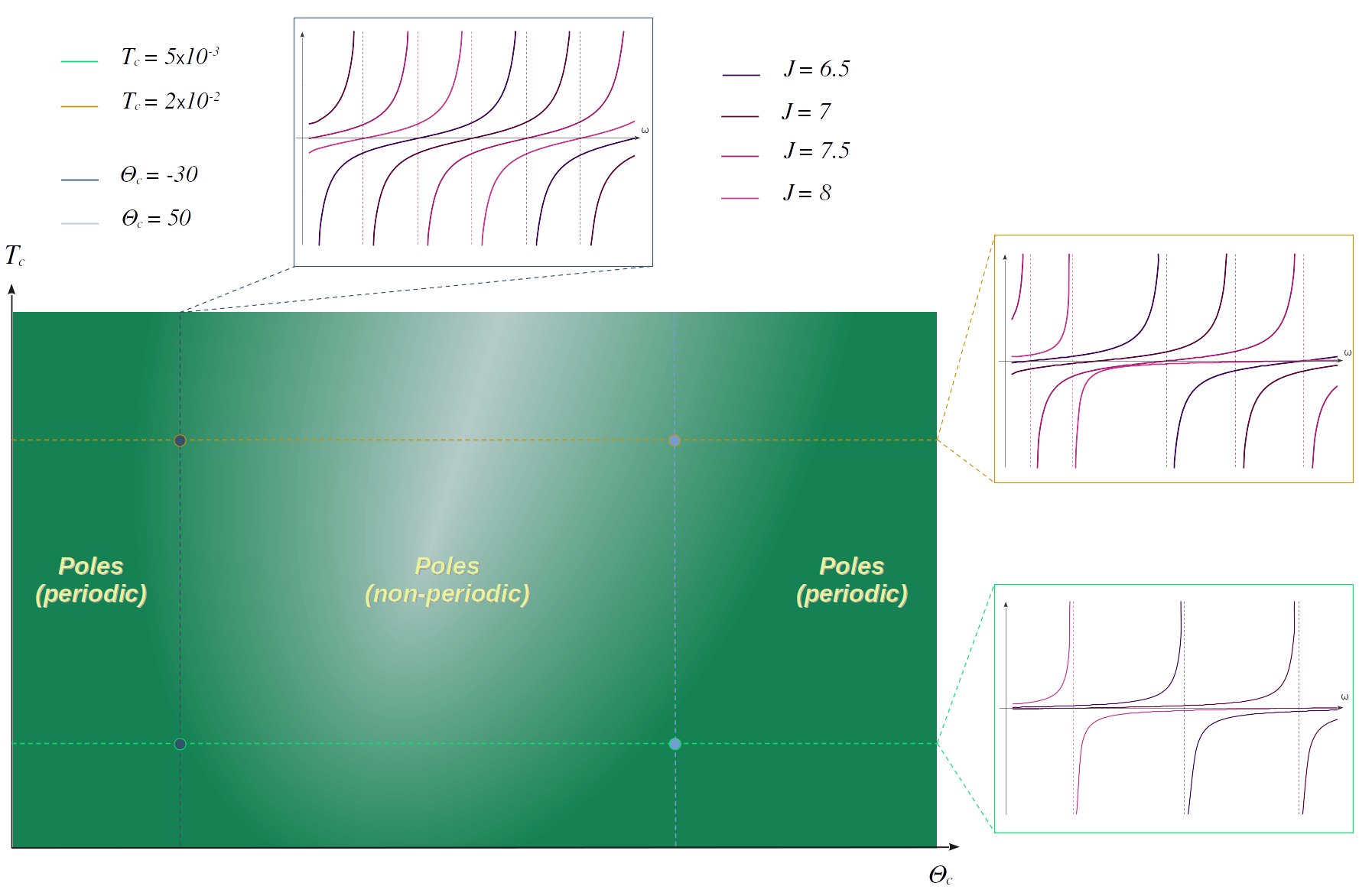}
 	\caption{\label{fig:phase.diagram.fermions} \small		Central temperature ${T}_c$ {\em vs.} central degeneracy $\Theta_c$ phase diagram. The dotted horizontal lines represent constant temperatures, the vertical ones represent constant values of $\Theta_c$. The fermionic correlators plotted inside each box at the top and and the right of the figure are located in the phase diagram by a dot with the same color sitting along the corresponding line.
    Compare with Fig.~\ref{fig:phase.diagram.metal} is suggestive.\normalsize
 }
\end{figure}

The information was included in the upgraded phase diagram in Fig.~\ref{fig:phase.diagram.fermions}.

\begin{tcolorbox}[colback=blue!5!white,colframe=blue!75!black,title=Hint \arabic{boxnew}]   
The fermionic two-point correlator is dominated by its poles, whose separation grows with the central degeneracy, with a sensitivity that is higher for higher temperatures. The response to $J$ seems to have an approximate periodicity, which breaks down as we move into the zone were the critical features were detected in the previous chapters. 
\end{tcolorbox}
\stepcounter{boxnew}

\newpage

%
%\begin{figure}[ht]
%	\vspace{.01cm}
%	\centering
%	\includegraphics[width=0.45\textwidth]{Chapters/Thesis6/thetam30_e_vs_imG_t50.png}
%	\hspace{0.05\textwidth}
%	\includegraphics[width=0.45\textwidth]{Chapters/Thesis6/thetam30_e_vs_imG_t200.png}
%	\caption{\label{fig:evsImG_m30} Plot of the two-point function for $\Theta_c = -30 $, temperatures $T_c=1/50$ (left) and $T_c=1/200$ (right) for different fixed lambdas.}
%\end{figure}
%
%\begin{figure}[ht]
%	\vspace{.01cm}
%	\centering
%	\includegraphics[width=0.46\textwidth]{Chapters/Thesis6/theta_m30_vs_theta_30_t1.png}
%	\hspace{0.05\textwidth}
%	\includegraphics[width=0.46\textwidth]{Chapters/Thesis6/theta_m30_vs_theta_30_t4.png}
%	\caption{\label{fig:t1_vs_t4} The left plot correspond to a fixed temperature $T_c=1/50$ but with two different thetas the small points indicates $\Theta_c = -30 $ and the bigger ones $\Theta_c = 30 $. The right plot is the same but for fixed $T_c=1/200$.}
%\end{figure}

%Thermodynamic inst
\chapter{Stability analysis}
\label{ch:holo_neutron_stars}

In this chapter we  study the thermodynamic stability of the holographic neutron star, using the Katz and turning point criteria for self-gravitating fermions. The turning point criterion is a standard tool on astrophysics, and it was explained in \ref{sec:turning} in its full general form \cite{cita39}. The Katz criterion, as also explained in section \ref{sec:katz}, was applied to self gravitating fermions for Newtonian gravity in \cite{cita44, cita45} and in \cite{cita46, cita47} for the general relativistic case. 
\section{Microcanonical ensamble}

The turning point criterion considers the total entropy as a function of the mass and the number of particles. This implies that it is naturally defined in the microcanonical ensamble. 

In order to be able to apply it to our neutron star, we need to define its total mass as the value of the mass function at the boundary \eqref{eq:mass}, in formulas
\begin{equation}
    M=M(r_b)\,.
    \label{eq:mass2}
\end{equation}
We will exploit this definition of the total mass to explore our domain of parameters.

\section{Grand canonical ensamble}
\label{subsec:canonicalpotential1}
\label{sec:potential}
The Katz criterion is defined for a generic thermodynamic potential. In the present holographic context, it is convenient to use it in the grand canonical free entropy, since our boundary theory is naturally defined at constant temperature $T$ and chemical potential $\mu$. 

The holographic grand canonical potential $\Omega(\mu,T)$ is calculated according to equation \eqref{eq:freeEnergy}, where the matter term $S_{\sf EM}$ is given by the integral of the fluid pressure (see Appendix \ref{sec:background}). By evaluating this on our Ansatz, we get
\begin{align}
	\Omega(\mu, T)
	&=
	\frac{L }{G\, }e^{-\frac{\nu_\infty}{2}}
	\left(
	 M-4\pi \int dr
	\,e^{\frac{\nu+\lambda}{2}}r^2
	(\rho+ P)
	\right)\,,
	\label{eq:on.shell.action}
\end{align}
where $M$ is the total mass as defined on \eqref{eq:mass2}, and $\nu_\infty$ corresponds to the asymptotic value of the function $\nu(r)$. With this, we can calculate the grand canonical free entropy $\Phi(T,\mu)$ according to the definition \eqref{eq:free.entropy}.

The on-shell action, and thus the grand canonical free entropy, are natural functions of the central parameters $T_c$ and $\Theta_c$. Since we are interested in the boundary physics, we want to write  them as functions of the boundary quantities $T$ and $\mu$, which depend on $T_c$ and $\Theta_c$ according to \eqref{eq:tolman.klein}. We get a parametric relation which might be multi-valued, a situation that is somewhat frequent in self gravitating systems. Then, we need a criterion to determine which of its many branches corresponds to a stable phase of the boundary theory at a given temperature $T$ and chemical potential $\mu$. We  identify the stable equilibrium state with the diluted configurations, for which the central degeneracy $\Theta_c$ is negative and near $-20$.

\section{Results}
\begin{figure}[H]
	\centering
	\includegraphics[width=.495\textwidth]{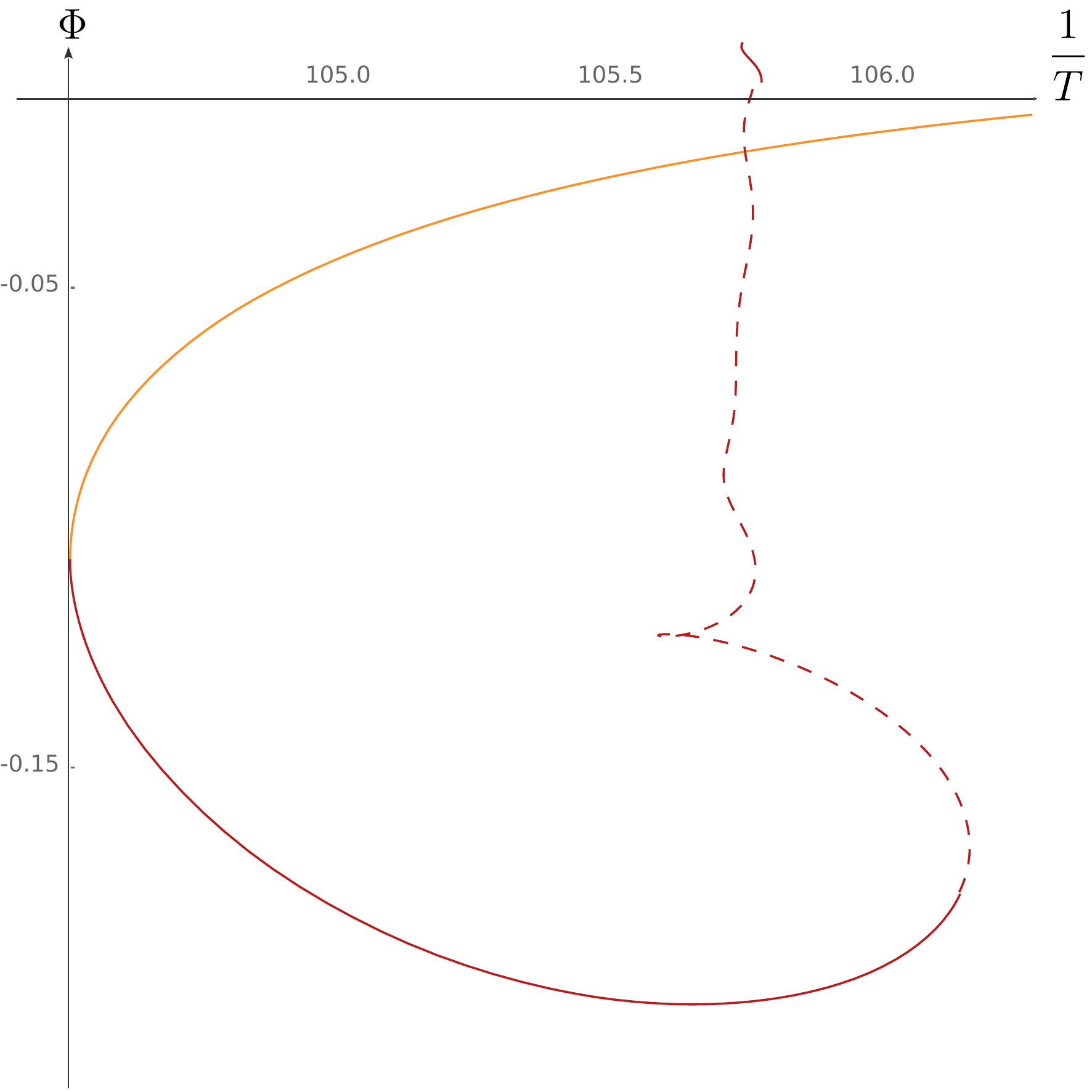}
	\hfill
	\includegraphics[width=.495\textwidth]{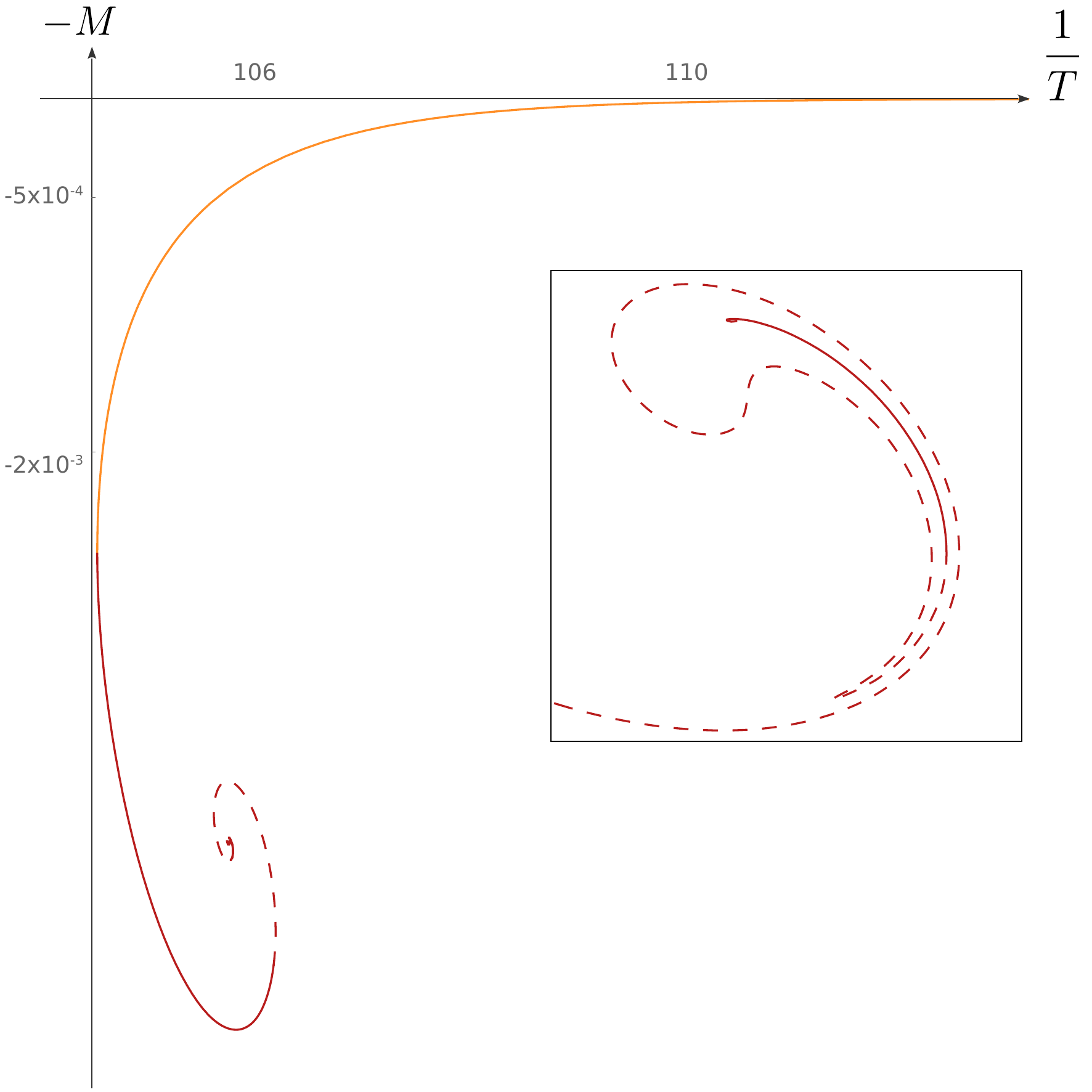}
	\\
	~
	\\	
	\centering
	\includegraphics[width=.495\textwidth]{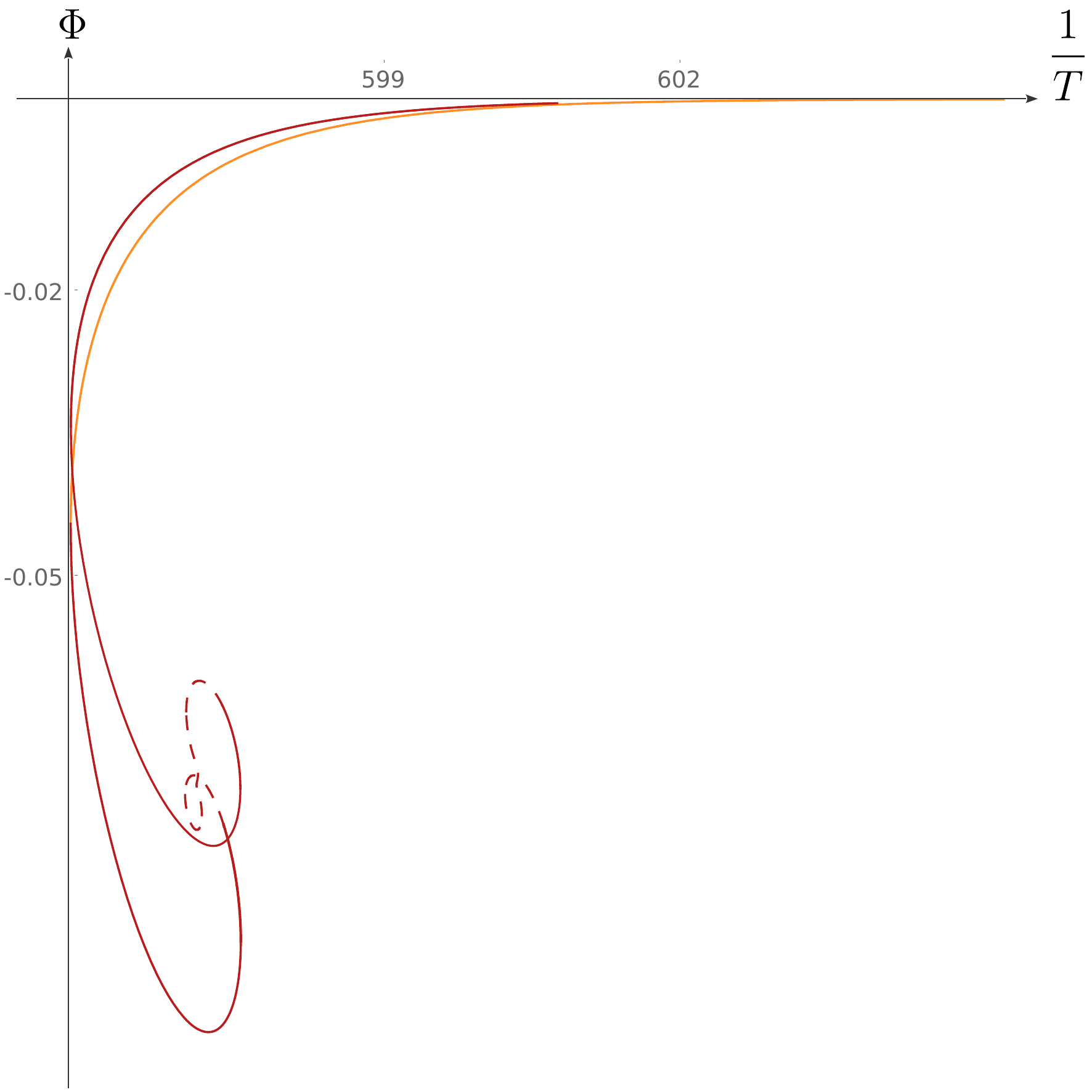}
	\hfill
	\includegraphics[width=.495\textwidth]{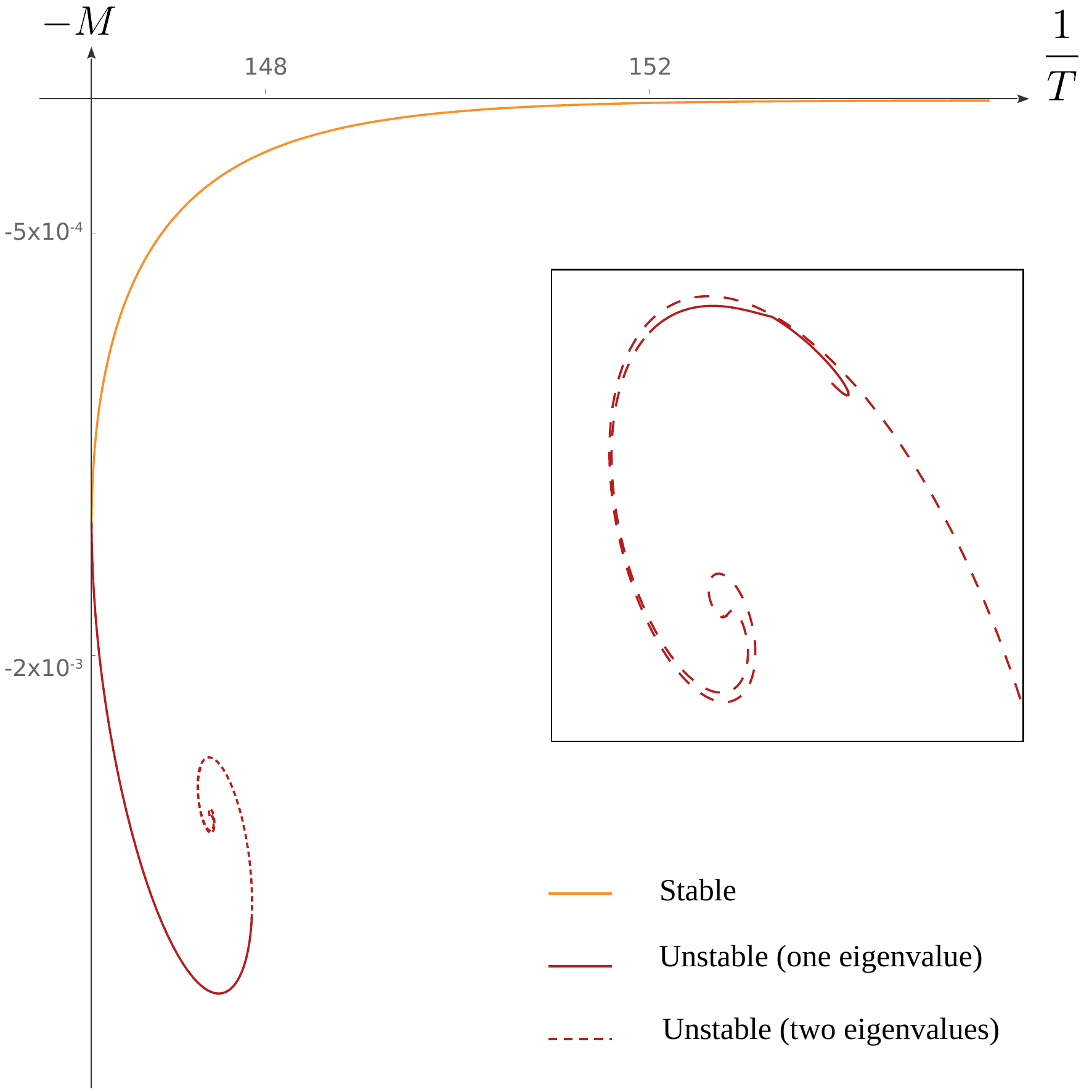}
	\caption{\label{fig:massvstemp06}\small
    Plots as functions of the inverse boundary temperature $1/{T}$, for a fixed value of $\mu/{T}=50$ (upper plots) and $\mu/\tilde{T}=80$ (bottom plots). \underline{Left:} Parametric plot of the grand canonical free entropy $\Phi$. \underline{Right:} Parametric plot of minus the total mass $-{M}$.
		At negative values of $\Theta_c$ all the eigenvalues are assumed to be negative (orange line). As the curve reaches a vertical asymptota, an eigenvalue changes sign from negative to positive (continuous red line). The process repeats on the second vertical asymptota, a second eigenvalue changing sign (dashed red line). It reverses st the center of the spiral (see inset) changing back the sign of the corresponding eigenvalues. However, the stability is not recovered.\normalsize}
\end{figure}

\begin{figure}[H]
	\centering
	\includegraphics[width=.495\textwidth]{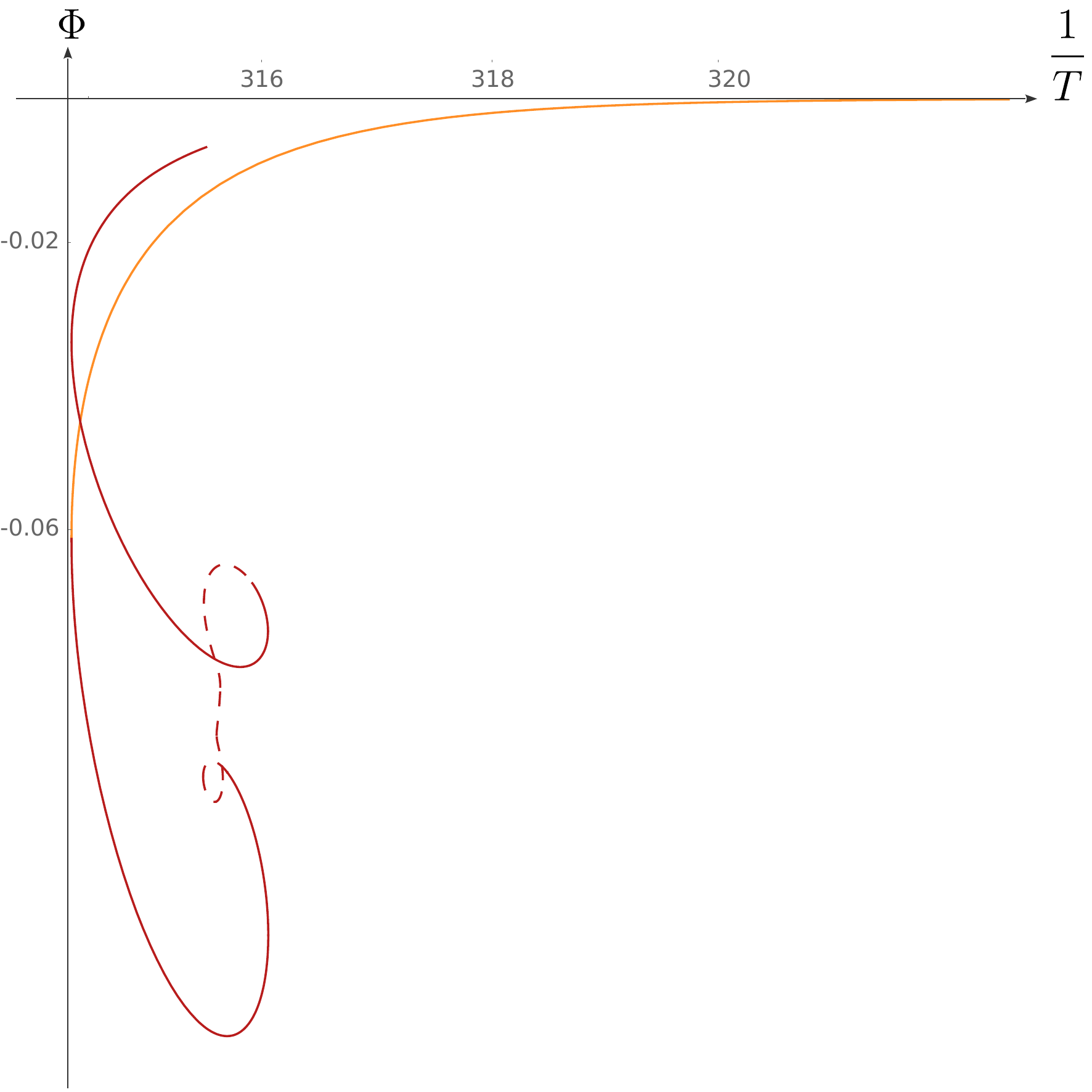}
	\hfill
	\includegraphics[width=.495\textwidth]{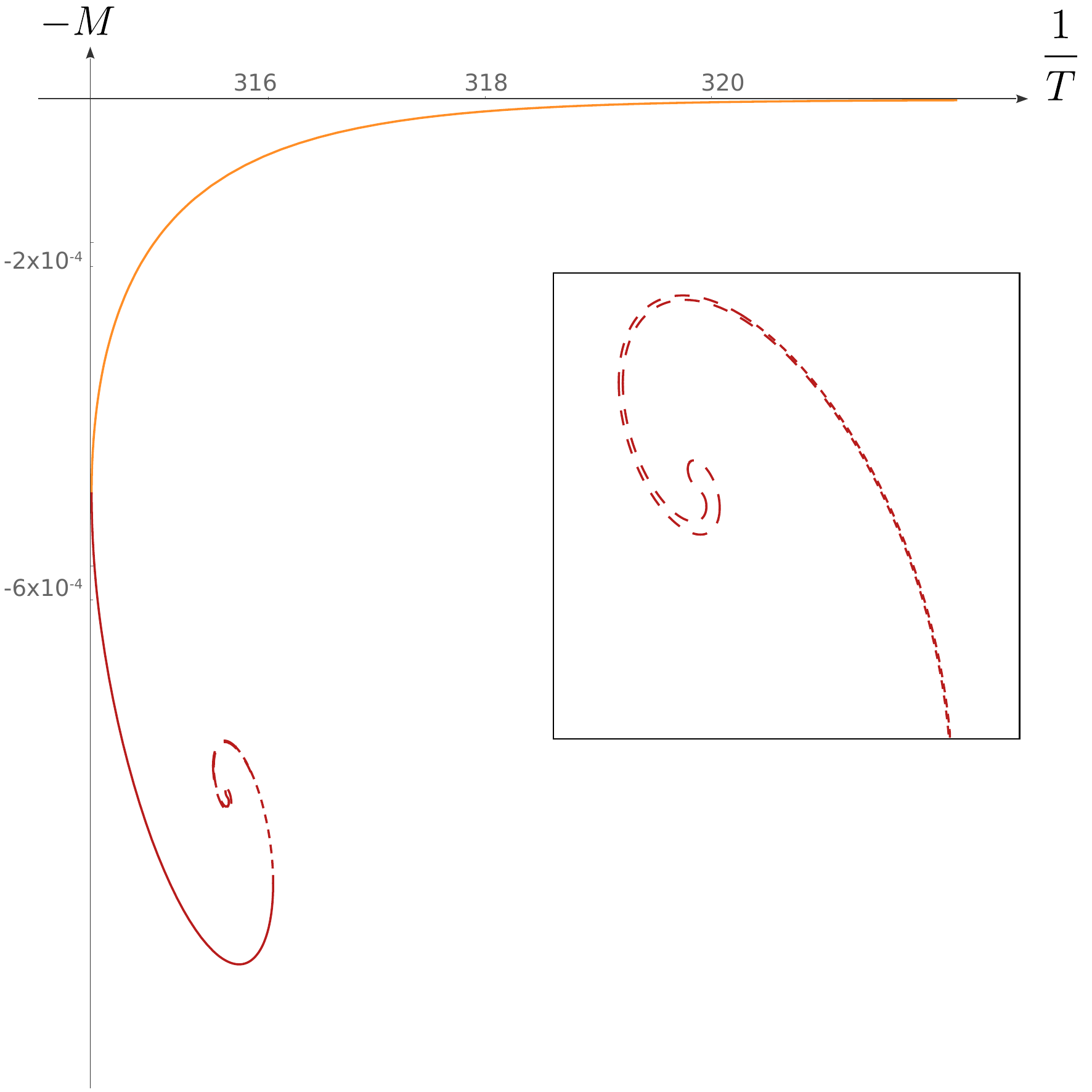}
	~
	\\	
	\includegraphics[width=.495\textwidth]{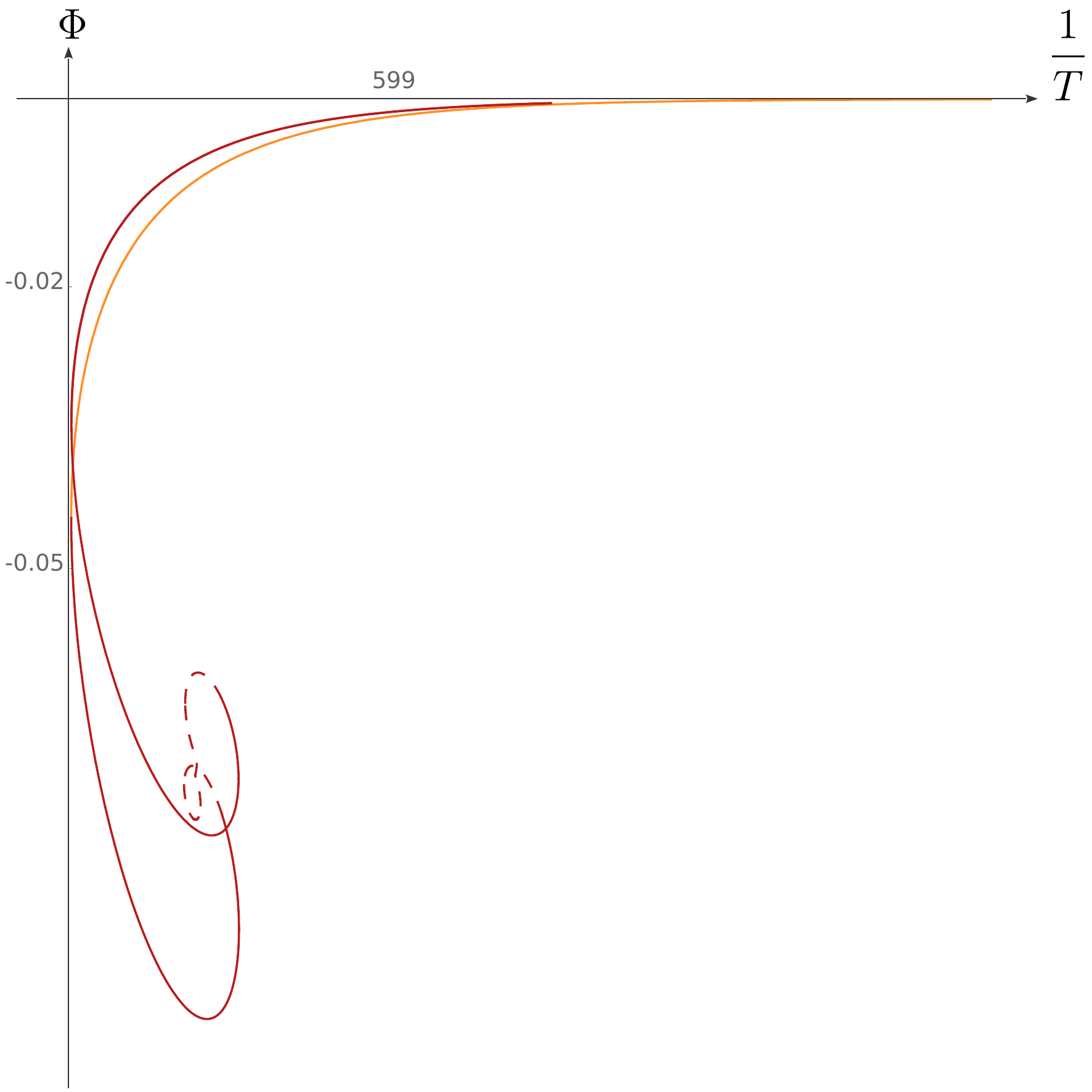}
	\hfill
	\includegraphics[width=.495\textwidth]{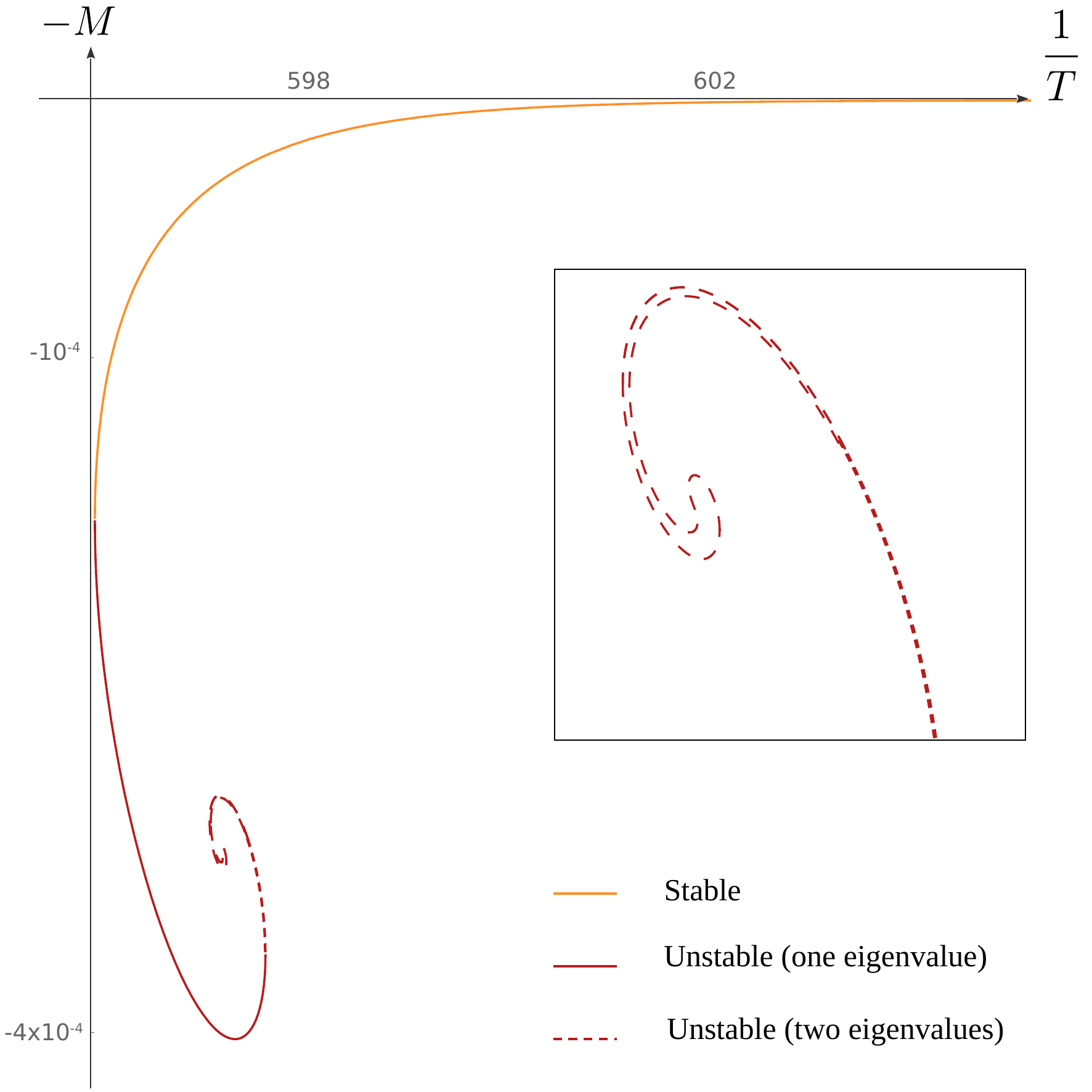}
	\caption{\label{fig:massvstemp062} \small Plots as functions of the inverse boundary temperature $1/{T}$, for a fixed value of $\mu/{T}=200$ (upper plots) and $\mu/\tilde{T}=400$ (bottom plots). \underline{Left:} Parametric plot of the grand canonical free entropy $\Phi$. \underline{Right:} Parametric plot of minus the total mass $-{M}$.
		At negative values of $\Theta_c$ all the eigenvalues are assumed to be negative (orange line). As the curve reaches a vertical asymptota, an eigenvalue changes sign from negative to positive (continuous red line). The process repeats on the second vertical asymptota, a second eigenvalue changing sign (dashed red line). It reverses st the center of the spiral (see inset) changing back the sign of the corresponding eigenvalues. However, the stability is not recovered.\normalsize}
\end{figure}

\vspace{-.15cm}
We evaluated the grand canonical free entropy $\Phi(T,{\mu})$ and the total mass $M(T,{\mu})$ for our numerical solutions. The resulting curves are shown in Figs.~  \ref{fig:massvstemp06}-\ref{fig:massvstemp062} and \ref{fig:turning_point}.

In Figs. ~\ref{fig:massvstemp06} and \ref{fig:massvstemp062} we see that, starting from the most negative value of $\Theta_c$ identified as the stable branch, the curve of $-{M}$ {\em vs.} $1/{T}$  spirals into a succession of vertical asymptota, in which the slope changes from positive to negative. This indicates that a succession of eigenvalues are becoming positive, increasing the degree of instability. At a certain point, the process reverses, and the curve spirals out. Nevertheless, the number of eigenvalues that change back to negative sign is not enough to restore local stability. At larger $\Theta_c$, the curve spirals again into gravitational collapse.

\begin{figure}[ht]
	\centering
	\includegraphics[width=.8\textwidth]{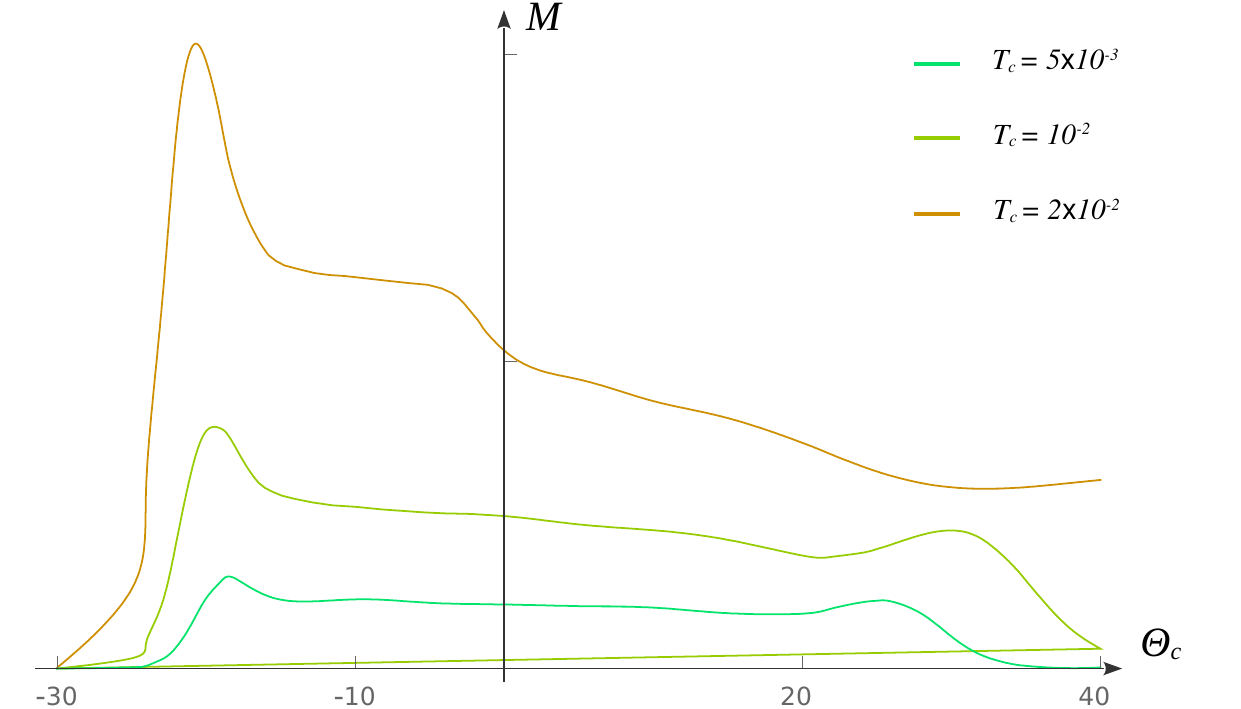}
	\vspace{.5cm}
 \caption{\label{fig:turning.point} 
  \small
 		Total mass $M$ {\it vs.} central degeneracy $\Theta_c$ plot, for different temperatures. As we increace the central degeneracy from the left, we reach a turning point where the system becomes unstable in the microcanonical description. Remarkably, there is a second turning point at higher degeneracies.    
   \normalsize}
\end{figure}

In Fig. ~\ref{fig:turning.point} we see a plot of the total mass as a function of the central density. We notice that it has two turning points. As the first one is reached when the central degeneracy grows from the stable negative values, an instability is triggered in the microcanonical description of the system.

\begin{figure}[ht]
	\centering
	\includegraphics[width=.95\textwidth]{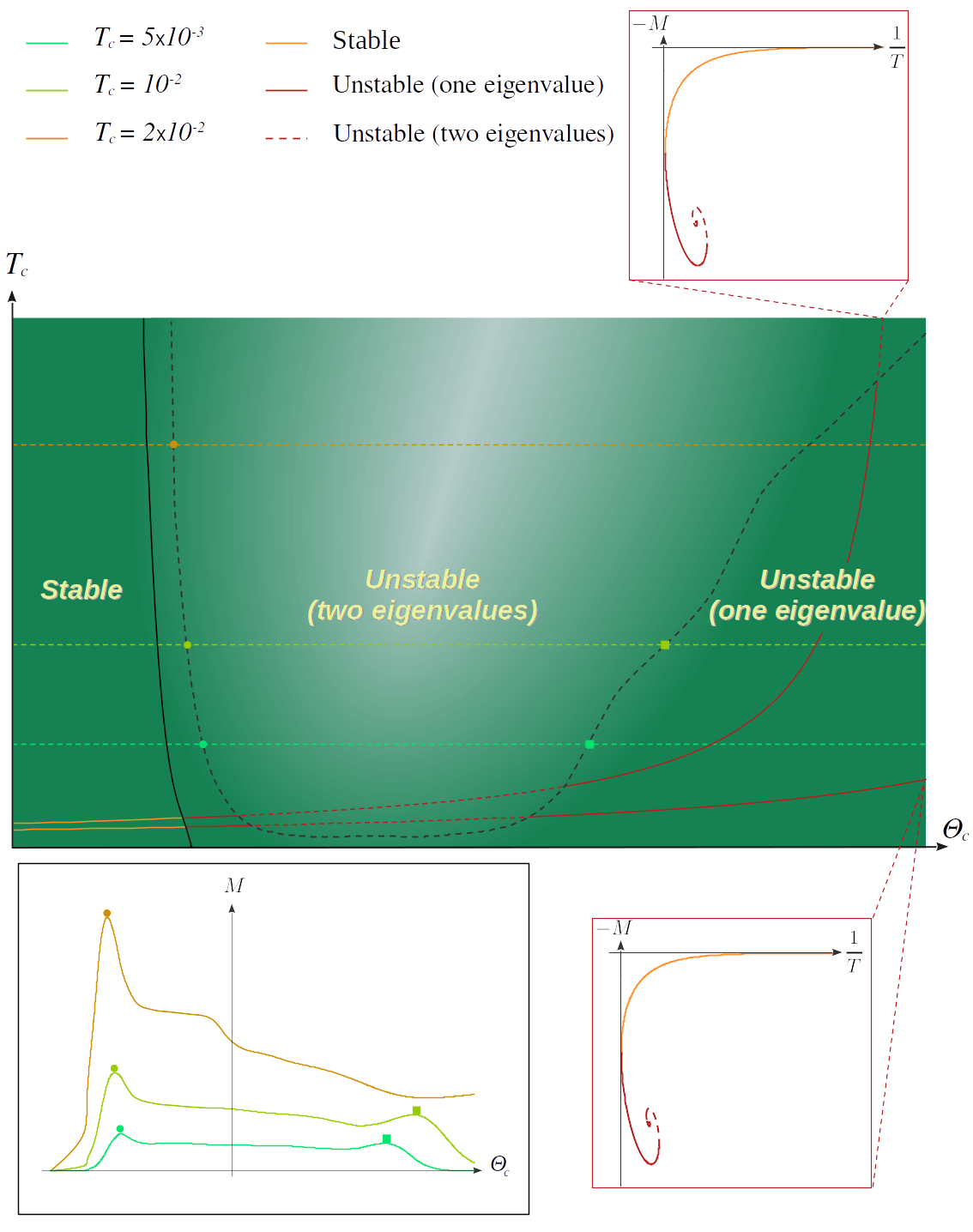}
	\caption{\label{fig:phase.diagram.instability} \small
 		Central temperature ${T}_c$ {\it vs.} central degeneracy $\Theta_c$ phase diagram. The dotted horizontal lines represent the constant temperatures for which the mass as a function of the central degeneracy is plotted in the box on the left bellow the diagram. The curved lines represent the constant values of $\mu_c/T_c$ for which minus the mass as a function of $1/T_c$ is plotted in the boxes on the right at the top and bottom. As we can see, the turning points in the mass function coincide with the change of sign of a second eigenvalue according to Katz criterion, as it was signaled with the colored circles and squares in the mass plot and in the diagram. The continuous black line separates the stable region at the left from the unstable one, the dotted black line delimits the zone where more than one eigenvalue is unstable. Again, the comparison with Fig.~\ref{fig:phase.diagram.metal} is suggestive.\normalsize}
\end{figure}

As the dotted lines in Figs.~ \ref{fig:massvstemp06}-\ref{fig:massvstemp062} and Fig.~\ref{fig:turning.point} show, the interval in temperatures where two of the eigenvalues are positive lies approximately  between the turning points in the mass as a function of the central density. Interestingly, this coincides with the region at which the critical features discussed in the previous chapters are present. This can be seen in the updated phase diagram of Fig.~\ref{fig:phase.diagram.instability}

\begin{tcolorbox}[colback=blue!5!white,colframe=blue!75!black,title=Hint \arabic{boxnew}]  As $\Theta_c$ grows from negative stable values, we observe the appearance of an unstable region in the phase diagram according to Katz criterion, as the first vertical asymptota is reached. The second vertical asymptota coincides with the turning points in the mass. It signals the change of sign of another eigenvalue, which remains positive in the region of phase space where the system  exhibits critical features.
\end{tcolorbox}
\stepcounter{boxnew}

\newpage

\part{Conclusions}
%Summary
\chapter{Holography and metallic criticality}
\label{ch:summary}

In this PhD thesis, we delved into the fascinating realm of holographic neutron stars at finite temperature. 

Our journey began by formulating a gravitational background that represents a neutron star within asymptotically global AdS space. To achieve this, we solved Einstein's equations while incorporating a stress-energy source that mimics a neutral perfect fluid at finite temperature. We studied the features of the resulting profiles for the energy density as a function of their radius.

Then, we probed the constructed backgrounds with a scalar operator, computing the corresponding two-point correlator. Our analysis started by computing it in the limit large conformal dimension, to explore later the finite conformal dimension case, enriching our understanding of the system's behavior. We further calculated the two-point correlator of a fermionic operator, in the WKB limit. This multifaceted approach allowed us to gain deeper insights into the properties of the holographic theory.

To conclude, we conducted extensive stability assessments of the solution, varying the system parameters. This was done by employing both the turning point and the Katz criteria.

%In the upcoming chapter, 

\section{Main results of the present thesis}

Here we summarize the key findings and insights gained throughout our exploration:

\begin{itemize}
    \item \underline{Holographic neutron star backgrounds:}

    The Tolman-Oppenheimer-Volkoff equations provide a three-parametric family of neutron star solutions featuring AdS asymptotics. These solutions are uniquely characterized by the central temperature $T_c$, the central degeneracy parameter  $\Theta_c$, and the number of local degrees of freedom  $\gamma$. In the holographic framework, the value of $\gamma$ is assumed to be very large, effectively reducing our moduli space to two dimensions. Within the resulting central degeneracy $\Theta_c$ versus central temperature $T_c$ plane, we can construct a phase diagram that encapsulates the fundamental traits of these neutron star solutions.

    Starting from the region of large negative $\Theta_c$, we encounter energy density profiles characterized by a central plateau that sharply terminates at the star's outer boundary. These solutions are characterized by extreme dilution and fall within the Boltzmannian regime, resisting against gravitational collapse primarily due to the thermal pressure.

    As we progress into   higher central degeneracy, a remarkable transformation occurs at the edge of the radial density profile. It degenerates into a power-law form, a trend that becomes more pronounced as the degeneracy increases.

    Continuing further into larger $\Theta_c$, the star develops a distinctive core-halo structure. A dense and highly degenerate core emerges, defying gravitational force through degeneracy pressure. Surrounding it there is a more diluted halo, sustained thermally. Both the core and the halo exhibit power-law edges.

    Finally, when the central degeneracy becomes sufficiently high, the configuration's edge recovers a sharp and distinct form.

    Notably, the central region of the $\Theta_c$ versus  $T_c$ plane where the power-law edge is present exhibits a wider profile at higher temperatures and narrows into a wedge-like shape as the temperature decreases.   

    All these features are summarized in the phase diagram \ref{fig:diagram.background}.

    \item \underline{Scalar two-point correlator:}

    We went further in our analysis by computing the two-point correlator of a scalar operator within the backgrounds we  derived.

    In the regime of a large conformal dimension $\Delta$, this calculation can be mapped to the motion of a massive Euclidean particle moving through the AdS bulk. The particle dives into the bulk up to a minimal distance from the neutron star's center and then is scattered, bouncing back to the boundary. The exponential of the corresponding Euclidean on-shell action furnishes the desired  correlator.
    
    Our calculations across different regions of the phase diagram revealed intriguing results. At both large negative and large positive values of the central degeneracy $\Theta_c$, the resulting correlator exhibits a smooth behavior. However, when dealing with intermediate values of $\Theta_c$, the on-shell action becomes multi-valued, with a distinctive swallow tail shape. This, in turn, yields a correlator that is non-smooth at antipodal separations on the sphere. Remarkably, this behavior aligns with the same wedge-like region in the $\Theta_c$ versus $T_c$ phase diagram where the density profiles exhibit a power-law edge.
     
    In the scenario involving a finite conformal dimension $\Delta$, the computation of the scalar correlator requires solving the Klein-Gordon equation for a neutral scalar field in the bulk with regular boundary conditions at the center of the star, and then capturing the leading and subleading behaviors of the solution at infinity. The resulting correlator diverges at the energies corresponding to the normalizable scalar modes.

    This computation led to a correlator that, as a function of the frequency, is primarily governed by the poles whenever the central degeneracy is negative enough. As the value of $\Theta_c$ increases, these poles gradually disperse, allowing the regular part of the correlator to take the forefront. This results in a distinctive power-law behavior. Remarkably, such behavior becomes conspicuous precisely at the left boundary of the previously mentioned wedge-like region. However, it is important to note that our calculations have yet to provide insights into the behavior at the right edge of this wedge-like region.

    These properties are schematized in the phase diagram \ref{fig:phasediagram}.

     \item \underline{Fermionic two-point correlator:}

    Computing the two-point correlator of a fermionic operator involves tackling a spinorial perturbation within the bulk. Consistently with the approximations made to obtain the background metric, the Dirac equation can be solved in the WKB limit. This approach allows to derive an analytical expression for the correlator, which can be expressed in terms of integrals computed on the numerical background.

    The resulting two-point correlator exhibits a distinctive pole structure, the behavior of which undergoes a significant transformation as the central degeneracy $\Theta_c$ transitions from large negative values into positive ones. For cases characterized by a large negative $\Theta_c$, the pole energies exhibit an approximately periodic behavior as functions of the angular momentum $J$, and are barely affected by a change in the central temperature $T_c$. However, this periodicity vanishes when dealing with large positive degeneracies. Moreover, in the same region the dependence on the central temperature is much stronger. 
 
    These results can be seen in the phase diagram \ref{fig:phase.diagram.fermions}.

~
 
    \item  \underline{Stability analysis:}

    We conducted a stability analysis of our neutron star solutions in the microcanonical ensemble, employing the turning point criterion. In this approach, we plotted the total mass of the configuration as a function of the central degeneracy $\Theta_c$ for various constant values of the central temperature $T_c$.

    Our investigation revealed a compelling pattern: as the central degeneracy increases, the mass function attains a peak value, followed by a subsequent decrease before embarking on an ascent towards a second peak at a larger value of the central degeneracy. According to the turning point criterion, our solutions are deemed unstable for all central degeneracy values exceeding the one corresponding to the first peak.

    Notably, when we map these results onto the $\Theta_c$ versus $T_c$ plane, the fascinating correlation emerges once again. The positions of the first and second turning points align precisely with the left and right boundaries of the previously discussed wedge-like region. It becomes evident that this region harbors a plethora of intriguing  phenomena.

    Furthermore, we conducted a comprehensive stability analysis within the grand canonical ensemble, employing the Katz criterion. This involved the calculation of the grand canonical free entropy, expressed in terms of the Euclidean on-shell action of the bulk fields. Subsequently, we plotted the derivative of this entropy, which corresponds to the negative of the total mass, as a function of the reciprocal boundary temperature.

    We identified the region with large negative central degeneracy $\Theta_c$ with the stable solutions. As the temperature increases, the free entropy embarks on a counterclockwise spiral trajectory, punctuated by a succession of vertical asymptotae. According to the Katz criterion, each passage through one of these asymptotae signals the emergence of a positive eigenvalue within the perturbative free entropy matrix, indicating an instability.
    
    After a certain number of counterclockwise revolutions, the behavior changes and the curve spirals back in a clockwise direction. Nevertheless, the number of turns results insufficient to reestablish stability.

    Remarkably, when we map these findings onto the central degeneracy $\Theta_c$ versus central temperature $T_c$ plane, a clear alignment with our wedge-like region of interest comes into view. The left boundary of the wedge coincides with the second turn of the free entropy curve, whereas the right boundary aligns with the point at which the curve unravels to a single remaining positive eigenvalue.

    This results are summarized in the diagram \ref{fig:phase.diagram.instability}.
 
\end{itemize}

In summary, one of the most compelling features that has surfaced is the emergence of a fascinating wedge-like region within the $\Theta_c$ versus $T_c$ plane, which harbours a multitude of intriguing properties.

\begin{figure}[ht]
	\centering 
	\includegraphics[width=.8\textwidth]{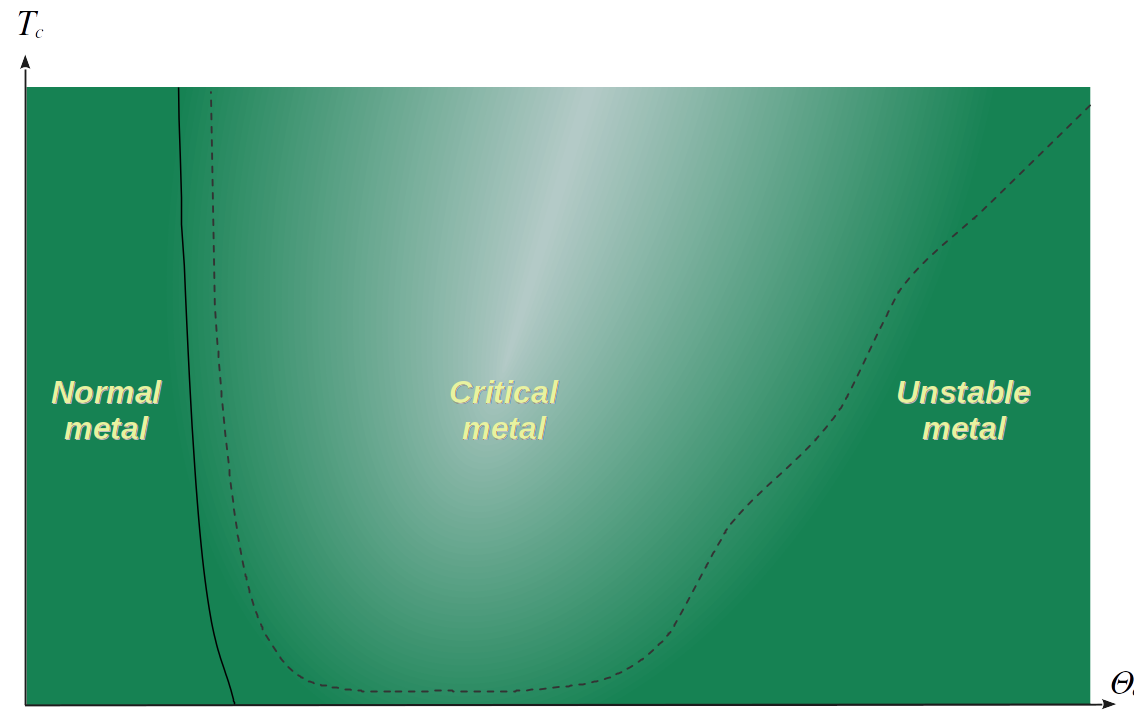} 
	\caption{\label{fig:phase.diagram.final}
\small		Central temperature ${T}_c$ {\em vs.} central degeneracy $\Theta_c$ phase diagram. Compare with the proposal of Fig.~\ref{fig:phase.diagram.metal}.\normalsize}
\end{figure}

\section{Discussion and further directions}
The findings just presented pave the way for engaging discussions within the realms of both the thermodynamics of self-gravitating systems and the critical behavior of strongly coupled metals. In particular:

\begin{itemize}
    \item \underline{Thermodynamics of self-gravitating systems:} 

    The wedge region, as depicted within the $\Theta_c$ versus $T_c$ plane, resides entirely within the realm of instability, a fact supported by both the turning point criterion and the Katz criterion: 
    \begin{itemize}
        \item     The left boundary of the wedge aligns with the location of the first maximum of the total mass as a function of the central degeneracy, indicating an instability within the microcanonical framework, as defined by the turning point criterion. 
        \item   The onset of the first unstable eigenvalue of the free entropy perturbation matrix occurs to the left of the left boundary of the wedge, signaling an instability within the grand canonical description, as established by the Katz criterion.
    \end{itemize}
    However, it's important to note that the criteria do not perfectly align: the unstable region identified by the turning point criterion is wholly encompassed by the one derived from the Katz criterion. This discrepancy can be recognized as an instance of the ``non-equivalence of ensembles'', a phenomenon well-documented in the thermodynamics of self-gravitating systems. This non-equivalence arises due to the long-range nature of gravity, which renders such systems non-extensive.

    Our findings can shed some light into the relationship between these two descriptions. Notably, both the left and right boundaries of the wedge correspond to the positions of the first and second maxima in the mass as a function of central degeneracy. Furthermore, they also align with the emergence of a second unstable eigenvalue within the perturbation matrix. These correlations suggest a potential connection between the presence of two unstable eigenvalues and the maxima of the mass function, giving some insight on the interplay between these properties.
    
    \item \underline{Quantum metallic criticality:}

    Viewed through the lens of holography, our system represents a highly degenerate state of strongly coupled fermions, existing at a finite temperature and chemical potential. This state is confined within a spherical vessel, and the radius of this vessel plays a crucial role by providing a scaling factor that effectively disentangles the temperature and chemical potential axes. Consequently, we are able to construct a two-dimensional phase diagram. 

    In the wedge-like region at intermediate central degeneracies (which correspond to intermediate chemical potentials in the boundary theory), the system manifests power-law relationships among some of its physical observables. In particular, these include the behavior of the density profile's edge as a function of radius and the two-point correlator of a scalar operator with finite conformal dimension as a function of frequency. These power-law phenomena can be traced back to the criticality of the underlying physics, and are  summarized in the phase diagram in Fig.~\ref{fig:phase.diagram.final}.
    
    This result holds significant importance and stands as one of the central outcomes of this thesis. It aligns with the expectations from the behavior of metallic degrees of freedom within a strongly coupled system, see Fig.~\ref{fig:phase.diagram.metal}. Moreover, it finds support in the context of High $T_c$ superconductors, see Fig.~\ref{fig:HihgTc} and the presumed characteristics of quark-gluon plasmas.
 
    Furthermore, this region also exhibits interesting behaviors in other observables. For instance, the two-point correlator of scalar operators with large conformal dimension shows a non-smooth behavior at antipodal separations. Additionally, the poles on the fermionic correlator become sparse. It would be a compelling avenue for further exploration to ascertain whether these features can also be attributed to criticality within the system.
    
\end{itemize}

Within the gravitational context, the instabilities we've uncovered are often interpreted as precursors to gravitational collapse. However, in our research, a critical question arises: What is the nature of the stable solution following the collapse? This is a conundrum that remains unresolved. Any stable phase must align with our boundary conditions, which explicitly include a non-vanishing chemical potential $\mu$. According to the Klein condition, the chemical potential will remain non-zero at any interior point where the lapse function is finite. Consequently, the requisite bulk solution cannot be a black hole, as adopting a black hole solution with our boundary condition for the chemical potential would lead to a non-zero (indeed, divergent) chemical potential at the horizon, resulting in a  non-equilibrium state.

One potential avenue is thermal AdS, which lacks a horizon where the non-zero chemical potential would cause complications. Another approach involves introducing additional degrees of freedom, such as extra bulk fields that vanish at the boundary but modify the bulk, permitting the emergence of a distinct phase. A vector field could facilitate the introduction of a finite charge within the fluid, enabling the calculation of boundary conductivity. A charged scalar field, on the other hand, would provide room for a superconducting phase. These are intriguing possibilities that warrant further exploration.

It's worth noting that the findings presented in this thesis could have practical applications. A highly degenerate fermionic state within a spherical context can be experimentally realized through the electronic degrees of freedom of an isolated fullerene molecule, C$_{60}$ \cite{cita60}. Such a system boasts a substantial rotational symmetry characterized by the point group $I_h$, and its phonon perturbations feature a transverse component described by a neutral scalar. Doped fullerene crystals exhibit a phase diagram akin to that of unconventional superconductors, like cuprates \cite{cita61}. Even as crystallization reduces rotational symmetry, intramolecular phonons may still influence the physics of fullerene superconductivity \cite{cita62}. A holographic construction based on the approach outlined in this work has the potential to contribute significantly to our understanding of the phase diagram of these materials.

\begin{figure}[ht]
	\centering 
	\includegraphics[width=.75\textwidth]{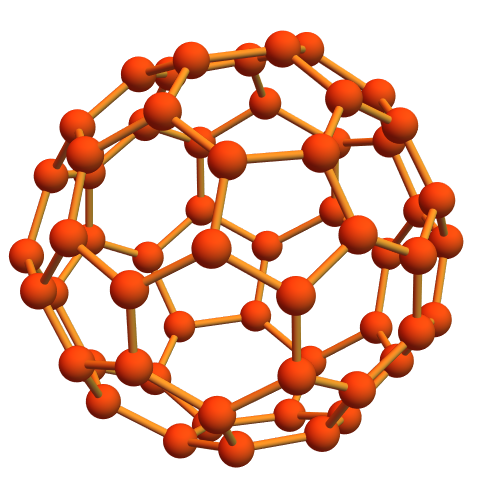} 
	\caption{\label{fig:fullerene}
\small		A fullerene molecule.\normalsize}
\end{figure}

\part{Appendices}
% Appendix A

\chapter{Appendix A} % Main appendix title

\label{AppendixA} % For referencing this appendix elsewhere, use \ref{AppendixA}

\section*{Renormalization of Euclid Action}
In this appendix we show the renormalization of Euclid Action \eqref{eq:action} for fluid and gravity using the metric \eqref{eq:schwarzchild.metric},
\begin{equation}
S=\frac{1}{2k^2}\int \left(\left( R+\frac{6}{L^2}\right)+L_{fluid}\right)\sqrt{g}d^{4}x\,,
\end{equation}
where $L_{fluid}=\frac{p(r)}{k^2 L^2}$.

Is easy to see that the euclidean action take this form:
\begin{equation}
S^{E}= \frac{L^2}{k^2}\int e^{\frac{\eta(r)}{2}}r^2 (p(r)+\rho(r))\sin(\theta)d^{4}x\,,
\end{equation}
plus this boundary extra terms,
\begin{equation}\label{bt1}
BT1(r)=-\frac{L^2}{2k^2}\int e^{\frac{\eta(r)}{2}} r(r+r^3-2M(r))\sin(\theta)\eta' d^{3}x \,,
\end{equation}
\begin{equation}\label{bt2}
BT2(r)=\frac{L^2}{k^2}\int e^{\frac{\eta(r)}{2}} r \sin(\theta) M'(r) d^{3}x \,,
\end{equation}
\begin{equation}\label{bt3}
BT3(r)=-\frac{L^2}{k^2}\int  \sin(\theta)  M(r) e^{\frac{\eta(r)}{2}}  d^{3}x \,,
\end{equation}
\begin{equation}\label{bt4}
BT4(r)=-\frac{L^2}{k^2}\int  \sin(\theta)  r^3 e^{\frac{\eta(r)}{2}}  d^{3}x \,,
\end{equation}
where $'=\frac{d}{dr}$. Now including Gibbons,Hawking and York(GHY) boundary term plus a counter term to save the divergences \eqref{bt1}, \eqref{bt2}, \eqref{bt3} and \eqref{bt4}.

\begin{equation}
S^{boundary}= S_{GHY}(\gamma_{\mu \nu})  +  S_{ct}(\gamma_{\mu \nu}) \,,
\end{equation}
using \cite{cita66} for AdS4.
\begin{equation}\label{sboundary}
S^{boundary}= \frac{1}{k^2} \int \sqrt{-\gamma}\textit{K} d^3x -  \frac{2}{k^2 L} \int \sqrt{-\gamma}\left( 1-\frac{L^2}{4}R \right) d^3x \,,
\end{equation}
where $\textit{K}$ is the trace of the extrinsic curvature of the boundary, and $R$ is the Ricci scalar refer to the boundary metric,
\begin{eqnarray}\nonumber
S_{GHY}(\gamma_{\mu \nu}) =
\frac{1}{k^2} \int\frac{d^3x}{2r} \!\!\!\!\!\! & \Bigg\{\Big(r(r^3+r-2M(r))\eta'(r)+6r^3-2M'(r)r+4r\\ & -6M(r)\Big) L^2e^{\frac{\eta(r)}{2}}\sqrt{r^2+1-\frac{2M(r)}{r}}\sin(\theta)\Bigg\} \,.
\end{eqnarray}
If we take the terms that kill the divergences BT1 and BT2, which are:
\begin{equation}
kBT1(r)=\frac{L^2}{2k^2} \int e^{\frac{\eta(r)}{2}}\frac{r(r^3+r-2M(r))\sin(\theta) \eta'(r)\sqrt{r^2+1-\frac{2M(r)}{r}}}{r} d^3x\,,
\end{equation}
\begin{equation}
kBT2(r)= \frac{-L^2}{k^2} \int\frac{e^{\frac{\eta(r)}{2}}\sqrt{r^2+1-\frac{2M(r)}{r}}r \sin(\theta) M'(r)}{r} d^3x \,.
\end{equation}
so when $r\rightarrow\infty$, $kBT1=-BT1$ and $kBT2=-BT2$. 

Therefore, the remaining terms of GHY are:
\begin{equation}\label{sghy}
S_{GHY}(\gamma_{\mu \nu}) =
\frac{L^2}{2k^2} \int\frac{(6r^3+4r-6M(r))e^{\frac{\eta(r)}{2}}\sqrt{r^2+1-\frac{2M(r)}{r}}\sin(\theta)}{r} d^3x \,.
\end{equation}
Looking to the counter term in \eqref{sboundary},
\begin{equation}\label{sct}
S_{ct}(\gamma_{\mu \nu})=- \frac{2L^2}{k^2} \int e^{\frac{\eta(r)}{2}} \sqrt{r^2+1-\frac{2M(r)}{r}} r^2 \sin(\theta)\left( 1-\frac{1}{2r^2} \right) d^3x\,,
\end{equation}
adding \eqref{sghy} to \eqref{sct},
\begin{equation}
S_{ct}(\gamma_{\mu \nu})+S_{GHY}(\gamma_{\mu \nu}) = \frac{L^2}{k^2} \int \frac{(r^3-3M(r)+3r)e^{\frac{\eta(r)}{2}}  \sin(\theta) \sqrt{r^2+1-\frac{2M(r)}{r}}}{r} d^3x\,,
\end{equation}
\begin{equation}
S_{ct+GHY}(\gamma_{\mu \nu})=\frac{L^2}{k^2} \int r^3\left(1-\frac{3M(r)}{r^3}+\frac{3}{r^2}\right)e^{\frac{\eta(r)}{2}}  \sin(\theta) \sqrt{1+\frac{1}{r^2}-\frac{2M(r)}{r^3}} d^3x \,,
\end{equation}
when $r\rightarrow \infty$, $ S_{ct+GHY}(\gamma_{\mu \nu})=-BT4$.  

And finally when $r\rightarrow \infty$, $BT3=M(R)$ give the total mass.

\chapter{Appendix B}\label{AppendixB}

\section*{Details on the bulk state}
\label{sec:background}
\paragraph{The model}
\label{sec:model}

We want to describe the thermodynamics of a very large number of neutral self-gravitating fermions in equilibrium in $3+1$ dimensions in global AdS space-time. We approximate the dynamics with a perfect fluid coupled to the gravitational field. We use the action
\begin{equation}
	S= S_{\sf Gravity}+S_{\sf Fluid}\,.
	\label{eq:action}
\end{equation}
Here the gravity part reads
\begin{equation}
	S_{\sf Gravity}= \frac{1}{16\pi G}\int d^4x\sqrt{-g}\left(R-2\Lambda\right)\,,
	\label{eq:action.gravity}
\end{equation}
where $\Lambda=-3/L^2$ being $L$ the AdS length, and we chose natural units such that $\hbar=c=k_{B}=1$ throughout this work. On the other hand the perfect fluid part can be written as a Schutz action
\begin{equation}
	S_{\sf Fluid}= \int d^4x\sqrt{-g}\left(-\rho +\sigma\, u^\mu\left(\partial_\mu\phi+\theta\,\partial_\mu s\right)+\lambda \left(u_\mu u^\mu+1\right)\right)\,,
	\label{eq:action.fluid}
\end{equation}
where $\rho$ is the fluid energy density taken as a function of the fluid particle density $\sigma$ and $s$ the entropy per particle, while $u_\mu$ is its four-velocity. The variables  $\phi$, $\theta$ and $\lambda$ are auxiliary Lagrange multipliers, enforcing that the particle number $\sigma$ and the entropy density $\sigma  s$ are conserved, and that the four-velocity $u_\mu$ is a time-like unit vector.
The resulting equations of motion read
\begin{eqnarray}
	\sigma\,\left(\partial_\mu\phi+\theta\,\partial_\mu s\right)&=&-\lambda u_\mu\,,
	\label{eq:delta.umu}
	\\
	u^\mu\left(\partial_\mu\phi+\theta\,\partial_\mu s\right)&=&\mu\,,
	\label{eq:delta.sigma}
	\\
	u_\mu u^\mu&=-1\,,
	&
	\label{eq:delta.lambda}
	\\
	\partial_\mu\left(\sigma u^\mu\right)&=&0\,,
	\label{eq:delta.phi}
	\\
	\partial_\mu\left(\sigma s\, u^\mu\right)&=&0\,,
	\label{eq:delta.theta}
	\\
	u^\mu\partial_\mu \theta\, &=&-T\,,
	\label{eq:delta.s}
	\\
	G_{\mu \nu}+\Lambda g_{\mu \nu}&=&8\pi G \left(P g_{\mu \nu}+(P+\rho)\,u_{\mu}u_{\nu}\right)\,.
	\label{eq:einstein}
\end{eqnarray}
In these equations we have defined the local chemical potential $\mu \equiv {d\rho}/{d\sigma}$, local temperature $\sigma \,T \equiv {d\rho}/{ds}$, and pressure $P \equiv-\rho +\mu\sigma$. The equations in the first two lines fix the auxiliary variables, implying $\lambda=\mu\sigma$ and $\partial_\mu\phi+\theta\,\partial_\mu s=-\mu \,u_\mu$. The third imposes that the four-velocity is unitary and time-like. The next two lines correspond to the conservation of the particle current $\sigma u^\mu$ and the entropy current $\sigma s u^\mu$. Then we define the temperature as the proper time derivative of the ``thermasy'' $\theta$. The last line contains the Einstein equations.

The above equations of motion need to be supplemented with a explicit dependence of $\rho$ in $\sigma$ and $s$. In the limit $mL \gg 1$, in which there is a large number of particles within one AdS radius \cite{cita16}, we can define such relation implicitly by
\begin{eqnarray}
	\label{eq:density}
	\rho&=&\frac{g}{8\pi^3}\int f(p)\,\sqrt{p^2+m^2}\,d^3p\,,
	\\
	\label{eq:pressure}
	P&=&\frac{g}{24\pi^3}\int f(p)\,\frac{p^2}{\sqrt{p^2+m^2}}\,d^3p\,,
\end{eqnarray}
where $g$ is the number of fermionic species or spin degeneracy, the integration runs over all momentum space, and $f(p)$ is the Fermi distribution for a fermion of mass $m$ with local temperature  $T$ and local chemical potential $\mu$
\begin{equation}
	f(p)=\frac{1}{e^{\frac{\sqrt{p^2+m^2}-\mu}{T}}+1}\,.
\end{equation}
\paragraph{The Ansatz}
\label{sec:ansatz}
We solved the above equations of motion with a stationary spherically symmetric ``neutron star'' Ansatz with the form
\begin{eqnarray}\label{metric}
	&&ds^{2}=L^{2}(-e^{\nu(r)}dt^{2}+e^{\lambda(r)}dr^{2}+r^2d\Omega^2_2)\,,
	\\
	&&u^\mu = u(r)\partial_t\,,
\end{eqnarray}
where $d\Omega^2_2=d\vartheta^2+\sin^2\!\vartheta\, d\varphi^2$ is a two-sphere. The local temperature $T$ and chemical potential $\mu$ are then radial functions.
Defining the functions $\tilde M$ and $\chi$  by
\begin{equation}
	e^{\lambda}=\left( 1-\frac{2\tilde{M}}{r} +r^2 \right)^{-1}\,,
\end{equation}
\begin{equation}
	e^{\nu}=e^{\chi } \left( 1-\frac{2\tilde{M}}{r} +r^2 \right)\,,
\end{equation}
equations \eqref{eq:delta.umu}-\eqref{eq:einstein} become the thermodynamic equilibrium conditions of Tolman and Klein
\begin{equation}\label{eq:tolman}
	e^{\frac{\nu}{2}}\tilde{T}={\rm constant}\,,
\end{equation}
\begin{equation}\label{eq:klein}
	e^{\frac{\nu}{2}}\tilde{\mu}={\rm constant}\,,
\end{equation}
and the Einstein equations
\begin{equation}\label{eq:M}
	\frac{d\tilde{M}}{d r}=4\pi r^2\tilde{\rho}\,,
\end{equation}
\begin{equation} \label{eq:nu}
	\frac{d {\chi}}{dr}=8\pi r\left(\tilde{P}+\tilde{\rho}\right)e^{\lambda}\,.
\end{equation}
In the above expressions to we have written the dimensionless combinations $\tilde{T}= T/m$ and $\tilde{\mu}=\mu/m$ for the temperature and chemical potential. The constants
are obtained by evaluating the expression
%$\tilde{\mu_c}$ and $\tilde{T_c}$ are the dimensionless chemical potential and temperature measured
at a reference point, conventionally taken at $r=0$. Moreover, we have re-scaled the density $\tilde \rho=GL^2 \rho$ and pressure $\tilde P=GL^2 P$. They are obtained from \eqref{eq:density} and \eqref{eq:pressure} rewritten in terms of the variable $\epsilon=\sqrt{1+p^2/m^2}$, resulting in the expressions
\begin{equation}
	\label{eq:density.dimensionless}
	\tilde \rho=\gamma^2\int_1^\infty
	\frac{\epsilon^2\sqrt{\epsilon^2-1}}{e^{\frac{\epsilon-\tilde{\mu}}{\tilde{T}}}+1}\,d\epsilon\,,
\end{equation}
\begin{equation}
	\label{eq:pressure.dimensionless}
	\tilde P=
	\frac{\gamma^2}3\int_1^\infty
	\frac{\left(\sqrt{\epsilon^2-1}\right)^3}{e^{\frac{\epsilon-\tilde{\mu}}{\tilde{T}}}+1}\,d\epsilon\,,
\end{equation}
where $\gamma^2=gGL^2m^4/2\pi^2$.

By expanding the equations \eqref{eq:M} and \eqref{eq:nu} around the center of the configuration $r=0$, we obtain the boundary conditions that correspond to a regular metric, as
\begin{eqnarray}
	\tilde{M}(0)=0,
	\label{eq:boundary.mass}\\
	\chi(0)=0,\\
	\tilde{T}(0)=\tilde{T_c},\\
	\tilde{\mu}(0)=\mu_c\equiv \Theta_c\tilde{T_c}+1 \,.
	\label{eq:boundary.mu}
\end{eqnarray}
Here $\Theta_c$ is called the ``central degeneracy'', and we used it as a way to parameterize the central chemical potential. Families of solutions are then indexed by the parameters $(\tilde T_c, \Theta_c. \gamma^2)$.

\chapter{Appendix C}\label{AppendixC}
\section*{Details on space-like geodesics}
\label{subsec:geodesics}

In order to probe the resulting gravitational background, we study the scattering of a massive particle with an Euclidean worldline. The trajectories are obtained by extremizing the Euclidean particle action
\begin{equation}
	S_{\sf EWL}  = {\sf m}\int d\tau_E\,\sqrt{g_{\mu\nu} \, {x'}^\mu {x'}^\nu}\,.
	\label{eq:action.particle.no.gauge}
\end{equation}
Here ${\sf m}$ is the mass of the particle, $x^\mu(\tau_E)$ describes the geodesic in terms of an Euclidean parameter $\tau_E$, and a prime ($'$) means a derivative with respect to ${\tau_E}$.
Evaluated on our Ansatz  \eqref{metric} this action  reads
\begin{equation}
	S_{\sf EWL} = {\sf m}L\int d\varphi\,\sqrt{r^2+e^{\lambda(r)}r'^{2}}\,,
	\label{eq:action.particle.gauge.fixed}
\end{equation}
where we specialized to constant time $t'=0$ and chose coordinates such that the trajectory lies in the equator of the sphere $\vartheta=\pi/2$. Moreover we fixed the reparametrization gauge as $\tau_E=\varphi$.
The resulting system is invariant under shifts of the Euclidean parameter $\varphi$, resulting in a conserved quantity
\begin{equation}
	r_* = \frac{r^2}{\sqrt{r^2+e^{\lambda(r)}r'^{2}}}\,.
	\label{eq:r.star}
\end{equation}
The value of $r_*$ coincides with the value of the coordinate $r$ at tip of the trajectory, defined as the point at which $r'=0$. It can be used to label different geodesics by their minimum approach $r_*$ to the center of the geometry.

%
%\begin{figure}[ht]
%\centering
%\vspace{-3cm}
%\includegraphics[width=0.8\textwidth]{sphere3.pdf}
%\vspace{-2.2cm}
%\caption{\label{fig:sphere} We study the scattering problem of a massive Euclidean particle entering the geometry from infinity, approaching the neutron star up to a tip radius $r_*$, and then moving again to the asymptotic region, spanning an angle $\Delta \varphi$.}
%\end{figure}
%

We are interested in geodesics starting at a very large radius $r_\epsilon$, falling into the geometry up to a minimum radius $r_*$, and then bouncing back into the asymptotic region, spanning a total angle $\varphi$ (see Fig.~\ref{fig:sphere}). Then, our boundary conditions are
\begin{equation}
	\left.r\right|_{\varphi=0}=\left.r\right|_{\varphi=\varphi}=r_\epsilon\,,
\end{equation}
here $\varphi$ is can be calculated according to
\begin{equation}
	\varphi = \int d\varphi=\int \frac{dr}{r'}\,.
\end{equation}
Solving \eqref{eq:r.star} for the velocity $r'$ and using the fact that the trajectory is symmetric around its tip, we get $\varphi$ as a function of $r_*$, in the form
\begin{equation}
	\varphi = 2r_*\int_{r_*}^{r_\epsilon} dr\,\frac{ e^{\frac{\lambda(r)}2}}{
		r \sqrt{
			r^2-r_*^2
		}
	}\,.
	\label{eq:delta.varphi}
\end{equation}
For $r_*=0$ we have no scattering and then $\varphi=\pi$. On the other hand in the limit of very large  $r_*$ we get $\varphi=0$, implying a backward scattering. In the intermediate region, the behaviour of $\varphi$ can be either monotonic or non-monotonic. In the last case, the same angle $\varphi$ is spanned by geodesics with different values of the minimum approach radius $r_*$.

\begin{figure}[ht]
	\centering
	\includegraphics[width=0.7\textwidth]{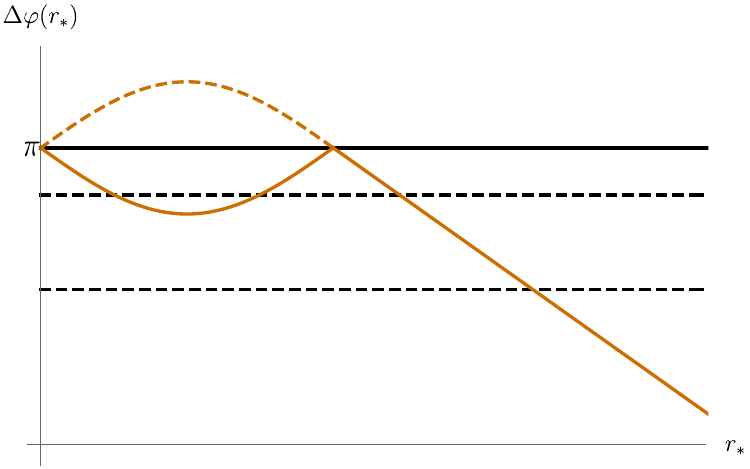}
	\put(-130,112){\scriptsize $\varphi$ with multiple values of $r_*$ \normalsize}
	\put(-281,81){\scriptsize $\varphi$  with a single value of $r_*$ \normalsize}
	\caption{\label{fig:angle-generic} The angle $\varphi$ as a function of the tip position $r_*$. Whenever the integral \eqref{eq:delta.varphi} returns a value larger than $\pi$ (dotted curve), it must be mapped to $2\pi -\varphi$ (continuous curve). Notice that this results in a non-monotonic behavior, with three different values of $r_*$ returning the same value of $\varphi$, for $\varphi$ close enough to $\pi$.
	}
\end{figure}

Due to the symmetries of the problem, a value of $\varphi$ larger than $\pi$ is not physically different from the value smaller than $\pi$ obtained by the transformation $\varphi\to 2\pi-\varphi$. This is evident in Fig.~\ref{fig:sphere}. Thus, in what follows we restrict $\varphi$ to values smaller than $\pi$, applying such transformation whenever the integral \eqref{eq:delta.varphi} returns a value larger than $\pi$. This may result in a non-monotonic behaviour of $\varphi$ as a function of $r_*$, see Fig.~\ref{fig:angle-generic}.

\chapter{Appendix D}\label{AppendixD}

\section*{Details on the correlator of a scalar operator}
\label{sec:correlators}
In order to probe the field theory on the boundary, we consider  a scalar operator, and we used the dictionary of the AdS/CFT correspondence to evaluate its two-point correlator. We consider two points in the boundary sphere, separated by an angle $\Delta \varphi$ at fixed Euclidean time. In the limit of a large conformal dimension $\Delta\equiv {\sf m}L$, the correlator is given as
\begin{equation}
	\langle{\cal O}(\varphi){\cal O}(0)\rangle=\lim_{r_{\epsilon}\rightarrow\infty}r_{\epsilon}^{2{\sf m}L}
	e^{-S^{E~ \sf on-shell}_{\sf Particle}(\varphi)}\,,
	\label{eq:correlator}
\end{equation}
where the exponent is the Euclidean particle action \eqref{eq:action.particle.no.gauge} evaluated on shell, and $r_{\epsilon}$ is an UV bulk regulator, whose power is included in the pre-factor in order to get a finite result. Solving \eqref{eq:r.star} for $r'$ and plugging back into \eqref{eq:action.particle.gauge.fixed}, we obtain the Euclidean action as
\begin{equation}
	S^{E~ \sf on-shell}_{\sf Particle}=2{\sf m}L\int_{r_{*}}^{r_{\epsilon}}dr
	\frac{r e^{\frac{\lambda(r)}{2}}}{r \sqrt{r^2-r_{*}^2}}\,,
	\label{eq:action.particle.on.shell}
\end{equation}
where we have included the same cutoff $r_{\epsilon}$.

Equation \eqref{eq:action.particle.on.shell} gives the on-shell action as a function of the position of tip if the geodesic $r_*$. Equation \eqref{eq:delta.varphi} on the other hand, provide $\varphi$ as a function of the same variable. This allows us to parametrically plot the the on-shell action as a function of the angular separation. Interestingly, for the cases in which $\varphi$ is non-monotonic as a function of $r_*$ that we studied in section \ref{subsec:geodesics}, there are three values of $S^{E~ \sf on-shell}_{\sf Particle}$ for a single value of $\varphi$, {\em i.e.} the Euclidean particle action become multi-valued. The correlator is then given by the absolute minimum of the Euclidean particle action, which corresponds to the smaller branch of the multi-valued function.

\chapter{Appendix E}\label{AppendixE}

\section*{Normal modes of a scalar field in global AdS$_4$}
In this appendix we determine the normal modes of a scalar field in global AdS$_4$,
\begin{align}
	ds^2
	=L^2\left\{
	-(1+r^2)\,dt^2+(1+r^2)^{-1}\,dr^2
	+r^2\,d\Omega^2\right\}\,.
\end{align}
The solutions of the equation of motion
\begin{align}
	\left(\Box-{\sf m}^2\right)\Phi(t,r,\Omega)=0\ ,
\end{align}
can be written as
\begin{align}
	\Phi(t,r,\Omega)=e^{-i\omega t}\,Y_{\ell m}(\Omega)\,R_{\omega\ell}(r)\,,
\end{align}
where the radial dependence is determined by
\begin{align}
	(1+r^2)\,R_{\omega\ell}''
	+\left(\frac2r+4r\right)\,R_{\omega\ell}'
	+\left(\frac{\omega^2}{1+r^2}
	-\frac{\ell(\ell+1)}{r^2}
	-L^2{\sf m}^2
	\right)R_{\omega\ell}=0\,.
\end{align}
Any solution to this equation which is regular at the origin is proportional to
\begin{align}
	R_{\omega\ell}(r)=
	\ \frac{r^\ell}{({1+r^2})^\frac{\omega}2}
	\ \mbox{}_2F_1\left(\tfrac{\ell+\Delta_-- \omega}2,\tfrac{\ell+\Delta_+- \omega}2;\ell+\tfrac32;-r^2\right)\,,
\end{align}
with
\begin{align}
	\Delta_\pm=\frac32\pm\sqrt{\frac94+{\sf m}^2{L}^2}\,.
\end{align}
The behavior of the radial part at large $r$ can be read from the relation
\begin{align}\label{relhyper}
	\mbox{}_2F_1(a,b;c;-z)
	&=\tfrac{\Gamma(b-a)\Gamma(c)}{\Gamma(b)\Gamma(c-a)}
	\ z^{-a}
	\ \mbox{}_2F_1(a,a-c+1;a-b+1;-z^{-1})+\mbox{}\nonumber\\[2mm]
	&\mbox{}+\tfrac{\Gamma(a-b)\Gamma(c)}{\Gamma(a)\Gamma(c-b)}
	\ z^{-b}
	\ \mbox{}_2F_1(b,b-c+1;b-a+1;-z^{-1})\,.
\end{align}
Therefore, as $r\to\infty$,
\begin{align}\label{behatinf}
	R_{\omega\ell}(r)\sim
	a_{\omega\ell}\, r^{-\Delta_-}
	+b_{\omega\ell}\, r^{-\Delta_+}\,,
\end{align}
where
\begin{align}
	a_{\omega\ell}&=\tfrac{\Gamma\left(\frac12(\Delta_+-\Delta_-)\right)\,\Gamma\left(\ell+\frac32\right)}
	{\Gamma\left(\frac12(\ell+\Delta_+-\omega)\right)\,\Gamma\left(\frac12(\ell+\Delta_++\omega)\right)}\,,
	\label{A}\\[2mm]
	b_{\omega\ell}&=\tfrac{\Gamma\left(-\frac12(\Delta_+-\Delta_-)\right)\,\Gamma\left(\ell+\frac32\right)}
	{\Gamma\left(\frac12(\ell+\Delta_--\omega)\right)\,\Gamma\left(\frac12(\ell+\Delta_-+\omega)\right)}\,.
\end{align}
Notice that we have omitted the subleading contributions of the first term in \eqref{relhyper}.

If the mass parameter ${\sf m}$ is real and non-vanishing, then $\Delta_-<0$ and the leading behavior in \eqref{behatinf} diverges as $r\to\infty$, unless $a_{\omega\ell}=0$. This determines the oscillation frequencies $\omega_n$ of the normal modes:
\begin{align}
	\pm\omega_n=\Delta_++\ell+2n
	\qquad n=0,1,2,\ldots
\end{align}

\chapter{Appendix F}\label{AppendixF}
\label{app:conventions}
\subsection*{The Dirac operator on $S^2$}
\label{app:DiracopS2}

The eigenvalue problem of the Dirac operator on the sphere was tackled in many references in the past. Here we give a brief review of the case of interest, namely the two-dimensional sphere $S^2$, following closely \cite{cita63}.

We start by writing the metric
\bea
d\Omega^2&=&d\vartheta^2+\sin^2\!\vartheta \,d\varphi^2\cr
&=&\delta_{IJ}\,\hat\omega^I\,\hat\omega^J \qquad\qquad\qquad I,J\in\{1,2\}
\eea
where in the second line we have introduced a local frame, {\em i.e.} a sweibein basis $\hat \omega^I$ and dual vector field $\hat e_I$, which satisfy $\hat \omega^I(\hat e_J) = \delta^I{}_J$. These objects can be written explicitly as
\begin{align}
\hat\omega^1 &=  d\vartheta\qquad  \qquad &\hat e_1 =  \partial_\vartheta\qquad \ \cr
\hat\omega^2  &=    \sin\vartheta\, d\varphi  \qquad  \qquad&\hat e_2 = \frac1{ \sin\vartheta}\,\partial_\varphi
\end{align}
The resulting spin connection reads 
\bea
\hat\omega^1{}_2=-\cos\vartheta\,d\varphi
\eea
A suitable choice of gamma-matrices satisfying $\{\gamma^I,\gamma^J\} = 2\,\delta^{IJ}$ is
\bea\label{eq:gamma12}
\gamma^1=\sigma_1\qquad\qquad,\qquad\quad 
\gamma^2=\sigma_2
\eea
With this choice, 
the Dirac operator reads
\be
\hat{\slashed{\nabla}} 
= \sigma_1 \,\left(\partial_\vartheta + \frac{\cot\vartheta}{2}\right)+ \sigma_2\,\frac1{\sin\vartheta}\,\partial_\varphi
\ee
resulting in the eigenvalue equation 
\be\label{eq:eigenDeq}
\hat{\slashed{\nabla}}\psi_\alpha(x,\varphi) = -i\,\alpha\;\psi_\alpha(x,\varphi)
\ee
It results convenient to introduce the coordinate: $x= \cos\vartheta \in[-1,+1]$. If we now write a generic spinor of defined Fourier mode $m\in\Z+{1}/{2}$ as,
\be\label{eq:Psim}
\psi_{\alpha m}(x,\varphi) = \frac{e^{i\,m\,\varphi}}{\sqrt{2\pi}}\;
\left(1-x^2\right)^{-\frac{1}{4}}\;
\left(\begin{array}{c}\phi_{\alpha m}^+(x)\\
	\phi_{\alpha m}^-(x)\end{array}\right)
\ee
we get the following coupled system for the components $\phi_{m}^\pm(x)$,
\be\label{eq:coupledDE}
\sqrt{1-x^2}\;\phi_{\alpha m}^\pm{}'(x) \pm \frac{m}{\sqrt{1-x^2}}\;
\phi_{\alpha m}^\pm(x) = i\,\alpha\;\phi_{\alpha m}^\mp(x)
%\cr\sqrt{1-x^2}\;\phi'{}_{\lambda m}^-(x) - \frac{m}{\sqrt{1-x^2}}\;
%\phi_{\lambda m}^-(x) &=& i\,\lambda\;\phi_{\lambda m}^+(x)
\ee
If $\alpha=0$ it is straight to integrate them, obtaining singular solutions at the poles of the sphere $x=\pm 1$ (even for $m=0$ the constant solution is singular, in view of (\ref{eq:Psim})). 
So there is no eigenfunction with null eigenvector\footnote{
This is a particular case of a general result known as the Lichnerowicz theorem, which roughly states that the Dirac operator does not admit zero modes on manifolds of positive curvature  \cite{cita64}.
}.
Then we can safely proceed to get second order differential equations, introducing in the down (respectively up) equation the value of 
$\phi_{\alpha m}^-(x)$ (respectively $\phi_{\alpha m}^+(x)$) giving by the up  (respectively down) equation. 
We obtain,
\be\label{eq:2order1}
(1-x^2)\,\phi_{\alpha m}^\pm{}''(x) - x\,\phi_{\alpha m}^\pm{}'(x)
+\left( \alpha^2 + \frac{m(-m\pm x)}{1-x^2}\right)\,
\phi{}_{\alpha m}^\pm(x) =0
\ee
As a last step, we put the above equation in a recognizable  hypergeometric form, 
through  the redefinitions
\be\label{eq:phiP}
\phi{}_{\alpha m}^\pm(x) \equiv (1+x)^{\frac{1}{4}+\frac{|m\pm\frac{1}{2}|}{2}}\;
(1-x)^{\frac{1}{4}+\frac{|m\mp\frac{1}{2}|}{2}}\;
P_{\alpha m}^\pm(x)
\ee

and then by inserting (\ref{eq:phiP}) in (\ref{eq:2order1}) we get, 

\be
\label{eq:2order2}
(1-x^2)\,P_{\alpha m}^\pm{}''(x) 
+ 2\,\left(a_\pm-b_\pm\,x\right)P_{\alpha m}^\pm{}'(x)
+\left(\alpha^2- \left(b_\pm-\frac{1}{2}\right)^2\right)\,
P_{\alpha m}^\pm(x)=0\nonumber
\ee
\be\ee
where we have defined,
\be
a_\pm\equiv\frac{1}{2}\left(|m\pm{1}/{2}|-|m\mp{1}/{2}|\right)\qquad,\qquad
b_\pm\equiv\frac{1}{2}\left(|m\pm{1}/{2}|+|m\mp{1}/{2}|\right)+1
\ee
These equations have as unique regular solutions the Jacobi polynomials 
$P_{j-|m|}^{\left(b_\pm-a_\pm-1,b_\pm+a_\pm-1\right)}(x)$ with half-integer index $j\geq |m|$, provided that the eigenvalue has the form
\be\label{eq:eigenvalue}
\alpha_{j\epsilon} = \epsilon\,\left(j+\frac{1}{2}\right)\qquad \qquad \epsilon^2=1
\ee
Summarizing, the most general regular solution (\ref{eq:Psim}) to the Dirac equation (\ref{eq:eigenDeq}) has the form
\be\label{eq:Psigral}
\psi_{jm\epsilon}(\vartheta,\varphi) = \frac{e^{i\,m\,\varphi}}{\sqrt{2\pi}}\;
\left(\begin{array}{c}
C_{jm\epsilon}^+\;(1-x)^\frac{|m-\frac{1}{2}|}{2}\;(1+x)^\frac{|m+\frac{1}{2}|}{2}\;
P_{j-|m|}^{(|m-\frac{1}{2}|,|m+\frac{1}{2}|)}(x)\\
C_{jm\epsilon}^-(1-x)^\frac{|m+\frac{1}{2}|}{2}\;(1+x)^\frac{|m-\frac{1}{2}|}{2}\;
P_{j-|m|}^{(|m+\frac{1}{2}|,|m-\frac{1}{2}|)}(x)
\end{array}\right)
\ee
with the eigenvalue given in (\ref{eq:eigenvalue}).
However the constants $C_{jm\epsilon}^\pm$ are not independent, 
because to arrive to \eqref{eq:2order1} from \eqref{eq:coupledDE} 
one has to express one component in terms of the other. 
Compatibility leads to the relation, 
\be\label{eq:C-C+}
C_{jm\epsilon}^- = i\,\sign(m)\,\epsilon\;C_{jm\epsilon}^+
\ee
The solution is then determined up to un overall constant $C_{jm\epsilon}^+$. 
Its modulus can be fixed by imposing normalization with respect to the usual scalar product
\be
(\psi_{jm\epsilon};\psi_{j'm'\epsilon'})= 
\int_{-1}^{+1}dx\;\int_{0}^{2\pi}d\varphi\;
\psi^\dagger_{jm\epsilon}(x,\varphi)\psi_{j'm'\epsilon'}(x,\varphi) 
\equiv \delta_{jj'}\;\delta_{mm'}\;\delta_{\epsilon\epsilon'}
\ee
In what the phase respects, we note that by adding $\gamma^0 = -i\,\sigma_3$ to \eqref{eq:gamma12} we get a Dirac representation in $2+1$ dimensions, 
$\{\gamma^\mu ;\gamma^\nu\}= 2\,\eta^{\mu\nu}$. 
By definition  the time reversal transformation obeys: 
$T\,\gamma^\mu\,T^{-1} = \pm\eta_{\mu\mu}\gamma^{\mu*}$, and so we can take it as  $T=-i\,\sigma_2$. 
The phase can be fixed by imposing the complex conjugation rule given by
\be
{}^T\psi_{jm\epsilon} = T\;\psi_{jm\epsilon}^*
\equiv (-)^{j+m}\;\psi_{j\,(-m)\epsilon}
\quad\longleftrightarrow\quad
-i\,\sigma_2\;\psi_{jm\epsilon}^*  
= (-)^{j+m}\;\psi_{j\,(-m)\epsilon}
\ee
As a final remark, we note that the solutions at fixed $\epsilon$, i.e.  positive or negative eigenvalues $\pm(j+\frac{1}{2})$, are encoded in the half-integer representations $R_j$ of the isometry group $SU(2)$. 
This is so because there exist operators $J_a$ such that 
$[\slashed\nabla; J_a] =0$. 
But there also exists the parity operator $P:\{P;\gamma_a\}=0$ ($P=\sigma_3$ in our conventions), such that $[P; J_a] =0$. 
So we can have irreducible representations of $SU(2)$ of definite parity. 
In fact, the up and down components of (\ref{eq:Psigral}) are.
However $\slashed\nabla$ and $P$ does not commute, instead  
$\{\slashed\nabla; P\}=0$. 
The action of $P$ is then changing the sign of 
$\lambda_{j\epsilon}$, 
(if you like, act with $P$ on (\ref{eq:Psigral}) and recall (\ref{eq:C-C+})). 
Thus, the doubling by parity is at the origin of the existence of eigenvalues positives as well as negatives in the same $R_j$-representation. 
See \cite{cita63} for details.

% References:
\include{Chapters/references}

\backmatter

% Bibliography:
%\bibliographystyle{unsrt}

%\addcontentsline{toc}{chapter}{Bibliography}
\bibliographystyle{unsrt}
\bibliography{refs}

% Summary
\chapter*{Summary}
\markboth{Summary}{Summary}
\addcontentsline{toc}{chapter}{Summary}

\begin{flushright}
\rightskip=0.8cm\textit{``Todo problema tiene una solución aunque esa solución este en otro problema.''} \\
\vspace{.2em}
\rightskip=.8cm---TC
\end{flushright}
\vspace{1em}

The PhD thesis presents my studies on self-gravitating systems and their relation with gauge gravity duality. This work has resulted in the publication of two papers in international journals, and a third paper is currently being completed. All three papers are discussed in Chapter~\ref{ch:carlos_nico}, Chapter~\ref{ch:nmodes}, and Chapter~\ref{ch:fermions} of this thesis.

Last but not least, during my PhD, I also worked on other astrophysical topics. I have published several papers in the bulletin of the Argentinean Astronomical Association.

\href{https://orcid.org/0000-0003-2285-3103}{Tobías Canavesi Orcid}

% Empty page
%\thispagestyle{empty}

\end{document}